\documentclass[11pt]{article}
\usepackage[dvipdfmx]{graphicx}
\usepackage{cite}

\newcommand{\aterm}[1]{{\bm a}_{#1}}
\newcommand{\avec}[2]{ A_{#1_{#2}}  }
\newcommand{\abar}[1]{ a_{\bar{#1}}  }
\newcommand{\ddt}[1]{\frac{d #1}{d ( \log \mu )}}

\newcommand{\yukawa}[1]{{\bm y}_{#1}}

\newcommand{\yukbar}[1]{ y_{\bar{#1}}}

\usepackage{color}
\usepackage{braket}

\newcommand{\beq}{\begin{equation}}
\newcommand{\eeq}{\end{equation}}
\def\lrp#1{\left( #1 \right)}

\newcommand{\dm}[1]{m_{\chi_{\Phi}}^{#1}}
\newcommand{\sfer}[2]{m_{\tilde{f}_{#1}}^{#2}}
\usepackage{bm}
\usepackage{amsmath,amssymb}
\usepackage[top=30truemm,bottom=30truemm,left=25truemm,right=25truemm]{geometry}
 \allowdisplaybreaks[4]

\makeatletter

\@addtoreset{equation}{section}
\makeatother
  \makeatletter
    
    \@addtoreset{figure}{section}
  \makeatother
   \makeatletter
    
   \@addtoreset{table}{section}
   \makeatother
\title{\bf Flavor structure, Higgs boson mass and dark matter in
supersymmetric model with vector-like generations}

\author{
	Tetsutaro Higaki$^a$\footnote{email: thigaki@rk.phys.keio.ac.jp},~
	Michinobu Nishida$^a$\footnote{email: blacky@a6.keio.jp}~ and
      Naoyuki Takeda$^b$\footnote{email: takedan@keio.jp}, 
      \\[3mm]
$^a${\it Department of Physics, Keio University,}\\{\it Yokohama 223-8522, Japan} \\
$^b${{\it Research and Education Center for Natural Sciences,
Keio University,}}\\
{{\it Yokohama 223-8521, Japan}}
      }

\date{}
\begin{document}
\baselineskip=17.3pt
\maketitle
\setcounter{page}{1}
\begin{abstract}
\normalsize 
\noindent 
We study a supersymmetric model in which the Higgs mass, the muon anomalous magnetic moment and the dark matter are simultaneously explained with extra vector-like generation multiplets. For the explanations, non-trivial flavor structures and a singlet field are required. In this paper, we study the  flavor texture by using the Froggatt-Nielsen mechanism, and then find realistic flavor structures
which reproduce the Cabbibo-Kobayashi-Maskawa matrix and fermion
masses at low energy.
Furthermore, we find that the fermion component of the singlet field becomes a good candidate of dark matter.
In our model, flavor physics and dark matter are explained with moderate size couplings through renormalization group flows,
and the presence of dark matter supports the existence of just three generations in low energy scales.
%it is found that the flavor physics are closely associated with dark matter physics.
We analyze the parameter region where the current thermal relic abundance of
dark matter, the Higgs boson mass and the muon $g-2$ can be explained simultaneously.
\end{abstract}
\thispagestyle{empty}

\newpage

\newpage

\section{Introduction} \label{sec:int}
The discovery of the Higgs boson by ATLAS and CMS collaborations of the
LHC gives big impact on particle physics \cite{Aad:2012tfa}. At the experiment, the Higgs mass is confirmed to be $m_h^2=125.09\pm0.21({\rm stat.})\pm0.11({\rm syst.})\,{\rm GeV}$~\cite{Aad:2015zhl}.
By the discovery, the particles predicted by the Standard Model (SM) are experimentally confirmed.
However, there still remains various problems to be solved beyond the SM. Among them, we focus on the muon anomalous magnetic moment (muon $g-2$) and dark matter (DM) in this paper. We reveal that flavor structures are important for the issues.

Experimentally, the muon $g-2$ is reported with $\Delta a_{\mu}=(26.1\pm8.0)\times10^{-10}$~\cite{Bennett:2006fi,hagiwara:2007}. For the explanation of this result, the extension of the SM is a possibility. In~\cite{Nishida:2016lyk}, one of the present authors (M.N.) showed that the supersymmetric model with vector-like generations explains experimental results. In the analysis, the authors solved the renormalization group (RG) flow of the Yukawa matrices, and then found that a certain flavor structure of the quark and lepton sectors are required for the explanation of the muon $g-2$ and the Higgs mass. Further, in the model, a singlet scalar field is required to give large masses to the vector-like generations enough to avoid electroweak precise measurements~\cite{Agashe:2014kda}. In this paper, we show that the Yukawa structure is determined by the Froggatt-Nielsen mechanism, and the single field is a good candidate of the DM. 

By astrophysical observations such as the Galaxy rotation curves~\cite{Oh:2015xoa}, collisions of bullet clusters~\cite{Markevitch:2001ri}, or Cosmic Microwave Background~\cite{Ade:2015xua}, it is confirmed that the matter contents of the present Universe is mainly dominated by DM with the abundance
$\Omega_{\rm DM}h^2=0.1198 \pm 0.0015$~\cite{Ade:2015xua}. 
The formation of large scale structures also requires DM since it is a significant source of gravitational potential. However, there does not exist a natural candidate of DM in the SM. Thus, the extension of particle contents is needed. In our model, the superpartner of the singlet scalar field, which is called singlino, is the lightest supersymmetric particle (LSP), and it could be DM in the presence of R-parity. We show that the thermal relic abundance of the singlino field explains the DM abundance. 

Another issue relevant to this paper is the origin of the flavor structure in the quark and lepton sectors. As in the case of the Cabbibo-Kobayashi-Maskawa (CKM) matrix in the SM, our model~\cite{Nishida:2016lyk} needs Yukawa structures for the mass matrices of the quark and lepton sectors extended with vector-like generations. Especially, there should exist an appropriate flavor structure for explaining experimental values of both the Higgs mass and the muon $g-2$ through quantum corrections simultaneously. However, as in the SM, Yukawa couplings are just free parameters. We try to explain the structure by the Froggatt-Nielsen (FN) mechanism.

Froggatt and Nielsen explained the structure by assuming additional $U(1)$ symmetry called flavor symmetry~\cite{Froggatt:1978nt}. The mechanism is realized also in SUSY models~\cite{Leurer:1993gy}. In this work, we reproduce the flavor structure of the model~\cite{Nishida:2016lyk} by the FN mechanism. Then we show a charge assignment to the chiral superfields for realizing a realistic Yukawa hierarchy, which explains 
the CKM matrix and fermion masses at low energy with the parameter $\epsilon \simeq 0.33$. 
Here, $\epsilon$ is the breaking scale of FN U(1) symmetry normalized by the cutoff scale.

With an appropriate assignment of the FN charge, a SM singlet superfield $\Phi$ plays two important roles. 
One is to fix the mass scales of vector-like generations with the vacuum expectation value (VEV) of the scalar field $\langle \Phi \rangle$.
Another is its fermion component becomes a candidate of DM with $R$-parity.
In this sense, the presence of the DM supports the existence of three generations in low energy scales within our model. As seen later, we show a parameter space where one obtains the right amount of the Higgs boson mass, the muon $g-2$, and the observed relic abundance of DM in our model.

The organization of this paper is as follows.
In Section \ref{sec:mod}, we introduce our model with the flavor symmetric superpotential.
We assign the U(1) charge to each field and give possible
Yukawa structure in both quark and lepton sectors.
Then the observed CKM matrix and fermion masses can be reproduced at the $M_{Z}$ scale. Here, $M_Z \simeq 91$ GeV~\cite{Olive:2016xmw} 
is the $Z$ boson mass.
In Section \ref{sec:dm}, we shall explain a candidate of DM in our model.
Next, we give analytic equation for calculating the thermal relic abundance
of the DM. In Section \ref{sec:res}, we show 
%the result of 
the parameter region where the DM abundance, Higgs boson mass and
the muon $g-2$ within $2\sigma$ level are simultaneously explained.
The final section is devoted to the conclusion and discussion.

%%%%%%%%%%
\section{Model}  \label{sec:mod}
%%%%%%%%%%
In this section, we first give an explanation of the model proposed in~\cite{Nishida:2016lyk}. Second, we extend the model by adding a $U(1)$ flavor symmetry with the FN mechanism~\cite{Froggatt:1978nt,Leurer:1993gy}, and then show that a nontrivial flavor structure is obtained through the symmetry breaking with an appropriate charge assignment. Sizable couplings between the supersymmetric SM sector and vector-like generations significantly contribute to the Higgs boson mass and the muon $g-2$.

The original model is a extension of minimal supersymmetric standard model (MSSM) by adding a pair of vector-like generations and a SM singlet field $\Phi$~\cite{Nishida:2016lyk}. We assume that the K\"ahler potential is canonical. As usual, the superfields of the MSSM sector are given by
\begin{align}
& Q_i, \; u_i, \; d_i, \; L_i, \; e_i, \quad (i=1,\cdots,3),  \\
& H_u, \; H_d,
\end{align}
where $Q_i$ and $L_i$ are the SU(2) doublets of quarks and leptons, $u_i, d_i$ and $e_i$ are the SU(2) singlets of up-type, down-type quarks and charged leptons, respectively. The Higgs doublets are denoted by $H_u$ and $H_d$. 
In addition to these, we have other superfields of the vector-like generations and a singlet as
\begin{align}
& Q_4, \; u_4, \; d_4, \; L_4, \; e_4,  \label{eq:fourth} \\
& \bar{Q}, \; \bar{u}, \; \bar{d}, \; \bar{L}, \; \bar{e}, \label{eq:bar} \\
& \Phi.  \label{eq:higgssec2}
\end{align}
The quantum charges of these superfields are summarized in~Table \ref{tb:newgeneration}.
The superfields of fourth generation in (2.3) have the same charges as those of matters in the MSSM, while those of fifth in (2.4) have the opposite ones. These pairs with opposite charges are called vector-like generations. 
The fields $\Phi$ is a SM gauge singlet field, and its vacuum expectation value gives mass scales of the vector-like generations. With these superfields, the Yukawa sector of the superpotential is written as~\cite{Nishida:2016lyk}
\begin{align}
W & = \sum_{i, j = 1, \cdots, 4} \Big(
  (y_u)_{ij}u_i Q_j H_u
+ (y_d)_{ij} d_i Q_j H_d
+ (y_e)_{ij} e_i L_j H_d \Big) \nonumber \\
& \hspace{20mm} 
+ y_{\bar{u}}  \bar{u} \bar{Q} H_{d}
+ y_{\bar{d}} \bar{d} \bar{Q} H_u
+ y_{\bar{e}} \bar{e} \bar{L} H_u
 \nonumber \\
& \quad + \sum_{i = 1, \cdots, 4} \Big(
 y_{Q_i} \Phi Q_{i}
\bar{Q}
+ y_{u_i} \Phi u_{i}
\bar{u}
+ y_{d_i} \Phi d_{i}
\bar{d} 
+ y_{L_i} \Phi L_{i} 
\bar{L}
+ y_{e_i} \Phi e_{i} 
\bar{e}
 \Big) \nonumber\\
& \hspace{20mm}
+ y \Phi^{3}
, \label{eq:oriY}
\end{align}
where $y$'s are dimensionless couplings for each generation of quark and lepton sectors.\footnote{%%%
The superpotential (\ref{superpotential44bar}) evokes us that the potential has $Z_3$ symmetry with respect to $\Phi$, but we can not assign the discrete charge. Thus, there is not domain wall problem.
}%%%
Each term of the interactions contributes to experimental results of the muon $g-2$, the Higgs mass and DM abundance as explained below. The Yukawa couplings in the first line give the flavor structure in the SM sector on top of the fourth generation matter, 
and they largely contribute to the flavor physics, i.e., the muon $g-2$~\cite{Nishida:2016lyk}. The second line shows the coupling of the fifth generations to the Higgs fields, and these interactions do not crucially contribute to the flavor structure in the SM sector but do to the muon $g-2$ in our model. The third line shows the coupling of vector-like generations to the SM singlet field $\Phi$. After the symmetry breaking of the scalar component of $\Phi$, the vector-like generations obtain each mass from these terms. Further, owing to these terms, the lower experimental bounds on the mass of vector-like
generations are avoided. In our model, the fermion component of $\Phi$ is a DM candidate whose mass is given by the last term $y \Phi^3$. We explain such structures by the FN mechanism.

The FN mechanism requires an additional SM singlet scalar field and it is supposed that the singlet field has interactions with quark and lepton sectors~\cite{Froggatt:1978nt}. Then, the effective Yukawa couplings are determined through the interactions by the singlet VEV. The magnitude of the couplings are controlled by an assignment of the FN charge to the quark and lepton sectors. We apply this mechanism to our model by introducing a SM singlet superfield
\beq
\Theta.
\eeq
Let us consider an example of the interaction based on FN charge with $W=(\Theta/\Lambda)^n u_1 Q_1 H_u$, where $n$ is an integer and $\Lambda$ is a cut off scale. Here we take the scale to $\Lambda\simeq10^{16}{\rm GeV}$. Under an assignment of the FN $U(1)$ charge, the integer $n$ is determined to satisfy $n q(\Theta) + q(u_1) + q(Q_1) + q(H_u) =0$, where $q(\Theta),~q(Q_1),~q(u_1),~$and $q(H_u)$ are $U(1)$ charges of respective fields. With this integer, the VEV such that $\braket{\Theta}\neq 0<\Lambda$ makes the effective Yukawa couplings as $(y_{u})_{11}\propto (\braket{\Theta}/\Lambda)^n$. By this way, the Yukawa structures of the quarks and leptons are determined.

Using the FN mechanism, we extend the Yukawa interaction part (\ref{eq:oriY}). We assign the FN U(1) charge to the FN field $\Theta$ as 
\begin{align}
q(\Theta) = -1.
\end{align} 
Together with the charge assignment exhibited in Table \ref{tb:newgeneration}, the Yukawa sector of the superpotential is written as
\begin{align}
W & = \sum_{i, j = 1, \cdots, 4} \Big(
  (Y_u)_{ij} \left( \frac{\Theta}{\Lambda} \right)^{n_u^{ij}} u_i Q_j H_u
+ (Y_d)_{ij} \left( \frac{\Theta}{\Lambda} \right)^{n_d^{ij}} d_i Q_j H_d
+ (Y_e)_{ij} \left( \frac{\Theta}{\Lambda} \right)^{n_e^{ij}} e_i L_j H_d \Big) \nonumber \\
& \hspace{20mm} 
+ Y_{\bar{u}} \left( \frac{\Theta}{\Lambda} \right)^{n_{\bar{u}}} \bar{u} \bar{Q} H_{d}
+ Y_{\bar{d}} \left( \frac{\Theta}{\Lambda} \right)^{n_{\bar{d}}} \bar{d} \bar{Q} H_u
+ Y_{\bar{e}} \left( \frac{\Theta}{\Lambda} \right)^{n_{\bar{e}}} \bar{e} \bar{L} H_u
 \nonumber \\
& \quad + \sum_{i = 1, \cdots, 4} \Big(
 Y_{Q_i}  \left( \frac{\Theta}{\Lambda} \right)^{n_{Q_i}} \Phi Q_{i}
\bar{Q}
+ Y_{u_i}  \left( \frac{\Theta}{\Lambda} \right)^{n_{u_{i}}} \Phi u_{i}
\bar{u}
+ Y_{d_i}  \left( \frac{\Theta}{\Lambda} \right)^{n_{d_{i}}} \Phi d_{i}
\bar{d} \nonumber \\
& \hspace{20mm}
+ Y_{L_i}  \left( \frac{\Theta}{\Lambda} \right)^{n_{L_{i}}} \Phi L_{i} 
\bar{L}
+ Y_{e_i}  \left( \frac{\Theta}{\Lambda} \right)^{n_{e_{i}}} \Phi e_{i} 
\bar{e}
 \Big)
+ Y \left( \frac{\Theta}{\Lambda} \right)^{n_{\Phi}} \Phi^{3}
, \label{superpotential44bar}
\end{align}
where magnitude of all Yukawa couplings is assumed to be of $\mathcal{O}(1)$. As explained, each power of $\Theta/\Lambda$ is determined from the charge assignment as 
\begin{align}
 n_{u}^{ij} & = q(u_{i}) + q(Q_{j}) + q(H_{u}),\
 n_{d}^{ij}   = q(d_{i}) + q(Q_{j}) + q(H_{d}),\
 n_{e}^{ij}   = q(e_{i}) + q(L_{j}) + q(H_{d}), \nonumber \\
 n_{\bar{u}} & = q(\bar{u}) + q(\bar{Q}) + q(H_{d}),\  
 n_{\bar{d}}   = q(\bar{d}) + q(\bar{Q}) + q(H_{u}),\ 
 n_{\bar{e}}   = q(\bar{e}) + q(\bar{L}) + q(H_{u}), \nonumber \\
 n_{Q_{i}} & = q(Q_{i}) + q(\bar{Q}) + q(\Phi),\ 
 n_{u_{i}}   = q(u_{i}) + q(\bar{u}) + q(\Phi),\ 
 n_{d_{i}}   = q(d_{i}) + q(\bar{d}) + q(\Phi),\ \nonumber \\
 n_{L_{i}} & = q(L_{i}) + q(\bar{L}) + q(\Phi),\
 n_{e_{i}}   = q(e_{i}) + q(\bar{e}) + q(\Phi),\ 
 n_{\Phi}    = 3 q(\Phi). \label{eq:total}
\end{align}
Then, by the superpotential (\ref{superpotential44bar}), the effective Yukawa couplings are given by
\beq
y_{x}\equiv Y_{x}\left(\frac{\braket\Theta}{\Lambda}\right)^{n_{x}},
\eeq
where $x$ represents each generation of quark or lepton sectors. Note that with present charge assignment the cubic coefficient is same as $y=Y$ because of $q(\Phi) = 0$. In this paper, we assume that the $U(1)_{\rm FN}$ is a gauged symmetry. With the D-term, the FN field obtains vacuum expectation value, but in this case, anomalies due to the symmetry could exist. Here let us comment about these issues. The case of the global symmetry is discussed at the end of this section.
%%%%%%%%%%%%%%%%%%%
%%%%%%%%%%%%%%%%%%%%
%%%%%%%%%%%%%%%%%%%%

\begin{table}[t]
\begin{center}
\begin{tabular}{|c|c|c|}
\hline
~~~~~
& $ (\text{SU}(3),\, \text{SU}(2),\,\text{U}(1) )$  \\ \hline \hline
$Q_4$ & (${\bm 3}  ,\,{\bm 2},\, \frac{1}{6} ) $   \\
$u_4$ & (${\bm 3^*},\,   {\bm 1},\, \frac{-2}{3} ) $  \\
$d_4$ & (${\bm 3}^*,\,   {\bm 1},\, \frac{1}{3} ) $  \\
$L_4$ & (${\bm 1}  ,\,   {\bm 2},\, \frac{-1}{2} ) $   \\
$e_4$ & (${\bm 1} ,\,  {\bm 1},\, 1 ) $  \\
$\bar{Q} \equiv \left(
\begin{array}{c}
 (u_{5R})^{C} \\
 (d_{5R})^{C}
\end{array}
\right)
$ & (${\bm 3}^*,\,   {\bm 2},\, \frac{-1}{6} ) $   \\
$\bar{u} \equiv u_{5L}$ & (${\bm 3},\,   {\bm 1},\, \frac{2}{3} ) $  \\
$\bar{d} \equiv d_{5L}$ & (${\bm 3},\,   {\bm 1},\, \frac{-1}{3} ) $  \\
$\bar{L} \equiv \left(
\begin{array}{c}
 (\nu_{5R})^{C} \\
 (e_{5R})^{C}
\end{array}
\right)
$ & (${\bm 1},\,   {\bm 2},\, \frac{1}{2} ) $   \\
$\bar{e} \equiv e_{5L}$ & (${\bm 1},\,   {\bm 1},\, -1 ) $  \\
$\Phi$  & (${\bm 1},\,   {\bm 1},\, 0 ) $  \\
$\Theta$ &  (${\bm 1},\,   {\bm 1},\, 0 ) $ \\
\hline
\end{tabular}
\caption{ The chiral superfields and their quantum number under the SM
gauge group.} 
\label{tb:newgeneration}
\end{center} \bigskip
\end{table}

\begin{table}[t]
\begin{center}
\begin{tabular}{|c|c|c|c|c|c||c|c|c|c|c||c|c|c|c|c|}
\hline
&  $Q_{1}$ & $Q_{2}$ & $Q_{3}$ & $Q_{4}$ & $\bar{Q}$ 
&  $u_{1}$ & $u_{2}$ & $u_{3}$ & $u_{4}$ & $\bar{u}$ 
&  $d_{1}$ & $d_{2}$ & $d_{3}$ & $d_{4}$ & $\bar{d}$ 
\\  \hline \hline
U(1)$_{\text{FN}}$
& 5 & 2 & 0 & -2  & 2
& 4 & 2 & 0 & -2  & 2
& 4 & 4 & 1 & -2  & -2
 \\
\hline
\end{tabular}
\end{center}
 \hspace{12mm}
\begin{tabular}{|c|c|c|c|c|c||c|c|c|c|c||c|c|c|c|c|c|}
 \hline
& $L_{1}$ & $L_{2}$ & $L_{3}$ & $L_{4}$ & $\bar{L}$ 
&  $e_{1}$ & $e_{2}$ & $e_{3}$ & $e_{4}$ & $\bar{e}$ 
&  $H_{u}$ & $H_{d}$ & $\Phi$  & $\Theta$
\\  \hline \hline
U(1)$_{\text{FN}}$
& 4& 1 & 0 & -1  & 0
& 5& 1 & 1 & 0  & 2
& 0 & 0 & 0  & -1 \\
\hline
\end{tabular}
\caption{The list for FN charge of each field.} 
\label{tb:FNcharge}
\end{table}
 
Now, the VEV of the FN field is given by the FI D-term of the anomalous $U(1)_{\rm FN}$. Under the charge assignment, U(1)$_{\rm FN}$ becomes anomalous in our model because of tr$(q) > 0$,
if there are not any additional chiral multiplets with negative U(1)$_{\rm FN}$ charges.
In such cases, theory is ill-defined. Based on the string theory, however,
such anomalies can be canceled by the gauged shift of string theoretic axions
(or p-from potentials) \cite{Green:1984sg}, and Fayet-lliopoulos term is naturally induced
in U(1)$_{\rm FN}$ D-term at one-loop level with $\xi\sim\lrp{{\rm tr}(q)M_{\ast}^2}/(16\pi^2)$ \cite{Fischler:1981zk,Dine:1987xk}.
(See also \cite{Blumenhagen:2006ci} for a review.) 
Here $M_{\ast}$ is the string scale. In this paper, we assume that only $\Theta$ develops
VEV \cite{Dreiner:2003hw} in the presence of U(1)$_{\rm FN}$ D-term potential by $D\sim\xi-|\Theta|^2 \sim0$ and that U(1)$_{\rm FN}$ anomalies are cancel-led by shifts of (multiple) axions which are coupled to the SM gauge fields in the viewpoint of generalized Green-Schwarz mechanism.\footnote{
In other words, a different gauge sector in the SM may come from a stack of D-branes which are wrapping on a different cycle on the internal extra dimension or contain different world volume fluxes on such a cycle. Then, GUT-like relation between gauge couplings, which is led by geometric properties, can be found \cite{Blumenhagen:2003jy} through moduli stabilization \cite{Kachru:2003aw,Choi:2006xt}.}
Then, the chiral superfield $\Theta$ will be eaten by the anomalous U(1)$_{\rm FN}$ vector superfield in a supersymmetric manner.
They become massive around the cutoff scale,
and hence we will neglect them and U(1)$_{\text{FN}}$ $D$-term
contribution to the SUSY breaking,
and focus only on the VEV in the followings.

As explained above, the effective Yukawa couplings are determined by the charge assignment of FN $U(1)$ and $\braket{\Theta}$. Especially, the charge assignment determines the flavor structure of our model. We explain the strategy to determine the assignment. First, we require the superpotential homomorphic for $\Theta$ at a perturbative level. To satisfy this requirement, the power of $\Theta/\Lambda$ needs to be positive. Under this constraint, we determine the FN charge, paying attention to two points: mixing between second and fourth generations and masses of fourth and fifth generations. In our vector-like generations model, low energy flavor structure in the SM sector is finally determined by RG equations. By solving RG equations, the authors of~\cite{Nishida:2016lyk} found that for the appropriate flavor structure of the SM, the mixing between the second and fourth generations needs to be large. Thus we have to determine the FN charge such that this large coupling is reproduced.
Another point is about the couplings between 
fourth and fifth generations. From the experimental constraint at Large Hadron Collider (LHC), 
the masses of vector-like generations are required to be relatively large~as $m_{{\rm q}_4}\gtrsim800\,{\rm GeV}$ and $m_{{\rm l}_4}\gtrsim100\,{\rm GeV}$\cite{Olive:2016xmw}. 
Thus the FN charge needs to be chosen to achieve these large masses.

Paying attention to those points, we have determined the assignment of the FN charge shown in Table \ref{tb:FNcharge}. We explicitly show the mass matrix of each sector under the assignment. Let us here define a parameter for the convenience of explanation as 
\begin{eqnarray}
\epsilon = \frac{\left< \Theta \right>}{\Lambda}.
\end{eqnarray}
With this parameter, the matrix elements of up-type quark mass $M_u$, down-type quark mass $M_d$ and charged lepton mass $M_e$ are given as
\begin{eqnarray}
M_{u} & \approx &
\bordermatrix{
   & u_{1R} & u_{2R} & u_{3R} & u_{4R} & u_{5R} \cr
 u_{1L} & \epsilon^{9} v_{u}  & \epsilon^{7} v_{u}  & \epsilon^{5} v_{u}  & \epsilon^{3} v_{u} & \epsilon^{7} V   \cr             
 u_{2L} & \epsilon^{6}v_{u}  & \epsilon^{4} v_{u} & \epsilon^{2} v_{u} & v_{u} & \epsilon^{4} V \cr
 u_{3L} &  \epsilon^{4} v_{u} & \epsilon^{2} v_{u} & v_{u}  & 0 & \epsilon^{2} V \cr
 u_{4L} &  \epsilon^{2} v_{u} & v_{u}  & 0  & 0 & V \cr
 u_{5L} & \epsilon^{6} V & \epsilon^{4} V  & \epsilon^{2} V & V & \epsilon^{4} v_{d}  \cr
}, \label{uflavor} \\
M_{d} & \approx &
\bordermatrix{
  & d_{1R} & d_{2R} & d_{3R} & d_{4R} & d_{5R} \cr
 d_{1L} & \epsilon^{9} v_{d}  & \epsilon^{9} v_{d}  & \epsilon^{6} v_{d}  & \epsilon^{3} v_{d} & \epsilon^{7} V  \cr             
 d_{2L} & \epsilon^{6} v_{d}  & \epsilon^{6} v_{d} & \epsilon^{3} v_{d} & v_{d} & \epsilon^{4} V \cr
 d_{3L} &  \epsilon^{4} v_{d} & \epsilon^{4} v_{d} & \epsilon^{1} v_{d}  & 0 & \epsilon^{2} V \cr
 d_{4L} & \epsilon^{2} v_{d} & \epsilon^{2} v_{d}  & 0  & 0 & V \cr
 d_{5L} & \epsilon^{6} V & \epsilon^{6} V  & \epsilon^{3} V & V & \epsilon^{4} v_{u}  \cr
}, \label{dflavor} \\
M_{e} & \approx &
\bordermatrix{
   & e_{1R} & e_{2R} & e_{3R} & e_{4R} & e_{5R} \cr
 e_{1L} & \epsilon^{9} v_{d}  & \epsilon^{5} v_{d}  & \epsilon^{5} v_{d}  & \epsilon^{4} v_{d} & \epsilon^{6} V   \cr           
 e_{2L} & \epsilon^{6} v_{d}  & \epsilon^{2} v_{d} & \epsilon^{2} v_{d} & \epsilon^{1} v_{d} & \epsilon^{3} V \cr
 e_{3L} &  \epsilon^{5} v_{d} & \epsilon^{1} v_{d} & \epsilon^{1} v_{d}  & v_d & \epsilon^{2} V \cr
 e_{4L} &  \epsilon^{4} v_{d} & v_{d}  & v_{d}  & 0 & \epsilon^{1} V \cr
 e_{5L} & \epsilon^{5} V & \epsilon^{1} V  & \epsilon^{1} V & V & \epsilon^{2} v_{u} \cr
},\label{eflavor}
\end{eqnarray}
where we neglected ${\cal O}(1)$ bare Yukawa couplings in the superpotential.
Here we defined the vacuum expectation values of $H_u,~H_d$ and $\Phi$ as
\beq
\braket{H_u}\equiv v_u,~\braket{H_d}\equiv v_d,~\braket{\Phi}\equiv V,
\eeq
and defined the fermion components of $\bar{Q},~\bar{u}$ and $\bar{d}$ as
\beq
\left.\bar{Q}\right|_{\rm fermion}\equiv
\left(
\begin{matrix}
(u_{5R})^c\\
(d_{5R})^c
\end{matrix}
\right),~
\left.\bar{u}\right|_{\rm fermion}\equiv	(u_{5L}),~
\left.\bar{d}\right|_{\rm fermion}\equiv(d_{5L}).
\eeq

As for three generations of the SM and fourth generations, the up-type Higgs $H_u$ couples to up-type quarks, and the down-type Higgs $H_d$ couples to down-type quarks and leptons in the Yukawa sector. (See the first line in Eq.~(\ref{superpotential44bar}).) These terms correspond to the $4\times4$ parts of the mass matrices (\ref{uflavor}), (\ref{dflavor}) and (\ref{eflavor}).  As for the fifth generations, the Higgs couples in the opposite way, that is, $H_u$ couples to down-type quark of fifth generation, and $H_d$ couples to up-type quark as shown in the fourth line in Eq.~(\ref{superpotential44bar}), which correspond to $5 - 5$ entry in the mass matrices. The other elements of the mass matrices are given by the gauge singlet field $\Phi$. (See the third and fourth lines in Eq.\ (\ref{superpotential44bar}).)

\begin{figure}[t]
\begin{center}
  \includegraphics[width=160mm]{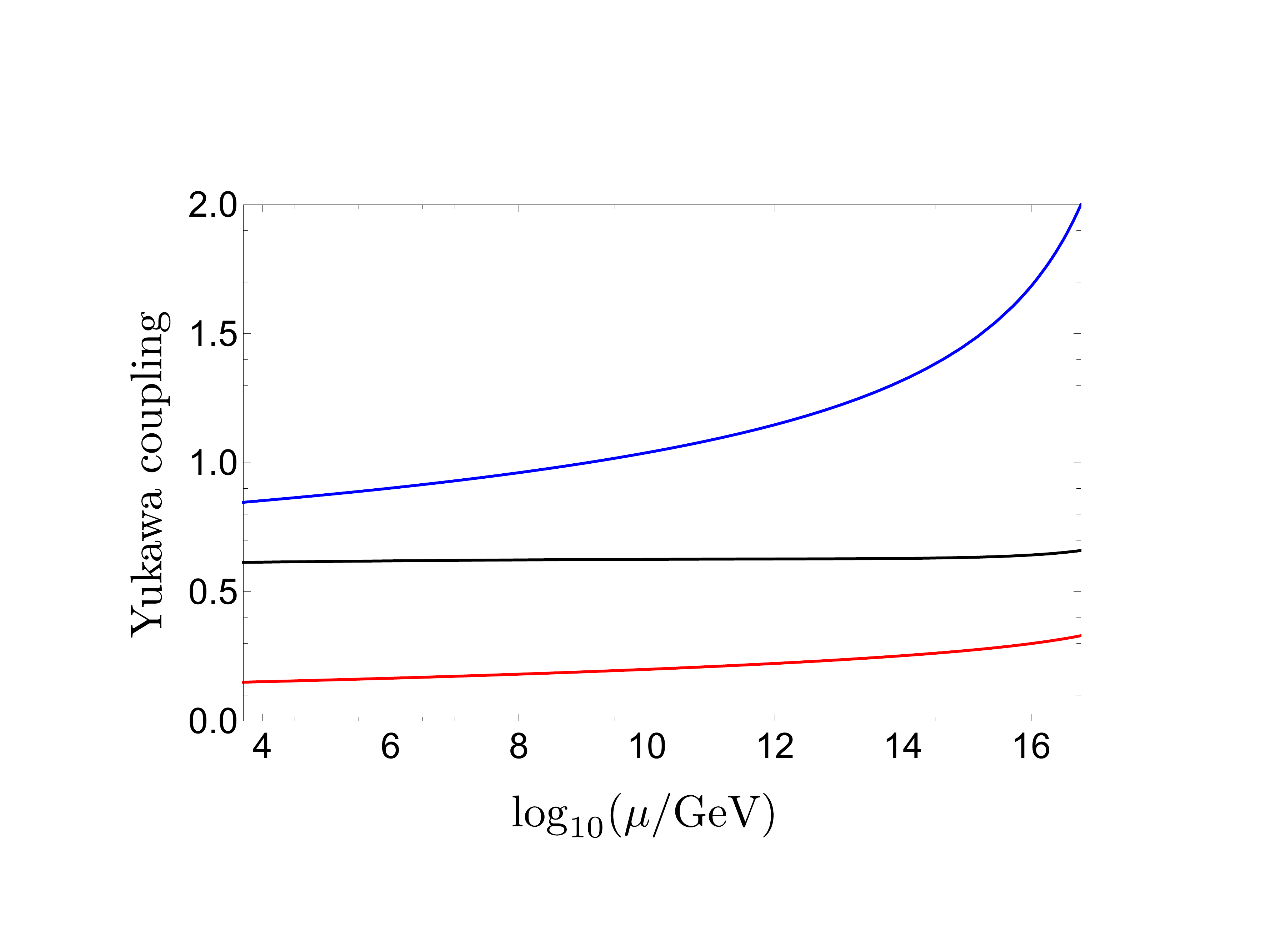}
\end{center}
\vspace{-10mm}
\caption{The result of RG running for the Yukawa couplings
of the third generation.
Horizontal axis is energy scale and vertical axis is the strength
of the Yukawa couplings.
Blue, black and red line correspond to the RG running of $(y_u)_{33}$, $(y_d)_{33}$ and $(y_e)_{33}$, respectively.}
\label{fig:yukawarge}
\end{figure}

Now we discuss about the observables of the CKM matrix and fermion masses in both quark and lepton sectors. In our model, the CKM matrix is extracted from $5 \times 5$ unitary matrix. Thus we need to check carefully that the unitarity of the CKM matrix is satisfied so that the CKM matrix elements other than ordinary $3 \times 3$ matrix are suppressed by the vector-like mass.
The CKM matrix is defined by the part of the upper-left $3 \times 3$ matrix of the product of $5 \times 5$ unitary matrices:
\begin{eqnarray}
 ( V_{\text{CKM}} )_{ij} = ( V_{uL}^{\dagger} V_{dL} )_{ij}
\hspace{3mm} (i,j = 1,2,3) ,
\end{eqnarray}
where $i,j$ are generation labels.
\begin{table}[t]
\begin{center}
\begin{tabular}{|c|c|c|c|c|c|}
\hline
$\epsilon$ & $\alpha_{\rm GUT}$ & 
$M_{\rm GUT}$ = $\Lambda$ & $M_{\rm SUSY}$  & $V$ & $\tan\beta$  \\ \hline \hline
0.33  &  0.10  &
$6.0\times 10^{16}$ GeV  &  5.0 TeV  &  2.0 TeV  &  40  \\
\hline
\end{tabular}
\caption{The set of input values.}
\label{tb:parameterset}
\end{center} 
%%%%%%%%%%%%%%%%%%%%
\begin{center}
\begin{tabular}{|c|c|c|c|c|c|c|c|c|c}
\hline
up-type quark Yukawa & down-type quark Yukawa & charged lepton Yukawa \\
\hline \hline
  $(Y_{u})_{11}$ = 2.000 & $(Y_d)_{11}$ = 0.500 & $(Y_e)_{14}$ = 2.000 \\
$(Y_u)_{23}$ = 2.000 & $(Y_d)_{21}$ = 1.930  & $(Y_e)_{12}$ = 2.000 \\
$(Y_u)_{33}$ = 2.000 & $(Y_d)_{22}$ = 1.200  & $(Y_e)_{21}$ = 2.000 \\
$(Y_u)_{41}$ = 0.500 & $(Y_d)_{23}$ = 0.900  & 
 $ (Y_e)_{22}$ = 0.500 \\
$Y_{u_3}$ = 2.000 &  $(Y_d)_{31}$ = 0.632   &  $(Y_e)_{24}$ = 2.000 \\
  &  $(Y_d)_{32}$ = 0.700   &  $(Y_e)_{34}$ = 0.500 \\
  &  $(Y_d)_{33}$ = 2.000   &  ${Y}_{41}$ = 2.000  \\
  &  $(Y_d)_{41}$ = 2.000   &  ${Y}_{42}$ = 2.000 \\
  &  $Y_{d_2}$ = 1.100    &  ${Y}_{L_2}$ = 2.000 \\
  &   & ${Y}_{L_3}$ = 0.500 \\
  &   & ${Y}_{L_4}$ = 0.500 \\
  &   & ${Y}_{e_2}$ = 0.500 \\
  &   & ${Y}_{e_3}$ = 2.000 \\
  &   & ${Y}_{\bar{e}}$ = 2.000 \\
\hline
\end{tabular}
\caption{The set of input values for coupling constants in
Eq.\ (\ref{superpotential44bar}).
Other coupling constants which are not written in this
table are set to be unity. Note that all couplings are of ${\cal O}(1)$.}
\label{tb:parayukawa}
\end{center} 
\end{table}
These $5 \times 5$ unitary matrices diagonalize the up-type and down-type quark mass matrices:
\begin{eqnarray}
 V_{uR} M_{u} V_{uL}^{\dagger}, \\
 V_{dR} M_{d} V_{dL}^{\dagger}.
\end{eqnarray}
With the structures of the mass matrices ({\ref{uflavor}}), (\ref{dflavor}) and  ({\ref{eflavor}}), we solve the RG equations. The input parameters for the equation are summarized in Table \ref{tb:parameterset} and \ref{tb:parayukawa}. By the calculation, we have confirmed that the CKM matrix and fermion masses at the $M_Z$ scale reproduce the observed ones.
The RG equations in our model are listed in Appendix \ref{sec:RGEs}.
In Table \ref{tb:parameterset}, 
$\alpha_{\text{GUT}}$ and $\Lambda$ are the initial condition of RG equations for the MSSM gauge couplings
obtained in \cite{Nishida:2016lyk}. 
It is found that the MSSM gauge couplings
unify at a certain scale $M_{\text{GUT}}$ which is slightly high compared to the MSSM without vector-like generations. Thus, we use these values as boundary conditions for
RG running.
The scale $M_{\text{SUSY}}$ is a typical threshold for supersymmetric
particles, and the ratio of the vacuum expectation values of Higgs
doublets is defined as $\tan\beta\equiv v_u/v_d$.
In Section \ref{sec:res},
we use the same input values for the numerical analysis of the 
DM abundance, the Higgs boson mass and the muon $g-2$.
Here, as an example, we show the result of RG running for the 
Yukawa couplings of the third generation in Fig. \ref{fig:yukawarge}. The horizontal axis is energy scale, and the vertical axis is the strength
of the Yukawa couplings.
The blue, black and red lines correspond to the RG running of $ (y_u)_{33}$, $(y_d)_{33}$ and $(y_e)_{33}$, respectively. In the Fig.\ \ref{fig:yukawarge}, each Yukawa coupling converges to a certain value at low energy, where these couplings do not depend on the initial values at $M_{\text{GUT}}$. The detailed analysis for the convergecy of Yukawa couplings are performed in \cite{Bando:1996in}. As for $(y_{d})_{33}$, it seems that this Yukawa coupling does not evolve, but the initial value of $(y_{d})_{33}$ just coincides with the infrared value.
Such RG runnings
reproduce the fermion masses and the CKM matrix at the scale of $M_{Z}$:
\begin{eqnarray}
 m_{t} \sim 170\ \text{GeV}, \hspace{3mm} m_{c} \sim 0.7\ \text{GeV},
 \hspace{3mm} m_{u} \sim 5.0\ \text{MeV}, \nonumber \\
 m_{b} \sim 3.0\ \text{GeV}, \hspace{3mm} m_{s} \sim 0.032\ \text{GeV},
 \hspace{3mm} m_{d} \sim 1.0\ \text{MeV}, \nonumber \\
 m_{\tau} \sim 1.6\ \text{GeV}, \hspace{3mm} m_{\mu} \sim 0.10\ \text{GeV},
 \hspace{3mm} m_{e} \sim 0.2\ \text{MeV},  \label{eq:fermionmass}
\end{eqnarray}
\begin{eqnarray}
 | V_{\text{CKM}} | \sim
 \left(
 \begin{array}{ccc}
   0.974 & 0.226 & 0.0035 \\
  0.225 & 0.973 & 0.040 \\
  0.0089 & 0.041 & 0.999 .
 \end{array}
\right) . \label{eq:CKM}
\end{eqnarray}
Fermion masses at the $M_{Z}$ scale are studied in \cite{Xing:2007fb}, and the CKM matrix takes its value within 2$\sigma$ level of observed CKM matrix~\cite{Olive:2016xmw} as
\begin{eqnarray}
\left(
\begin{array}{ccc}
0.97403 - 0.97449 & 0.22406 - 0.22606 & 0.00327 - 0.00387 \\
0.22392 - 0.22592 & 0.97325 - 0.97377 & 0.04084 - 0.04136 \\
0.00815 - 0.00939 & 0.0377 - 0.0429 & 0.9991 - 0.9992 
\end{array}
\right).
\end{eqnarray}
These are consistent with Eq.\ (\ref{eq:fermionmass})
and (\ref{eq:CKM}).

%%%%%%%%%%%%%
%\subsection{Global FN symmetry}
%\label{subsec:globalFN}.
%%%%%%%%%%%%%
Here let us discuss the case that the FN symmetry is global.~\footnote{%%%
In the case of global $U(1)_{\rm FN}$, we can realize the symmetry breaking by F-term of the superpotential such as $W=S\lrp{\Theta\bar{\Theta}-\braket{\Theta}^2}$ where $S$ and $\bar{\Theta}$ are singlet fields but $q(\bar{\Theta})=1$~\cite{Ema:2016ops}.
}%%%
In this case, there exists a (pseudo) NG boson associated with the spontaneous symmetry breaking of $U(1)_{\rm FN}$. We call it FN axion. As in the case of the QCD-axion, this FN axion could be another DM candidate. Further, it has interactions with the quarks and leptons, and through the anomaly effects it couples to gluons and electric-magnetic fields with an effective decay rates. The interactions give experimental and cosmological constraints on the VEV of FN field~\cite{Ema:2016ops}. In the basis that the mass matrices of quarks and leptons are diagonalized, the Yukawa interactions are written by
\beq
-{\cal L}
=\sum_{f=u,d,l}\left[
m_i^f\bar{f}_{Li}^{'}f'_{R_i}
+\kappa_{ij}^f\frac{s+ia}{\sqrt{2}v_{\Theta}}\bar{f}_{Li}^{'}f'_{R_j}
\right],
\eeq
where the coupling $\kappa^f_{ij}$ is determined by the mass matrix of fermions and FN charge as
\beq
\kappa_{ij}^{f}\equiv
\lrp{V^{f_L+}}_{ik}\lrp{M_{kn}^fn_{kn}^f}V^{f_R}_{nj}.
\eeq
Here we have expanded the FN field around the VEV as
\beq
\Theta=v_{\Theta}+\frac{s+ia}{\sqrt{2}}.
\eeq
Thus, the interactions of the axion with quarks and leptons are given by
\beq
-{\cal L}_{\rm int}=
\frac{ia}{\sqrt{2}v_{\Theta}}\sum_{f=u',d',l'}
\left[
\lrp{\kappa_H^f}_{ij}\bar{f}_i\gamma_{5}f_j
+
\lrp{\kappa_A^f}_{ij}\bar{f}_if_j
\right],
\eeq
where we have redefined the coupling of axion as
\beq
\left\{
\begin{split}
\lrp{\kappa^f_H}_{ij}&\equiv\frac{\lrp{\kappa^f+\kappa^{f\dagger}}_{ij}}{2},\\
\lrp{\kappa^f_A}_{ij}&\equiv\frac{\lrp{\kappa^f-\kappa^{f\dagger}}_{ij}}{2}.
\end{split}
\right.
\eeq
Further, through the critical rotation as in the case of the QCD-axion, the axial  interaction of the FN axion with gluons is given by
\beq
-{\cal L}_{{\rm gluon-axion}}
=\frac{g_s^2}{32\pi^2}\left[
\sum_{i=1...5;f=u,d}\frac{\lrp{\kappa_H^f}_{ii}}{m_i^f}
\right]
\frac{a}{\sqrt{2}v_{\Theta}}G_{\mu\nu}^a\tilde{G}^{\mu\nu a}
=\frac{g_s^2}{32\pi^2}\frac{a}{f_a}G_{\mu\nu}^a\tilde{G}^{\mu\nu a},
\eeq
where we have defined  the domain wall number $N_{DW}$ and the effective decay constant $f_a$ as
\beq
N_{DW}\equiv\sum_{i=1...5;f=u,d}\frac{\lrp{\kappa_H^f}_{ii}}{m_i^f}
={\rm tr}\lrp{
n_u^{ij}+n_d^{ij}+n_{\bar{u}}
+n_{\bar{d}}+n_{Q_i}+n_{u_i}+n_{d_i}
},
\eeq
and
\beq
f_a\equiv\frac{\sqrt{2}v_{\Theta}}{N_{DW}}.
\eeq
Now, the domain wall number for our FN charge assignment is calculated as $N_{\rm DW}=54$. In our model, the effective decay constant is typically $f_a\simeq10^{14}{\rm GeV}$.
Among them, the interactions of the FN axion to the up type and down type quarks give a sizable contribution to the decay of the charged kaon via $K^{+}\rightarrow\pi^{+}a$. The decay rate for this process is evaluated by
\beq
\Gamma\left(K^{+}\rightarrow\pi^{+}a\right)=\frac{m_K^3}{32\pi v_{\Theta}^2}\left(1-\frac{m_{\pi}^2}{m_K^2}\right)^3
\left|\frac{\lrp{\kappa_A^d}_{12}}{m_s-m_d}\right|.
\eeq
The last term is of order unity as $\lrp{\kappa_A^d}_{12}=(3/2)\lrp{m_s-m_d}$. Thus, the branching ratio for this process is given by
\beq
{\rm Br}\lrp{K^{+}\rightarrow\pi^{+}a}
\simeq10^{-11}\lrp{\frac{10^{11}{\rm GeV}}{v_{\Theta}}}^2.
\eeq
With experimental bound ${\rm Br}(K^{+}\rightarrow\pi^{+}a)\lesssim7.3\times10^{-11}$~\cite{Adler:2008zza}, we obtain a constraint 
\beq
v_{\Theta}\gtrsim10^{11}{\rm GeV}.
\eeq
Through the interaction of the axion with quarks, the FN axion takes away the energy of the supernovae explosion. From the observation of the SM1987A by Kamiokande, this interaction is constrained in terms of the effective decay rate as $f_a\gtrsim10^{9}{\rm GeV}$. With our definition for the effective decay rate, this constraint is reduced to
\beq
v_{\Theta}\gtrsim
\frac{N_{DW}}{\sqrt{2}}
10^9{\rm GeV}\sim10^{10}{\rm GeV}.
\eeq
Therefore, in the case that the FN $U(1)$ symmetry is global, the VEV of the FN field needs to be $v_{\Theta}\gtrsim10^{11}{\rm GeV}$. As in the case of the QCD-axion, energy density of the coherent oscillation of this FN axion could explain the DM abundance given by~\cite{Turner:1985si}
\beq
\Omega_{a_{\Theta}}h^2
=0.18\theta_i^2\lrp{\frac{f_a}{10^{12}{\rm GeV}}}^{1.19},
\eeq
where $\theta_i$ is the initial misalignment of the FN axion  $\theta_i\equiv a_{\Theta i}/f_a$. When the FN symmetry breaks, there could occur the formation of domain walls. We can assume that the breaking takes place before inflation. In this case, the domain wall problem is avoided, while in this case iso-curvature problem occurs. If the energy scale of inflation, that is, the Hubble parameter is small, this iso-curvature problem is avoided. The detail analysis and constraints on the parameters are studied in~\cite{Ema:2016ops}.

%%%%%%%%%%%%%%%%%%%%%%%%%%%%%%%%%%%%%%%
%%%%%%%%%%%%%%%%%%%%%%%%%%%%%%%%%%%%%%%

\section{Dark matter}  \label{sec:dm}
In this section, we calculate the abundance of thermal relics of a DM candidate,
which is the fermion component of the singlet $\Phi$. 
%and then reveal that the abundance accords with that of dark matter.
The singlet superfield $\Phi$ is expanded as
\begin{eqnarray}
 \Phi = V + \theta \chi_{\Phi} + \cdots,
\end{eqnarray}
where $V$ is the VEV of the scalar component, $\chi_{\Phi}$ is the fermion component
and $\theta$ is the fermionic coordinate on the superspace.
In our model, the VEV of the singlet field $V$ gives masses to vector-like generations, 
and then experimental constraints on them at LHC are avoided. 
%In this section and next, we further calculate the thermal abundance of the singlet field,

First, we discuss the scale of the DM mass $m_{\chi_{\Phi}}$.
In our model, the mass scale is determined by the 
VEV of $V$ and coupling constant of the cubic term $y~(=Y)$ as $m_{\chi_{\Phi}} = y V$.
The RG running of $y$, whose RG equation is given in Eq.\
(\ref{eq:Yrge}), is only governed by Yukawa couplings
related to $\Phi$ and, it is insensitive to gauge sector.
Thus, $y$ is pushed down to $\mathcal{O}(10^{-2})$ at low energy. With $V$ of a few TeV, $m_{\chi_{\Phi}}$ takes around $100\, {\rm GeV}$. This mass scale is smaller than other neutralino masses: bino and Higgsino masses. As for bino, its mass was studied in the previous study~\cite{Nishida:2016lyk}, and then it was revealed that the bino-like mass should be around $200\,{\rm GeV}$ to accord with the Higgs mass confirmed by the LHC. As for Higgsino-like neutralinos, those masses are determined from the electroweak symmetry breaking.
In our model, we have assigned no FN charge to Higgs fields. Thus, in the superpotential, $H_u$ and $H_d$ have the so-called 
$\mu$ 
term as
\begin{eqnarray}
 W = \mu_{H}~H_{u} H_{d},
\end{eqnarray}
where $\mu_H$ is a constant of dimension unity.
For the electroweak symmetry breaking, the mass parameter is required to $\mu_H\simeq 2\,{\rm TeV}$. Thus, the Higgsino-like neutralinos are heavier than the singlino. Therefore, $\chi_{\Phi}$ is the LSP in our model and could be a DM candidate
because $R$-parity symmetry is imposed to our model. Let us comment on an operator $\Phi H_{u} H_{d}$ which can be written by gauge invariant in the superpotential. In this paper, we focus on $\Phi^{3}$ term in the superpotential because we would like to discuss the minimal model of $\chi_\Phi$, which does not couple to the Higgs sector. Such situation might be realized by imposing some symmetry.
%Such situation is realized by imposing $R$-charge as $R(\Phi) = 2/3$ and $R(H_{u} H_{d}) = 2$, so that under $R$ symmetry $\Phi^{3}$ and $\mu H_{u} H_{d}$ terms become invariant and $\Phi H_{u} H_{d}$ term is forbidden in the superpotential.

In order to evaluate the DM abundance, we use the mass
eigenstate basis for squarks, sleptons, quarks and charged leptons.
For quark and lepton sectors, we diagonalize their mass matrices as
\footnote{ 
The definition of Eq.\ (\ref{uflavor})
, (\ref{dflavor}) and (\ref{eflavor}) are written except
$\mathcal{O} (1)$ couplings in Eq.\ \ref{superpotential44bar}.
However, as for the diagonalization, we use the mass matrices
including $\mathcal{O}(1)$ couplings.
}
\begin{align}
( V_{u_R} M_u V_{u_L}^\dagger )_{ij} & = 
{m_U}_i \delta_{ij},  \hspace{5mm} (i,j=1,\dots,5)~ \text{and}~
( m_{U_{i}} < m_{U_{j}},~ \text{if}~ i < j),
\label{eq:uferdia}  \\
( V_{d_R} M_d V_{d_L}^\dagger )_{ij} & = 
{m_D}_i \delta_{ij},  \hspace{5mm} (i,j=1,\dots,5)~ \text{and}~
( m_{D_{i}} < m_{D_{j}},~ \text{if}~ i < j),
\label{eq:dferdia}  \\
( V_{e_R} M_e V_{e_L}^\dagger )_{ij} & = 
{m_E}_i \delta_{ij},  \hspace{5mm} (i,j=1,\dots,5)~ \text{and}~
( m_{E_{i}} < m_{E_{j}},~ \text{if}~ i < j).
\label{eq:eferdia}
\end{align}
We denote the mass eigenvalues $m_{U_{i}}$, $m_{D_{i}}$ and
$m_{E_{i}}$ for the mass eigenstates of
up-type quark ($U_{i}$), down-type quark ($D_{i}$) and
charged lepton ($E_{i}$), respectively. 
As for squarks and charged slepton
sectors, we also diagonalize their mass matrices as
\begin{align}
 ( U_{\tilde{u}} M^2_{\tilde{u}} U_{\tilde{u}}^\dagger )_
{\alpha \beta} & = 
m^2_{\tilde{U}_{\alpha}} \delta_{\alpha \beta},  \hspace{5mm} 
(\alpha, \beta = 1,\dots,10)~ \text{and}~
( m_{\tilde{U}_{\alpha}}^{2} < m_{\tilde{U}_{\beta}}^{2}
,~ \text{if}~ \alpha < \beta),
\label{eq:uscaldia} \\
 ( U_{\tilde{d}} M^2_{\tilde{d}} U_{\tilde{d}}^\dagger )_
{\alpha \beta} & = 
m^2_{\tilde{D}_{\alpha}} \delta_{\alpha \beta},  \hspace{5mm}
(\alpha, \beta = 1,\dots,10)~ \text{and}~
( m_{\tilde{D}_{\alpha}}^{2} < m_{\tilde{D}_{\beta}}^{2}
,~ \text{if}~ \alpha < \beta),
\label{eq:dscaldia} \\
 ( U_{\tilde{e}} M^2_{\tilde{e}} U_{\tilde{e}}^\dagger )_
{\alpha \beta} & = 
m^2_{\tilde{E}_{\alpha}} \delta_{\alpha \beta},  \hspace{5mm} 
(\alpha, \beta = 1,\dots,10)~ \text{and}~
( m_{\tilde{E}_{\alpha}}^{2} < m_{\tilde{E}_{\beta}}^{2}
,~ \text{if}~ \alpha < \beta),
\label{eq:escaldia}
\end{align}
where $M_{\tilde{u}}^{2}$, $M_{\tilde{d}}^{2}$ and 
$M_{\tilde{e}}^{2}$ are up-type squark, down-type squark and charged
slepton mass matrices defined in \cite{Nishida:2016lyk}.
We denote the mass eigenvalues $m_{\tilde{U}_{\alpha}}^{2}$,
$m_{\tilde{D}_{\alpha}}^{2}$ and $m_{\tilde{E}_{\alpha}}^{2}$
for the mass eigenstates of
up-type squark ($\tilde{U}_{\alpha}$), down-type 
squark ($\tilde{D}_{\alpha}$) and
charged slepton ($\tilde{E}_{\alpha}$), respectively.

The interaction terms of $\chi_{\Phi}$
can be read from the third to fourth line of Eq.\ (\ref{superpotential44bar}). 
With the diagonalized basis in Eq.\ (\ref{eq:uferdia}) 
$-$ (\ref{eq:escaldia}), the interaction terms of $\chi_{\Phi}$
are given by
\begin{eqnarray}
 \mathcal{L} & =
 \tilde{\bar{\chi}}_{\Phi}
 ( O_{u R j \alpha} P_{R} + O_{u L j \alpha} P_{L} ) U_{j} \tilde{U}_{\alpha}^{*} +
 \tilde{\bar{\chi}}_{\Phi}
 ( O_{d R j \alpha} P_{R} + O_{d L j \alpha} P_{L} ) D_{j} \tilde{D}_{\alpha}^{*} \nonumber \\ & +
 \tilde{\bar{\chi}}_{\Phi}
 ( O_{e R j \alpha} P_{R} + O_{e L j \alpha} P_{L} ) E_{j} \tilde{E}_{\alpha}^{*} +
{\rm h.c.} , \label{eq:dminteraction}
\end{eqnarray}
where $P_{L} = ( 1 - \gamma_{5})/2$, $P_{R} = ( 1 + \gamma_{5})/2$
and coefficients are 
\begin{eqnarray}
 O_{e R j \alpha} & = & (y_{e})_{i} (V_{eR})_{ji} (U_{\tilde{e}})_{\alpha 5}, \hspace{3mm}
 O_{e L j \alpha}  =  (y_{L})_{i} (V_{eL})_{ji} (U_{\tilde{e}})_{\alpha 10}, \\
 O_{u R j \alpha} & = & (y_{u})_{i} (V_{uR})_{ji} (U_{\tilde{u}})_{\alpha 5}, \hspace{3mm}
 O_{u L j \alpha}  =  (y_{Q})_{i} (V_{uL})_{ji} (U_{\tilde{u}})_{\alpha 10}, \\
 O_{d R j \alpha} & = & (y_{d})_{i} (V_{dR})_{ji} (U_{\tilde{d}})_{\alpha 5}, \hspace{3mm}
 O_{d L j \alpha}  =  (y_{Q})_{i} (V_{dL})_{ji} (U_{\tilde{d}})_{\alpha 10},
\end{eqnarray}

\begin{figure}[t]
\begin{minipage}{0.5\hsize}
\begin{center}
 \includegraphics[width=60mm]{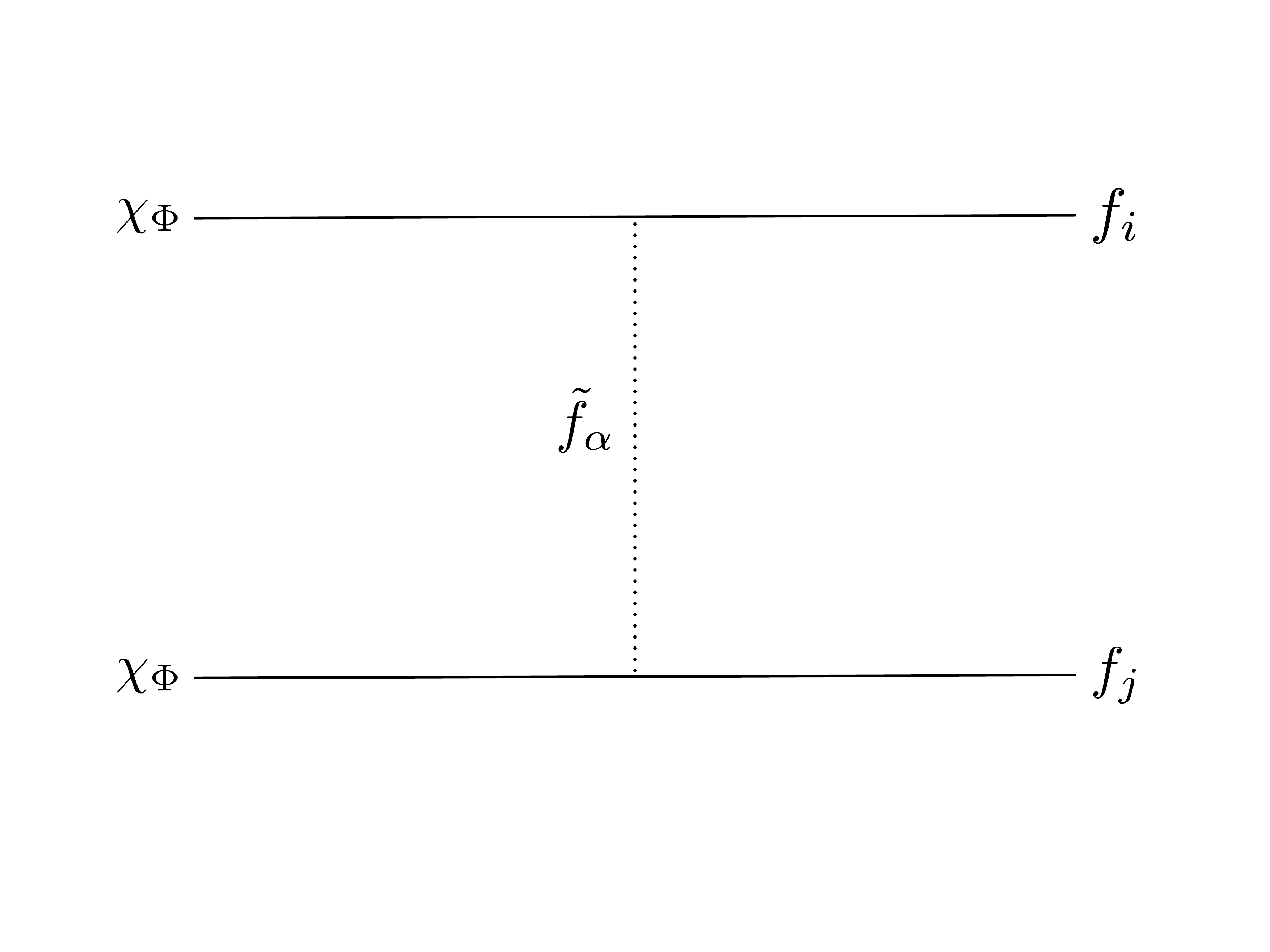}
\end{center}
\end{minipage}
%%%%%%%%%%%%%%
\begin{minipage}{0.5\hsize}
\begin{center}
 \includegraphics[width=60mm]{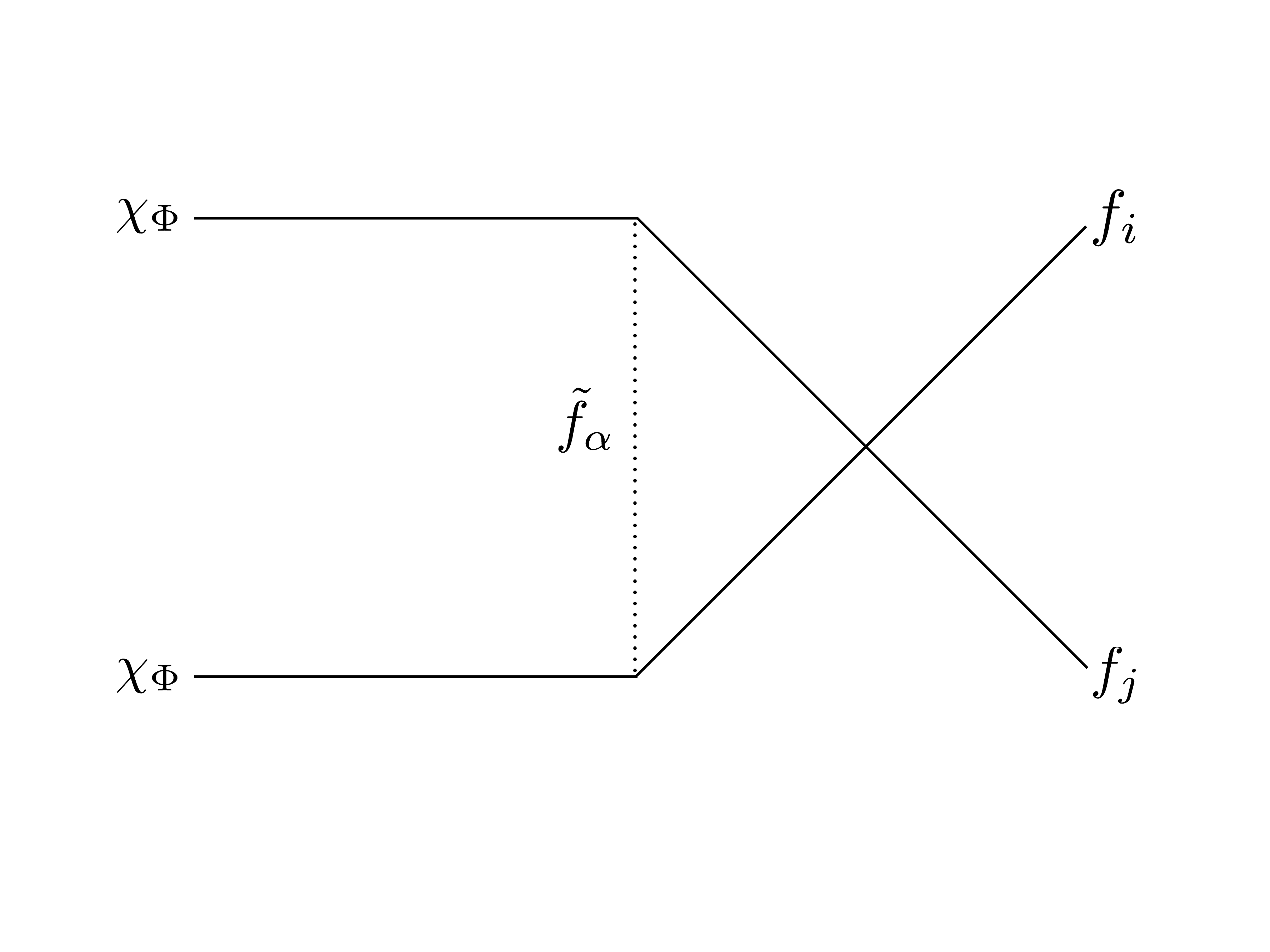}
\end{center}
\end{minipage}
\caption{The diagram of annihilation cross section for singlino DM.
The left panel is the t channel process and the right panel is
u channel process.
$\chi_{\Phi}$ is singlino. $f_{i,j}$ ($f = U, D, E$) are the SM fermions
 whose masses are below freeze-out temperature $T_{\text{F}}$ 
So, we neglected top quark in the external lines.
$\tilde{f}_{\alpha}$ ($\tilde{f} = \tilde{U}, \tilde{D}, \tilde{E}$)
 corresponds to mass eigenstate sfermions which are given in Eq.\
 (\ref{eq:dminteraction}). The results of calculation for
thermal averaged cross section are given in Appendix \ref{sec:tac}.
}
\label{fig:crosec}
\end{figure}

The thermal abundance of the singlino DM is determined by their pair annihilation into the SM particles as shown in Fig. \ref{fig:crosec}. This process ceases when the cosmic expansion rate drops bellow the annihilation rate:
\begin{eqnarray}
 \left< \sigma_{\text{ann}} v_{\text{rel}} \right> n_{\chi_{\Phi}} \simeq
 H(T_{\text{F}}), \label{eq:freezeout} 
\end{eqnarray}
where $\sigma_{\text{ann}}$ is the annihilation cross section, $v_{\text{rel}}$ is their relative velocity, $n_{\chi_{\Phi}}$ is the number density of $\chi_{\Phi}$ and $\left< ... \right>$ represents thermal averaged cross section. 
We defined $T_{\rm F}$ as the freeze out temperature.
After the freeze out, the number density of singlino drops at the same rate of the entropy density by the cosmic expansion. Thus, from Eq.(\ref{eq:freezeout}), we can estimate the ratio of the number density to the entropy density at $T_F$ as
\footnote{In order to calculate the abundance of DM accurately,
we have to solve Boltzmann equations. However, 
it is sufficient to the estimation of the order of the DM abundance.
}
:
\begin{eqnarray}
 \frac{n_{\chi_{\Phi}}}{s} \biggl|_{T_{\text{F}}} \simeq
 \frac{H(T_{\text{F}})}{ \left< \sigma_{\text{ann}} v_{\text{rel}} \right> s } \biggl|_{T_{\text{F}}}
= \frac{1}{4} \left( \frac{90}{ \pi^{2} g_{*}(T_{\text{F}}) } \right)
^{1/2} \frac{1}{ \left< \sigma_{\text{ann}} v_{\text{rel}} \right> T_{\text{F}}
M_{\text{pl}}}, \label{eq:abun}
\end{eqnarray}
where $s$ is the entropy density, $g_{*}(T_{\text{F}})$ is the effective
degrees of freedom of the radiation at the freeze-out and $M_\text{pl} = 2.43 \times 10^{18}$ GeV is the Planck mass.

The diagrams to calculate annihilation cross section are 
shown in Fig.\ \ref{fig:crosec}, where $f_i$ is the SM fermion and $\tilde{f}_{\alpha}$ is sfermion exchanged in the process. Since the singlino is the SM gauge singlet and does not couple to the Higgs boson, the bosons do not appear in the process. 
The SM fermions entering in the process are the ones whose masses are bellow the freeze-out temperature $T_{\text{F}}$. 
%So, we neglected top quark in the external lines. 
In the calculation, one can expand 
$\left< \sigma_{\text{ann}} v_{\text{rel}} \right> = a + b / x_{\text{F}} + \mathcal{O}((1/x_{\text{F}})^{2})$ in the inverse of the power of
$x_{\text{F}} \equiv m_{\chi_{\Phi}}/T_{\text{F}}$ and approximates $\left< \sigma_{\text{ann}} 
v_{\text{rel}} \right>$ by the coefficient $a$ and $b$.
We take $x_{\text{F}} = 20$ for the numerical analysis;
$T_{F}$ becomes around 5 GeV for $m_{\chi_{\Phi}} = 100$ GeV, and hence top quark contribution is neglected in the external lines.
We calculate coefficient $a$ and $b$ for the case of Fig. \ref{fig:crosec} referring to \cite{Roszkowski:1994tm}. The explicit forms of $a$ and $b$ are given in Appendix \ref{sec:tac}. With the values of $x_{\text{F}}$, $a$ and $b$,
the analytic form of the DM relic abundance is given by
\begin{align}
 \Omega_{\chi_{\Phi}} h^{2} \equiv
 \frac{\rho_{\chi_{\phi}}}{\rho_{c} / h^{2} }
 = \frac{ m_{\chi_\Phi} n_{\chi_{\Phi}} / s }{ \rho_{c} / (h^{2} s) }
& = \frac{1.07 \times 10^9 / \text{GeV} x_{\text{F}} }
{ \sqrt{g_{*}} M_{\text{Pl}} (a +  b / x_{\text{F}} ) } 
%\nonumber \\
%&
\approx 0.1 
\times \left( \frac{8.0 \times 10^{-9} \text{GeV}^{-2} }{
\left< \sigma_{\text{ann}} v_{\text{rel}} \right>} \right),
%\left(\frac{m_{\chi_{\Phi}}}{100 \text{GeV}}\right),
\label{eq:omegadm}
\end{align}
where $h$ is the re-scaled Hubble constant, $\rho_{c}$ is
the critical density of Universe, the ratio of critical density to
the entropy density today is $\rho_{c} / s \simeq 1.8 \times 10^{-9}$ GeV and we take $g_{*}(T_F) = 100$.
In our model, the $p$-wave contribution is dominant to 
the thermal averaged cross section.

%%%%%%%%%%%%%%%%%%%%%%%%%%%%%%%%%%%%%%
%%%%%%%%%%%%%%%%%%%%%%%%%%%%%%%%%%%%%%
\section{Higgs boson mass, muon $g-2$ and DM abundance}  \label{sec:res}
%%%%%%%%%%%%%%%%%%%%%%%%%%%%%%%%%%%%%%
%%%%%%%%%%%%%%%%%%%%%%%%%%%%%%%%%%%%%%
In this section, we show the results of numerical calculation for Higgs boson mass, the muon $g-2$ and the DM abundance. We have evaluated the Higgs boson mass and the muon $g-2$ at one-loop level.
In this analysis, we determine the $\mu_{H}$ and $b$ terms in order that the electroweak symmetry breaking is triggered at Fermi scale.
% Here, let us explain how we evaluate the Higgs mass and the muon $g-2$ in our model.First, we consider the Higgs boson mass. 
For the Higgs boson mass, we use the effective potential method \cite{Coleman:1973jx} as usual for the MSSM case \cite{Okada:1990gg}. The quantum corrections to the Higgs mass from vector-like generations are calculated in the literature \cite{Moroi:1991mg,Martin:2009bg}. In our model, the lightest Higgs boson mass $m_{h}^2$ is evaluated by
\begin{eqnarray}
 m_{h}^{2} = m_{h_{\text{tree}}}^{2} + \Delta m_{h}^{2}, \label{eq:formulahiggs}
\end{eqnarray}
where $m^{2}_{h_{\text{tree}}}  = M_{Z}^{2} \cos^{2}( 2\beta )$ 
and $\Delta m_{h}^{2}$ is one-loop corrections which are defined in Eq.\ (\ref{eq:higgsmasscal}).
In the case of MSSM, the mass correction is mainly given by the stop field as $\Delta m_{h,~{\rm MSSM}}^2\simeq (3/4\pi^2)y_t^2m_t^2\log(m_{\rm stop}^2/m_t^2)$, but in our model the squark fields of the vector-like generation also give correction, and it is dominant. For the muon $g-2$, in our model, the one-loop contributions to the muon $g-2$ is given by
\begin{equation}
\Delta a_{\mu} = \Delta a_{\mu}^{\rm SUSY} + \Delta a_{\mu}^{\rm non-SUSY}, \label{eq:formulag2}
\end{equation}
where $\Delta a_{\mu}^{\rm SUSY}$ and $\Delta a_{\mu}^{\rm non-SUSY}$ are SUSY contributions including both the MSSM \cite{Lopez:1993vi} and vector-like \cite{Nishida:2016lyk}  sector and non-SUSY contributions including only vector-like sector \cite{Dermisek:2013gta}, respectively.
$\Delta a_{\mu}^{\rm SUSY}$ and $\Delta a_{\mu}^{\rm non-SUSY}$ are defined in Eq.\ (\ref{eq:g2susycont}) and (\ref{eq:g2nonsusycont}), respectively. The non-SUSY term is approximately given by the ratio of the muon mass to the  charged lepton mass of the vector like generations as $\Delta a_{\mu}^{\rm non-SUSY}\simeq(\alpha/4\pi)~m_{\mu}^2/m_{L'}^2$, where $\alpha$ is the fine structure constant of $SU(2)$ as $\alpha=g^2/(4\pi)$, and $m_{L'}$ is the charged lepton mass of the vector-like generations. In addition to this term, in our model, the smuon, charged sleptons of the vector-like generations and their mixing in the mass matrix gives corrections to the muon $g-2$. Among of them, the charged leptons and sleptons of the vector-like generations gives sizable contributions.
 For the numerical calculation of the Higgs mass and the muon $g-2$, we have used Eq.\ (\ref{eq:formulahiggs}) and Eq.\ (\ref{eq:formulag2}).

As for the experiments, the current situation of the Higgs mass and muon $g-2$ are shown in the followings.
The recent combined result of the Higgs boson mass 
$m_{h}^{\text{Exp}}$ reported by ATLAS and CMS collaborations \cite{Aad:2015zhl} is given
by
\begin{eqnarray}
 m_{h}^{\text{Exp}} = 125.09 \pm 0.21 (\text{stat.}) \pm 0.11(\text{syst.})~\text{GeV} .
\end{eqnarray}
%}
%
The discrepancy of the muon $g-2$ between the SM predictions and
experimental value is above $3\sigma$ and quantified as \cite{Bennett:2006fi,hagiwara:2007}
\begin{eqnarray}
\Delta a_{\mu} \equiv a_{\mu}(\text{Exp}) - a_{\mu}(\text{SM}) =
(26.1 \pm 8.0) \times 10^{-10} .
\end{eqnarray}
%
%Here, we comment about the contributions of the SM and the SUSY-breaking sectors to the muon $g-2$.
%In the supersymmetric model with vector-like generations, 
%the contributions to the muon $g-2$ are comes from non-SUSY sector
%and SUSY-breaking sector
%\footnote{
%In Ref. {\cite{Nishida:2016lyk}}, since non-SUSY contributions becomes $\mathcal{O}(10^{-12})$ we ignore that contributions.
%}.
%The contributions to the muon $g-2$ highly depend on
%the texture of flavor.
%Under the texture of Eq. (\ref{eflavor}), 
%the contributions from non-SUSY sector becomes the same 
%order as SUSY-breaking sector.
%Thus, in this paper, we take into account the contributions
%form non-SUSY sector in addition to the contributions form
%SUSY-breaking sector. The analytic formula 
%for calculating the Higgs boson mass and the muon 
%$g-2$ are given in Ref. \cite{Dermisek:2013gta}

In the analysis, we assume the minimal gravity
mediation~\cite{Chamseddine:1982jx} as the boundary condition of the SUSY-breaking scenario. 
In this scenario, there are five parameters: $m_{1/2},~m_0,~A_0,~\tan\beta$ and sigh of $\mu_H$.
In this mediation model, we assume that the mass scale of gaugino, soft scalar and trilinear scalar coupling are universal respectively $m_{1/2},~m_0,~A_0$ at the unification scale.
$\tan \beta$ is fixed to reproduce the fermion masses at low energy
and the sigh of $\mu_H$ is fixed plus so that the contributions to the
muon $g-2$ become positive. 
Thus between the five parameters, there remain three free SUSY-breaking
parameters: $m_{1/2}$, $m_{0}$ and $A_{0}$.
In the following analysis, for the sake of simplicity, $A_{0}$ is 
fixed to be 0 \text{GeV}
\footnote{
Let us comment on $A_{0} \neq 0$. In general, the Higgs boson mass, 
the muon $g-2$ and the DM abundance depend on $A_{0}$ parameter.
However, since these values are determined by mass spectrum
at low energy, the dependence for these values
does not drastically change. Thus, it is sufficient that the
analysis $A_{0}$ is zero.
}.
In addition to the parameters of the minimal gravity mediation, our model have another parameter $y(=Y)$, 
which is exhibited in the superpotential (\ref{superpotential44bar}) and is related to the DM mass.
Yukawa couplings except for $y$ are
determined so that the observed CKM matrix and fermion mass are
reproduced. Since $y$ is insensitive to these observable, 
$y$ can be treated as a free parameter.
After all, there are three parameters:
\begin{eqnarray}
 m_{1/2}, \qquad m_{0},  \qquad  y.
\end{eqnarray}

In the analysis, we have calculated the Higgs boson mass at one-loop level. Inclusion of higher effects gives slight corrections for the mass. At two loop level, the Higgs mass correction in MSSM are studied such as in~\cite{Heinemeyer:1998yj,Heinemeyer:1998np,Carena:2000dp,Degrassi:2002fi,Brignole:2002bz,
Frank:2006yh,Hahn:2013ria,Bagnaschi:2014rsa,Vega:2015fna,Yanagida:2016kag}. 
%\bem{
The dominant contributions of the two-loop effects are from stop fields in loops by the superpotential
\beq
W=(y_u)_{33}u_3Q_3H_u.
\eeq
%
% shown in fig.~\ref{fig:diagram}. 
% The SU(3) gauge fields also give sizable contributions.
% as shown in fig.~\ref{fig:diagramgauge}.
The amount of $\mathcal{O}(\alpha_s \alpha_t)$ and $\mathcal{O}(\alpha_t \alpha_t)$-contributions is typically estimated as
\beq
\begin{split}
\Delta m_h^2
&\simeq
-3\frac{G_F\sqrt{2}}{\pi^2}\frac{\alpha_s}{\pi}
\bar{m}_t^4
\ln^2\lrp{\frac{\bar{m}_t^2}{\bar{M}_S^2}}
+
3\frac{G_F\sqrt{2}}{16 \pi^2}\frac{\alpha_t}{\pi}
\bar{m}_t^4
3 \ln^2\lrp{\frac{\bar{m}_t^2}{\bar{M}_S^2}}
\\
&\simeq
-(16.6{\rm GeV})^2 + (7.2{\rm GeV})^2,
\end{split}
\label{2loopmh}
\eeq
where $G_F$ is the fermi coupling constant, $\alpha_s = g_3^2/(4\pi)$, $\alpha_t = (y_u)_{33}^2/(4\pi)$, $m_t$ is the top mass and $M_{S}$ is the stop mass.
In case of the vector-like generations, scalar components of the fourth, fifth generations and singlet field mediating in the loops additionally contribute to the two-loop corrections.
% the loops of the diagrams in fig.~\ref{fig:diagram}.
In the assignment of the FN charge shown in the Table~\ref{tb:FNcharge}, the superpotential of the Yukawa interaction is given by
\beq
\begin{split}
W_{\rm Yukawa}
&\simeq%%
(y_u)_{24}u_2Q_4H_u+(y_u)_{42}u_4Q_2H_u+(y_d)_{24}d_2Q_4H_d
\\
&~~+%%
(y_e)_{34}e_3L_4H_d + (y_e)_{43} e_4L_3H_d+(y_e)_{42}e_4L_2H_d,
\\
&~~+y_{Q_4}\Phi Q_4\bar{Q} + y_{u_4} \Phi u_{4} \bar{u} + y_{d_4} \Phi d_{4} \bar{d} + y_{e_4} \Phi e_{4} \bar{e},
\end{split}
\eeq
where the Yukawa couplings smaller than 
order of  $\epsilon$
are neglected. Among them, the up-type Higgs interactions together with the interaction of the singlet field
\beq
W_{\rm Yukawa}\ni
(y_u)_{24}u_2Q_4H_u+(y_u)_{42}u_4Q_2H_u
+y_{Q_4}\Phi Q_4\bar{Q}
+y_{u_4}\Phi u_4\bar{u}
\label{eq:yukawapartial}
\eeq
give comparable amplitude. It turns out that the down-type Higgs interactions produce relatively 
small corrections to the Higgs mass in our numerical computations with a large $\tan \beta$ and $\mu_H \simeq 2$TeV. Some of these are similar to the sbottom contribution \cite{Brignole:2002bz}, whereas the remainings are suppressed by 
$1/\tan \beta$.
%This is because in the case where the external lines are down-type Higgs the factor $1/\tan \beta$ is multiplied at vertices.
With interactions Eq.~(\ref{eq:yukawapartial}), we roughly evaluate the contributions from vector-like generations in a diagrammatic way. As for gluon exchange diagrams, 
we obtain new contributions comparable to $\mathcal{O}(\alpha_s \alpha_t)$ in the Higgs mass
%the same order of $\mathcal{O}(\alpha_s \alpha_t^2)$-contributions are obtained 
by replacing stops with the up-type scalar components of second or fourth generations. It is noted that top Yukawa coupling is also replaced with appropriate Yukawas for vector-like generations in Eq.~(\ref{eq:yukawapartial}).
%with appropriate replacement of $\mathcal{O}(1)$ Yukawa couplings in Eq.~(\ref{eq:yukawapartial}). 
By these replacements, it turns out that the number of new diagrams
%, which are evaluated by the same order of $\mathcal{O}(\alpha_s \alpha_t^2)$, 
becomes twice as much as the MSSM ones. 
%Thus, with inclusion of the MSSM contribution, the two-loop corrections from gluon exchange diagrams are approximately evaluated as $3 \times \mathcal{O}(\alpha_s \alpha_t)$ in size.
As for the diagrams given only by Yukawa interactions, 
we obtain new contributions comparable to 
$\mathcal{O}(\alpha_t \alpha_t)$ in the Higgs mass
by replacing stops and up-type Higgs with the up-type scalar components of 
second, fourth, fifth generations, up-type Higgs and the scalar component of $\Phi$. 
Similarly to gluon exchange contributions,
top Yukawa coupling is replaced with appropriate Yukawas in Eq.~(\ref{eq:yukawapartial}).
% Since there are six ways replacements with these fields in Eq.~(\ref{eq:yukawapartial})
%The two-loop corrections from the diagrams given by only Yukawa couplings become six times larger than the MSSM ones since six times more new diagrams than the MSSM ones can be written with fields in Eq.~(\ref{eq:yukawapartial}). 
By these replacements, 
we can see that the number of new diagrams are
five times larger than the MSSM ones. 
%, which are evaluated by the same order of $\mathcal{O}(\alpha_t \alpha_t^2)$, 
%becomes five times larger than the MSSM ones. 
%Thus, the two-loop corrections from diagrams given by only Yukawa interactions are approximately evaluated as $ 6 \times \mathcal{O}(\alpha_t \alpha_t)$ in size.
For a rough estimation, we set all squark masses to be the order of 1 TeV and the mass of the scalar component of $\Phi$ to be the same order of the mass of the sleptons.
In our model, the slepton masses are a few hundred GeV.
Since we calculate the Higgs mass at one-loop level, we treat the two-loop correction as a theoretical uncertainty
and show parameter space with this uncertainty for the Higgs mass as $m_h=122-129$ GeV.
%\footnote{I would like to say that we have $125$GeV at two-loop level when the Higgs mass is $122-129$GeV at one-loop level.} 
%Then, its contribution changes the Higgs mass additionally about 3 GeV. 
(The loop correction with extension to the vector-like generations is also studied in~\cite{Choudhury:2017fuu}.)
For instance, in the case that
%When 
these new diagrams produce the similar sign to the MSSM case (the minus sign for the gluino exchange diagrams and the plus sign for Yukawa interaction diagrams), the Higgs mass at one-loop level can be estimated as
$127$~GeV.

% We show parameter space with this uncertainty for the Higgs mass.

Let us comment on the lepton flavor violation processes.
In the mass matrix Eq.(\ref{eflavor}), the muon has sizable couplings to the vector-like generations, which may induce flavor-changing rare processes. 
%It is well known that these rare processes are strictly constrained by experiments. 
Among them, we focus on the tau decay $\tau \rightarrow \mu \gamma$ and the muon decay $\mu \rightarrow e \gamma$.
Experimental bounds on the branching ratios are given by \cite{Aubert:2009ag,Mori:2016vwi}
\begin{align} 
 {\rm Br}(\tau \rightarrow \mu \gamma) &< 4.4 \times 10^{-8} ~({\rm EXP}), \\
 {\rm Br}(\mu \rightarrow e \gamma) &< 4.2 \times 10^{-13} ~({\rm EXP}).
\end{align}
In a model with vector-like generations, the lepton flavor violating processes are studied in~\cite{Kitano:2000zw,Ibrahim:2012ds,Ibrahim:2015hva}. We calculate the branching ratio Br($\tau \rightarrow \mu \gamma$) and  Br($\mu \rightarrow e \gamma$), following \cite{Ibrahim:2012ds,Ibrahim:2015hva}. 
The decay amplitudes of $l_{i} \rightarrow l_{j} \gamma$ are generally written as
\begin{equation}
\langle l_{j} (p') | J_{\alpha} | l_{i} (p) \rangle = \bar{u}_{e} (p') \Gamma_{\alpha} u_{\mu} (p) \,,
% T (\mu \rightarrow e \gamma) = e \epsilon^{\alpha *} \bar{u}_{e} (p - q)
%\left[ i \sigma_{\alpha \beta} q^{\beta} (A_{L} P_{L} + A_{R} P_{R})  \right]
%u_{\mu}(p) .
\end{equation}
where indices of $i,j(=1,2,3)$ denote the generation, $J_{\alpha}$ is an electromagnetic current for leptons 
and the corrected vertex $\Gamma_{\alpha}$ is given by
\begin{equation}
 \Gamma_{\alpha} (q) =\frac{F^{l_{i} l_{j}}_2 (q) i \sigma_{\alpha \beta} (p^{\prime} - p)^{\beta}}{m_{l_{i}} +m_{l_{j}}}
+\frac{F^{l_{i} l_{j}}_3 (q)  \sigma_{\alpha \beta} \gamma_5 (p^{\prime} - p)^{\beta}}{m_{l_{i}} +m_{l_{j}}}+\cdots.
\end{equation} 
With the these, the branching ratios are given by
%Then, the branching ratio of $l_{i} \rightarrow l_{j} \gamma$ is given by
% Then, the decay rate of $\mu \rightarrow e \gamma$ is given by
\begin{align} 
 {\rm Br} (\tau \rightarrow \mu \gamma) & = \frac{24 \pi^{2}}
{5 G_{F}^{2} m^{2}_{\tau} ( m_{\tau} + m_{\mu})^{2} }
(|F_{2}^{\tau \mu}(0)|^{2} + |F_{3}^{\tau \mu}(0)|^{2}),
 \label{eq:branchingtau} \\
 {\rm Br} (\mu \rightarrow e \gamma) & = \frac{24 \pi^{2}}
{ G_{F}^{2} m^{2}_{\mu} ( m_{\mu} + m_{e})^{2} }
(|F_{2}^{\mu e}(0)|^{2} + |F_{3}^{\mu e}(0)|^{2}),
 \label{eq:branchingmu}
\end{align}
where $G_{F}$ is the fermi coupling constant and hadronic decay modes are included in $\tau$ decay.
The form factors $F_{2,3}^{l_{i} l_{j}}(0)$ are defined in (\ref{eq:form2}) and (\ref{eq:form3}).
In our model, both non-SUSY sector and SUSY one contribute to Br($l_{i} \rightarrow l_{j} \gamma$). 
The former contributions come from one-loop diagrams mediated by W-boson and vector-like neutrinos, and
by Z-boson and vector-like charged leptons.
The latter contributions come from one-loop diagrams mediated by 
charginos (charged Winos and higgsinos) and sneutrinos, and
by neutralinos (neutral Wino and higgsinos) and charged sleptons.
We take into account these lepton flavor violating processes 
as constraints on our model. 
%%%%%%%%%
%We find that
%our model numerically satisfies experimental upper bounds of both ${\rm Br}(\tau \rightarrow \mu \gamma)$ and ${\rm Br} (\mu \rightarrow e \gamma)$ 
%are smaller than the upper bound of the experiment 
%in all parameter regions in Fig. \ref{fig:m12vsDM} and \ref{fig:DMhiggsg2}. 
%For $\mu \rightarrow e \gamma$,
%a red line shows ${\rm Br} (\mu \rightarrow e \gamma) = 1.0\times 10^{-14}$ as a reference value in both Figures. Above the line, the fraction becomes smaller. 
%For $\tau \rightarrow \mu \gamma$, 
%any values of the fraction are not shown because it is too small to reach the current experimental sensitivity. 
%Instead, we show a value of the fraction at a sample point in our model in Table \ref{tb:samplepoint}. 
%($\mu \rightarrow e \gamma$は1.0$\times$ $10^{-14}$をFig. 4.1, 4.2ではred lineで示すと記述 ?).
%From experiments, the current upper bounds of $\tau \rightarrow \mu \gamma$ \cite{Aubert:2009ag} and $\mu \rightarrow e \gamma$ \cite{Mori:2016vwi} are given by
%
%
% Our model avoids these present experimental bounds, but future experiments such as MEG $II$ \cite{Mori:2016vwi} could test the parameter regions.
%
% Thus, we show a constraint of $\mu \rightarrow e \gamma$ process by a red line.

%%%%%
\begin{figure}[t]
\begin{center}
 \includegraphics[width=100mm]{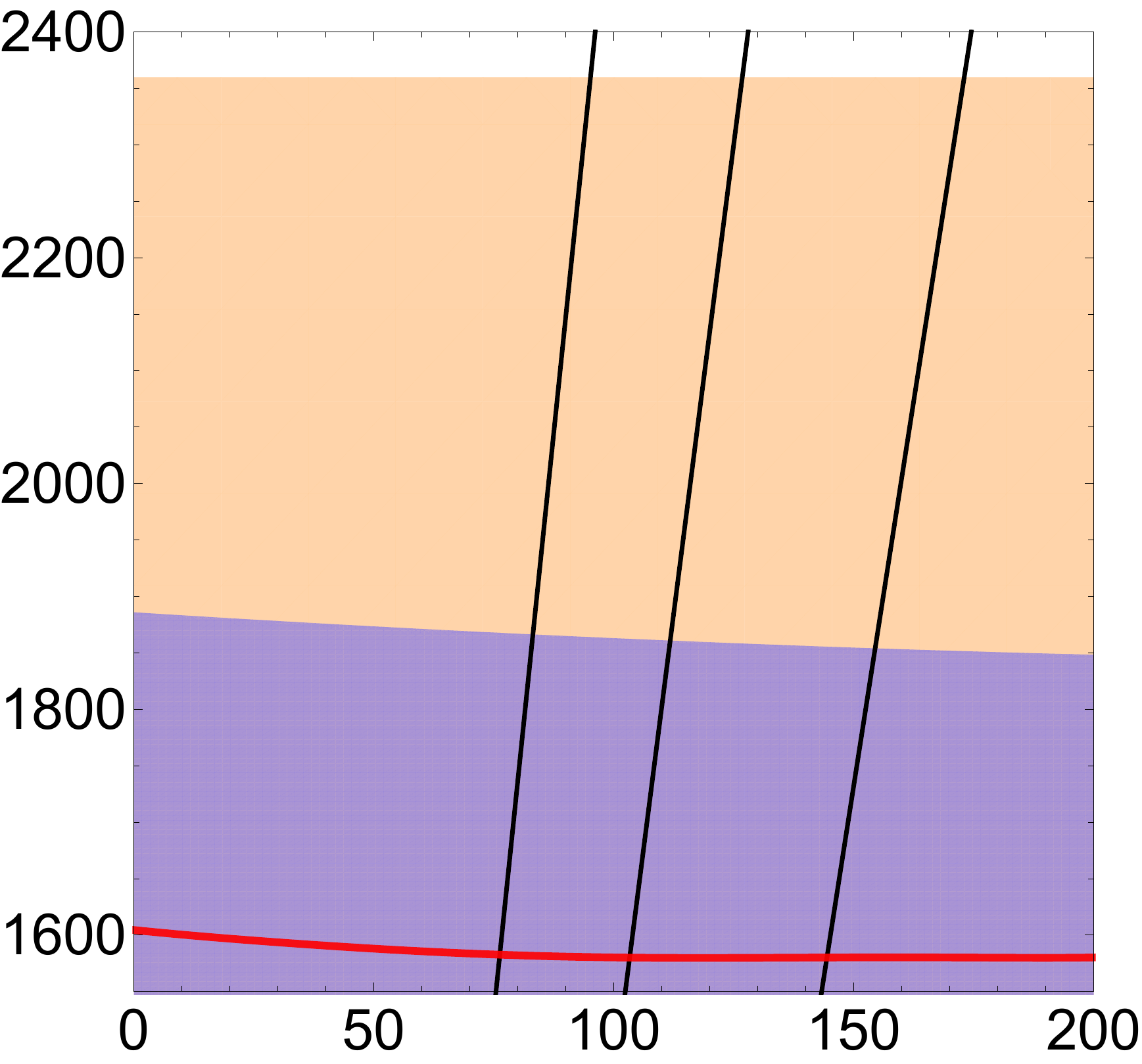}
\end{center}
\caption{The DM abundance, the Higgs boson mass and the muon $g-2$
in $m_{\chi_{\Phi}} - m_{1/2}$ plane with $m_0 = m_{1/2}/20$.
Three black contours correspond to values of $\Omega_{\chi_{\Phi}} h^{2} = 0.01,\ 0.10,\ 1.0$ from left to right.
The Higgs boson with a mass between 122 \text{GeV} and 129 \text{GeV}
is shown in the orange region.
The discrepancy of the muon $g-2$ within 2$\sigma$ level
is explained in the blue region. 
The red line shows ${\rm Br}(\mu \rightarrow e \gamma) = 
1.0 \times 10^{-14}$.
}
\label{fig:m12vsDM}
\end{figure}
\begin{figure}[t]
\begin{center}
 \includegraphics[width=100mm]{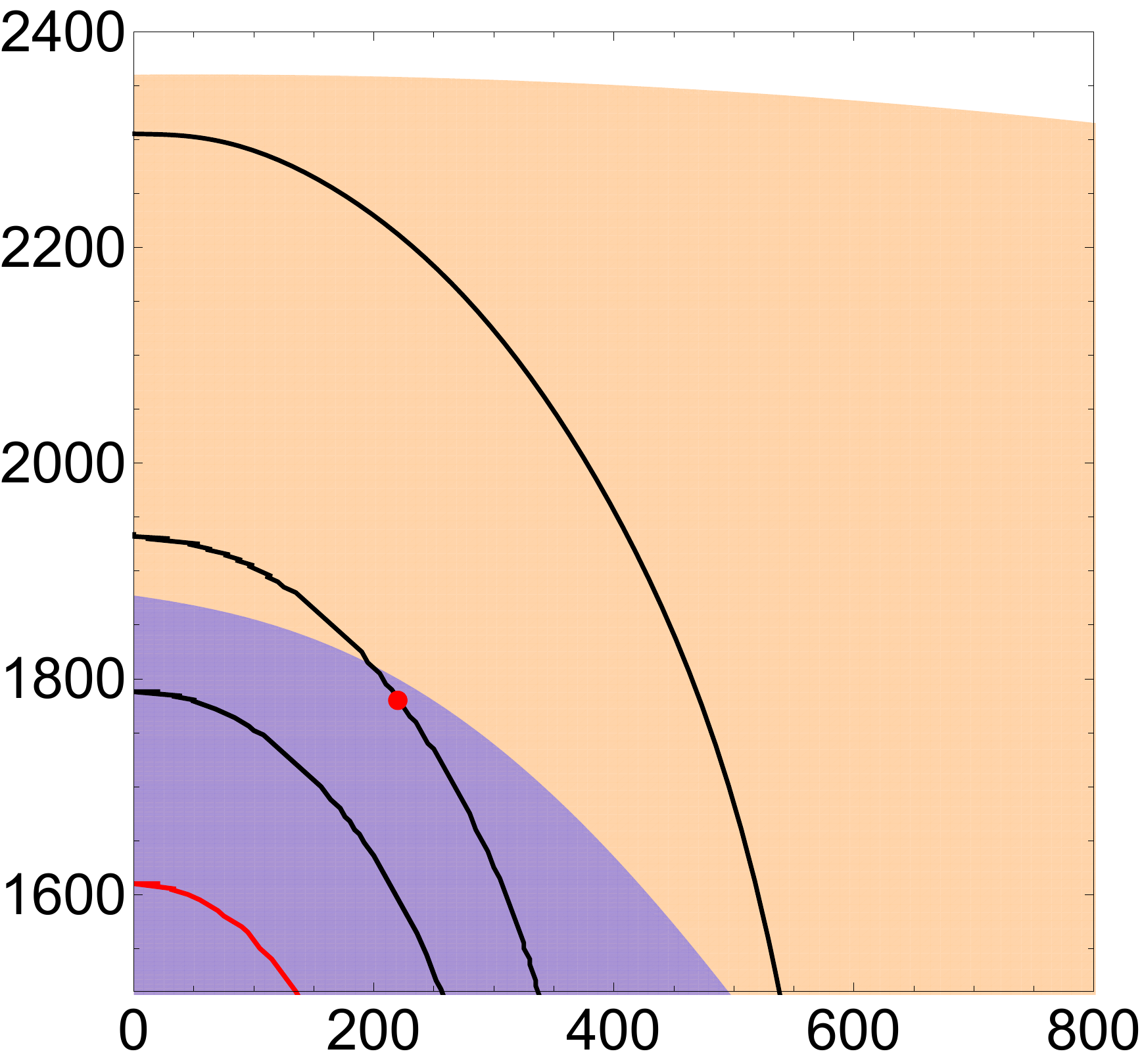}
\end{center}
\caption{
The DM abundance, the Higgs boson mass and the muon $g-2$
in $m_{0} - m_{1/2}$ plane. 
The DM mass is fixed to be 111 GeV ($y(=Y) = 2.1$).
The Higgs boson with a mass between 122 to 129 \text{GeV}
is shown in the orange region.
The discrepancy of the muon $g-2$ within 2$\sigma$ level
is explained in the blue region.
The black contours
show values of $\Omega_{\chi_{\Phi}} h^{2} = 0.08, 0.11, 0.30$
from left to right.
The red line shows ${\rm Br}(\mu \rightarrow e \gamma) = 
1.0 \times 10^{-14}$.
The red circle point is the sample point of our
model.
}
\label{fig:DMhiggsg2}
\end{figure}

In Fig.\ \ref{fig:m12vsDM}, we plot the contours of the Higgs boson
mass, the muon $g-2$, the DM abundance 
and the branching ratio of $\mu \rightarrow e \gamma$
in $m_{\chi_{\Phi}} - m_{1/2}$ plane. 
We fix $m_{0}$ to be $m_{1/2}/20$ 
because the small ratio of $m_{0}$ to $m_{1/2}$ is preferred to explain the muon $g-2$ \cite{Nishida:2016lyk}.
It is known that such mass spectra can be obtained in gaugino mediation scenarios \cite{Kaplan:1999ac}.
The orange region shows the Higgs boson
mass in the range from 122 GeV to 129 GeV.
The blue region explains the muon $g-2$ anomaly within 2$\sigma$
level.
Three black contours show $\Omega_{\chi_{\Phi}} h^{2} =
0.01,\ 0.1,\ 1$ from left to right.
We find also that
our model numerically satisfies experimental upper bounds of both ${\rm Br}(\tau \rightarrow \mu \gamma)$ and ${\rm Br} (\mu \rightarrow e \gamma)$ 
%are smaller than the upper bound of the experiment 
in all parameter regions of Fig.\ref{fig:m12vsDM} and \ref{fig:DMhiggsg2}. 
For $\mu \rightarrow e \gamma$,
a red line shows ${\rm Br} (\mu \rightarrow e \gamma) = 1.0\times 10^{-14}$ as a reference value in both Figures. Above the line, the fraction becomes smaller. 
For $\tau \rightarrow \mu \gamma$, 
any values of the fraction are not shown because it is too small to reach the current experimental sensitivity. 
%Instead, we show a value of the fraction at a sample point in our model in Table \ref{tb:samplepoint}.
%\bem{
%The red line is the contour of ${\rm Br}(\mu \rightarrow e \gamma) = 1.0 \times 10^{-14}$.
%}
As seen in the figure, there exits the regions where the Higgs boson mass, the muon $g-2$ and the DM abundance can be explained at the same time. It is noted that the DM abundance highly depends on the DM mass. In our model, the DM mass is mostly determined with only ${\cal O}(1)$ parameters, whereas
%Nevertheless the order of the DM mass which can 
%explain the DM abundance is naturally obtained by the FN charge assignment, where 
realistic values of the CKM matrix and fermion masses are reproduced. On top of those, 
the masses of the vector-like generations are determined by the VEV of the DM (singlet) multiplet $\Phi$.
In this sense, the presence of the DM supports the existence of three generations in low energy scales.
%Thus, the flavor structure are closely associated with the 
%DM physics in our model.

In Fig.\ \ref{fig:DMhiggsg2} we similarly show the Higgs boson mass, the muon $g-2$, the contours of the DM abundance
and the branching ratio of $\mu \rightarrow e \gamma$
in $m_{0} - m_{1/2}$ plane.
We set the DM mass $m_{\chi_{\Phi}}$ to be $111$ GeV which means that
$y= 2.1$ in other words.
The orange and blue regions are the same as in Fig.\ \ref{fig:m12vsDM}.
The black contours shows values of $\Omega_{\chi_{\Phi}} h^{2} = 0.08, 
0.11, 0.3$ form left to right.
As seen in Fig.\ \ref{fig:DMhiggsg2},
the Higgs boson mass, the muon $g-2$ and
$\Omega_{\chi_{\Phi}} h^{2} = 0.11$ can be explained simultaneously where
$m_{1/2}$ is lying in the range between 1500 GeV to 1800 GeV and $m_{0}$ is around 300 GeV.
%This region gives predictions of mass spectra in our model
The red circle point in Fig.\ \ref{fig:DMhiggsg2} is the sample point of our model: $(m_{1/2}, m_{0}, y) = (1780, 250, 2.1)$
\footnote{
In the region where $m_{1/2}$ is below 1500 GeV,
there is likely the region where the muon $g-2$ anomaly is
explained within $1\sigma$, but the gluino mass in such regions
is below 1 TeV which is excluded region by LHC \cite{Olive:2016xmw}.
}.
We show typical mass spectrum, the Higgs boson
mass, the muon $g-2$, the DM abundance
and the branching ratio of $\tau \rightarrow \mu \gamma$ and $\mu \rightarrow e \gamma$
in Table \ref{tb:samplepoint}.
In Table \ref{tb:samplepoint}, $M_{3}$ is the gluino mass,
$m_{\tilde{N}_{1}}$ is the lightest neutralino in the MSSM sector, $m_{\tilde{C}_{1}}$ is
the lightest chargino, $m_{\text{stop}}$ is lighter stop mass,
$m_{t^{\prime}}$ is the mass of vector-like generations for up-type quark,
$m_{\text{charged slepton}}$ is mass of the lightest charged slepton which is the vector-like generations one, 
$m_{L^{\prime}}$ is the mass of vector-like generations for charged lepton.
\begin{table}[t]
\begin{center}
\begin{tabular}{|c|c|c|c|} \hline
      &  Sample Point (1)  \\ \hline
 $m_{1/2}$               &     1780       \\
 $m_0$                   &      250       \\
 $y(=Y)$                     &      2.1        \\
 \hline
 $M_3$                   &     1202      \\
 $m_{\tilde{N}_{1}}$     &     347.4      \\
 $m_{\tilde{C}_{1}}$     &     587.5      \\
 $m_{\text{stop}}$    &     1394      \\
 $m_{t^{\prime}}$ &  1157  \\ 
 % $m_{\tilde{u}_{4L,4H,5L,5H}}$  &  $2715-3973$ \\
 % $m_{\tilde{e}_{2L}}$,$m_{\tilde{e}_{2H}}$
 %                         & 952.2, \ 1221 \\
 $m_{\text{charged slepton}}$    &     324.0   \\
 $m_{L^{\prime}}$ &  274.4  \\ 
 % $m_{\tilde{e}_{4H,5L,5H}}$ & $1129-1860$   \\
 % $m_{\tilde{\nu}_2}$     &     1227         \\
 % $m_{\tilde{\nu}_{4,5}}$ &  816, \ 1773    \\
 \hline
 $m_{h^0}$               &     125.2         \\
 $\Delta a_\mu$  &  $10.3\times 10^{-10}$ \\ 
 $\Omega_{\chi_{\Phi}} h^2$ &  0.116 \\
${\rm Br}(\tau \rightarrow \mu \gamma)$ & $5.5 \times 10^{-13}$ \\ 
${\rm Br}(\mu \rightarrow e \gamma)$ & $6.9 \times 10^{-15}$ \\ 
\hline
\end{tabular}
\caption{The sample points in our model. All the mass parameters are
given in unit of GeV\@.
$m_{\tilde{N}_{1}}$ is the lightest neutralino in the MSSM sector, $m_{\tilde{C}_{1}}$ is the lightest chargino, $m_{t^{\prime}}$ is the mass of fourth generations for up-type quark, $m_{L^{\prime}}$ is the mass of fourth generations for charged lepton. $\mu_{H}(\sim 2~\text{TeV})$ and $b$ terms are determined so that the electroweak symmetry breaking occurs.
}
\label{tb:samplepoint}
\end{center} 
\end{table}

Mixings of the vector-like generations with the three generations make a electric dipole moment (EDM). We show typical values of the electron EDM $d_e$ and then discuss the constraints from experiments. The contributions are mainly from four sources: the chargino exchange, the neutralino exchange, the $W$ boson exchange and the $Z$ boson exchange.  Each contribution is estimated as following~\cite{Ibrahim:2014oia}.
\begin{itemize}
 \item chargino contribution
\begin{align}
 | d_{e}^{\chi^{+}} | \sim 8.7 \times 10^{-29} e\ {\rm cm}
\left( \frac{ m_{\tilde{W}} }{ 600\ {\rm GeV}  }  \right)
\left( \frac{ 900\ {\rm GeV} }{ m_{\tilde{\nu}_{e}}  }  \right)^{2}
\left( \frac{ | c_{L \tilde{\nu}_{e} \tilde{W}} 
c_{R \tilde{\nu}_{e} \tilde{W}} |  } { 10^{-5} } \right)
\left( \frac{ {\rm arg}\lrp{c_{L \tilde{\nu}_{e} \tilde{W}} 
c_{R \tilde{\nu}_{e} \tilde{W}}}} {1.1 \times 10^{-4}}  \right),
\end{align}
where $m_{\tilde{\nu}_{e}}$ is the mass of snutrino, $m_{\tilde{W}}$ is wino mass, and $c_{L\tilde{\nu_e}\tilde{W}}$ and $c_{R\tilde{\nu_e}\tilde{W}}$ are couplings of mixings defined in appendix.~\ref{subsec:muong-2}. 
 \end{itemize}
\begin{itemize}
 \item neutralino contribution
\begin{align}
 | d_{e}^{\chi^{0}} | \sim 8.7 \times 10^{-29} e\ {\rm cm}
\left( \frac{ m_{\tilde{B}} }{ 300\ {\rm GeV}  }  \right)
\left( \frac{ 1000\ {\rm GeV} }{ m_{\tilde{e}_{1}}  }  \right)^{2}
\left( \frac{ | n_{L \tilde{e}_{1} \tilde{B}} 
n_{R \tilde{e}_{1} \tilde{B}} |  } { 10^{-5} } \right)
\left( \frac{{\rm arg}\lrp{n_{L \tilde{e}_{1} \tilde{B}} 
n_{R \tilde{e}_{1} \tilde{B}}}} {2.2 \times 10^{-4}}  \right),
\end{align}
where $\tilde{e}_{1}$ is the mass of lighter selectron, $m_{\tilde{B}}$ is bino mass, and $n_{L\tilde{e}_i\tilde{B}}$ and $c_{R\tilde{e}_i\tilde{B}}$ are couplings of mixings.
 \end{itemize}
\begin{itemize}
 \item W boson contribution
\begin{align}
 | d_{e}^{W} | \sim 5.9 \times 10^{-30} e\ {\rm cm}
\left( \frac{ m_{L^{\prime}} }{ 300\ {\rm GeV}  }  \right)
\left( \frac{ 80\ {\rm GeV} }{ m_{W}  }  \right)^{2}
\left( \frac{ | g^{W L}_{\nu_{4} e}  g^{W R}_{\nu_{4} e} |  } { 10^{-11} } \right)
\left( \frac{{\rm arg}\lrp{g^{W L}_{\nu_{4} e}  g^{W R}_{\nu_{4} e}}} {10^{-1}}  \right),
\end{align}
where $m_{L^{\prime}}$ is the vector-like lepton mass, and $g^{WL}_{\nu_4e}$ and $g^{WR}_{\nu_4e}$ are couplings of mixings.
 \end{itemize}
\begin{itemize}
 \item Z boson contribution
 \begin{align}
 | d_{e}^{Z} | \sim 4.6 \times 10^{-30} e\ {\rm cm}
\left( \frac{ m_{L^{\prime}} }{ 300\ {\rm GeV}  }  \right)
\left( \frac{ 90\ {\rm GeV} }{ m_{Z}  }  \right)^{2}
\left( \frac{ | g^{Z L}_{e_{4} e}  g^{Z R}_{e_{4} e} |  } { 10^{-11} } \right)
\left( \frac{{\rm arg}\lrp{g^{Z L}_{e_{4} e}  g^{Z R}_{e_{4} e} }} {10^{-1}}  \right),
\end{align}
where $g^{ZL}_{\nu_4e}$ and $g^{ZR}_{\nu_4e}$ are couplings of mixings.
 \end{itemize}
Among of them, the chargino and neutralino exchanges are the main contributions to $d_e$. Since the $\mu$ term of our model is around $2~{\rm TeV}$, the Higgsinos decouple from the mixing and its couplings are smaller than those of the gauninos. Thus, $d_e$ is manly given by the exchange of the gauginos. From experiments, its value is constrained as~\cite{Baron:2013eja}
\beq
 | d_{e} | < 8.7 \times 10^{-29} e\ {\rm cm}~~({\rm EXP}).
\eeq
Therefore, the CP phase of the couplings should  satisfy 
\beq
\theta<1.1\times10^{-4},
\eeq
where $\theta$ represents each phase.
%%%%%%%%%
\section{Conclusion and Discussion} \label{sec:con}
%%%%%%%%%
In this paper we studied the flavor structure in
a model with vector-like generations by using Froggatt-Nielsen
mechanism.
It is notable that the assignment of FN charges can be determined so that the CKM matrix and fermion masses 
at the $M_{Z}$ scale are reproduced.
Furthermore, under such FN charge assignments,
the fermion component of the gauge singlet superfield becomes a candidate of DM.
The DM mass is induced through the RG flow including the flavor textures which
can explain observed flavor physics of muon $g-2$.
With such flavor textures,
it is found that there exists parameter regions
where we can explain the Higgs boson mass, the muon $g-2$ and the DM abundance simultaneously.
The singlet plays two roles: One is to fix the vector-like mass by the VEV, and another is that
its fermion component is a DM candidate.
In this sense, the presence of the DM supports the existence of three generations in low energy scales.

Let us here discuss about the prospect for discovering the DM particle $\chi$ at direct/indirect detection experiments. In our model, the singlino $\chi$ couples to the first generation of the quark field mediated by the scalar component of the vector-like particle. By integrating out the mediator field, we obtain the effective interaction between DM and first generation as ${\cal L}_{\rm int}\simeq (y_1^2/m^2)\bar{\chi}\bar{Q}_1\chi Q_1$. Here $m$ is the  mass of the mediator field of order ${\cal O}({\rm TeV})$, and $y_1$ is the Yukawa coupling constant. 
With this interaction, $\chi$ scatters nucleons spin-independently with the cross section $\sigma_{\rm SI}^N=y_1^4\times{\cal O}(10^{2})\,{\rm pb}\,(\mu_{\chi}/10^2\,{\rm GeV})^2/(m/{\rm TeV})^4$~\cite{Belanger:2008sj}, where $\mu_{\chi}$ is the reduced mass defied as $\mu_{\chi}=m_Nm_{\chi}/(m_{\chi}+m_N)$ with nuclei mass $m_N$.  
Since the Yukawa coupling of the singlino to the first  generation of the quark is set to the small value by the FN mechanism as 
$y\simeq10^{-3}(\epsilon/0.33)^6$,
 we see that the magnitude of the spin independent cross section is reduced to
$
\sigma_{SI}^N\simeq10^{-10}{\rm pb}.
$
Thus, in the future direct detection experiments such as XENON1T~\cite{Aprile:2012zx} or DARWIN~\cite{Aalbers:2016jon}, our DM model will be  tested.
In the lepton sector, $\chi$ interacts with third generations by larger coupling 
$y^2\simeq10^{-1}(\epsilon/0.33)^2$
mediated by the vector like particle. Thus, through the annihilation of the DM particles in the Galactic Center or in the dwarf spheroidal galaxies, a significant excess of gamma rays might be produced. The excess of the energetic gamma ray might  be detected in indirect detection experiments such as Fermi-LAT~\cite{Ahnen:2016qkx,Ackermann:2015zua} or CTA~\cite{Wood:2013taa}. 

In the present paper, we have studied the thermal production of the DM through the interactions with the quarks and leptons, respecting the original model~\cite{Nishida:2016lyk} shown in (\ref{eq:oriY}), but from the arguments based on symmetries, there can be allowed other terms such as $W\propto \Phi,~\Phi^2$ or $\Phi H_uH_d$. In this paper, we have just dropped these terms by hands following~\cite{Nishida:2016lyk}, but the terms might affect the DM abundance. We will investigate the effects in the future work, but here briefly discuss the issues. Among the interactions, the coupling of the singlet fields with the Higgs fields
\beq\label{eq:PHH}
W=\lambda\Phi H_uH_d
\eeq
might give a large contributions to the experimental results.
This interaction makes the effective $\mu$-term, and contributes to the cross section of the dark matter with nucleons. Thus, to reproduce the result of our analysis and avoid the bound from direct detection experiments, we need to appropriately choose the coupling constants $\lambda$, trilinear coupling $Y$ and the VEV of the singlet field $\braket{\Phi}=V$, but on the other hands it affects muon $g-2$ and Higgs mass corrections. Here we discuss these prospects. With the interaction, the DM components $\chi_{\Phi}$ could have a sizable interaction with nucleons through the t-channel process of the neutral Higgs boson, and it affects the scattering process of the dark matter and nucleons~\cite{Cerdeno:2004xw,Ellwanger:2009dp}. Especially interactions with the strange-quark mainly contribute to the scattering, and its cross section is given in~\cite{Ellwanger:2009dp} by
\beq\label{eq:SICrossSection}
\sigma_{\rm SI}^{N} \sim \mu_{\chi} \frac{Y^2 h_{s}^2}{M_{Ha}^4} S_{a3}^2 S_{a1}^2,
\eeq
where $\mu_{\chi}$ is the reduced mass, $h_{s}$ is Yukawa coupling of the strange quark, $M_{Ha}$ is the Higgs mass,  $S_{a3}$ and $S_{a1}$ are diagonalization matrix elements for CP-even neutral Higgs boson (See~\cite{Ellwanger:2009dp} for the definition), and subscript $a$ runs from $1$ to $3$. The subscript represents  the mass eigenstates of the neutral Higgs with  order of $M_{H_1} < M_{H_2} < M_{H_3}$. Here the lightest Higgs $H_1$ is identified as the detected one of 125 GeV mass \cite{Cerdeno:2007sn}, and its contribution would be largest. To avoid current bound from the direct detection, we need to choose small value of $Y$, but to reproduce DM mass $m_{\chi_{\Phi}}=YV\simeq10^2{\rm GeV}$, $V$ is required to be larger than $10^2{\rm GeV}$. On the other hand, this singlet VEV also gives effective $\mu$-term as $\mu_{\rm eff}=\lambda V$. From the prospect of the Landau pole, $\lambda$ needs to be smaller than unity, and it requires $V$ to be larger than $10^{2}{\rm GeV}$ to satisfy $\mu_{\rm eff}=2{\rm TeV}$. Further, since the scattering of the DM with the SM quarks is mediated through the interaction(\ref{eq:PHH}), the smaller value of $\lambda$ is also required to avoid the experimental bounds on the spin-independent cross section (\ref{eq:SICrossSection}). This large value of $V$ leads to the decoupling limit of vector-like generations (i.e., the MSSM-like limit), and then the contributions from the vector-like generations to the Higgs mass and the muon $g-2$ might be too small to explain the experimental values simultaneously. Moreover, the Higgs mass has to be evaluated by taking into account corrections from a coupling $\lambda$. As for the thermal production process, the DM becomes more likely to annihilates into the SM particles mediated by the Higgs fields due to the interaction with the Higgs fields. Thus, we can expect that larger DM mass is required to explain the observed abundance. 
On the other hand, the interaction might be forbidden by some symmetry such as the R-symmetry. However, at the same time, several terms of the Yukawa interactions (\ref{superpotential44bar}) would be also inevitably absent by the assignment, and the absence would change the muon $g-2$. We investigate these issues in the future work.

Within the FN charge assignment in Table~\ref{tb:FNcharge}, flavor violating processes induced by the mixing between $\tau_R\leftrightarrow\mu_R$ or $d_R\leftrightarrow s_R$ could be allowed. 
% In this case, the SUSY flavor problem is not avoided by the assignment, but there would be another choice such that the off-diagonal elements of squark mass matrix are forbidden and we avoid the problem.
In this case, the SUSY flavor problem is not avoided by the assignment without a high scale SUSY breaking or a flavor blind mediation mechanism,
but there would be other choices such that the off-diagonal elements of squark mass matrix is suppressed and we avoid the problem.
% In other aspects, the problem can be avoided by high scale SUSY or flavor blind mediation models.
% In general, these flavor violations make SUSY flavor problem. 
In our model, since the mSUGRA scenario is assumed, the Yukawa matrices of the SUSY breaking sector are diagonal. Thus, the SUSY flavor problem is avoided.

We have not taken into account the presence of right-handed neutrino, the relevant flavor textures, and
collider physics in this paper. These are important to test our model. We will reveal these prospects in the future work.

\section*{Acknowledgement}
We would like to Koichi Yoshioka and Wen Yin for fruitful discussions.
This work is supported by MEXT-Supported Program for the Strategic Research Foundation at Private Universities,``Topological Science, Grant Number S1511006 
(T.H and N.T.) and JSPS KAKENHI Grant Number 26247042 (T.H.).

\appendix

\section{Annihilation cross section} \label{sec:tac}
In this Appendix, we give the expressions of $a$ and $b$ in Eq.\ (\ref{eq:omegadm}) which are needed to calculate the DM abundance.
We calculate $a$ and $b$ in the following way.
First, as in Fig.\ \ref{fig:crosec},
we consider the DM annihilation into the SM fermions, which
is denoted by $f_{i}$ (except top quark) by 
exchanging sfermions, which is denoted by $\tilde{f}_{\alpha}$.
Next, we evaluate square of the scattering amplitude, which
is given by t-channel and u-channel process, as the 
annihilation cross section $\sigma_{\text{ann}}$.
Lastly, we derive the thermal averaged cross section 
$\left< \sigma_{\text{ann}} v_{\text{rel}} \right> = a + b / x_{F}$
followed by \cite{Roszkowski:1994tm}.
In the limit where the mass of final state in DM annihilation
can be ignored, compared with the DM mass,  
the coefficients $a$ and $b$ are defined by
\begin{eqnarray}
 a &=& a^{0}, \\ \nonumber
 b &=& -3 a_{0} + b_{0}.
\end{eqnarray}
 After straightforwardly calculations, 
 $a_{0}$ and $b_{0}$ are given by 

\begin{eqnarray}
a_{0} &=& \frac{\dm{2}}{2^{5} \pi}
\left[
\sum_{\alpha, \beta, i, j}
\frac{ ( O_{R i \alpha} O_{R j \beta} + 
O_{L i \alpha} O_{L j \beta} )^{2} }
{ \Delta_{\tilde{f}_{\alpha}}  \Delta_{\tilde{f}_{\beta}}  } 
-
\sum_{\alpha, \beta, i, j}
\frac{ O_{R i \alpha}^{2} O_{R j \beta}^{2}  +
       O_{L i \alpha}^{2} O_{R j \beta}^{2}  }
     {  \Delta_{\tilde{f}_{\alpha}} \Delta_{\tilde{f}_{\beta}}  }
\right], \\
b_{0}  &=& \frac{ \dm{2} }{ 2^{5} \pi}
\sum_{\alpha, \beta, i, j}
\frac
{O_{R i \alpha} O_{R j \beta} O_{ L i \alpha } O_{ L j \beta}}
{ \Delta_{\tilde{f}_{\alpha}}^{3} \Delta_{\tilde{f}_{\beta}}^{3} }
\times \\ \nonumber 
 &{}&  \left\{ 
\Delta_{\tilde{f}_{\alpha}}  \Delta_{\tilde{f}_{\beta}} 
\left(
2 \dm{4} + 10 \dm{2} \sfer{\alpha}{2} 
 - 30 \dm{2} \sfer{\beta}{2}
- 30 \sfer{\alpha}{2} \sfer{\beta}{2}
\right) +
12 \dm{8} - 2 \dm{6} \sfer{\alpha}{2} \right. \\ \nonumber
&{}& \left.
+ \dm{6} \sfer{\beta}{2} -16 \dm{4}\sfer{\alpha}{4} + 
\dm{4} \sfer{\beta}{4} - 4 \dm{4} \sfer{\alpha}{2} \sfer{\beta}{2}
-13 \dm{2} \sfer{\alpha}{4} \sfer{\beta}{2}
+3 \dm{2} \sfer{\alpha}{2} \sfer{\beta}{4} \right. \\ \nonumber
&{}& \left.
+ 
2 \sfer{\alpha}{4} \sfer{\beta}{4}
\right\} \\ \nonumber
&{}& +  \frac{ \dm{2} }{ 2^{5} \pi}
\sum_{\alpha, \beta, i, j}
\frac{ (O_{R i \alpha} O_{R j \beta}+ O_{L i \alpha} O_{L j \beta})^{2} }
{ \Delta_{\tilde{f}_{\alpha}}^{3} \Delta_{\tilde{f}_{\beta}}^{3}} \times
\\ \nonumber
&{}& \left\{
 -24 \Delta_{\tilde{f}_{\alpha}} \Delta_{\tilde{f}_{\beta}} 
(\dm{4} - \sfer{\alpha}{2} \sfer{\beta}{2} )
+ 10 \dm{8}
+6 \dm{6} \sfer{\alpha}{2} + 6 \dm{6} \sfer{\beta}{2}
- 26 \dm{4} \sfer{\alpha}{2} \sfer{\beta}{2} \right. \\ \nonumber
&{}& \left.
+ 4 \dm{4} \sfer{\alpha}{4} + 4 \dm{4} \sfer{\beta}{4} 
- 8 \dm{2} \sfer{\alpha}{4} \sfer{\beta}{2} 
- 8 \dm{2} \sfer{\alpha}{2} \sfer{\beta}{4}
\right\} \\ \nonumber
&{}& +
\frac{ \dm{2} }{ 2^{5} \pi}
\sum_{\alpha, \beta, i, j}
\frac{ ( O_{R i \alpha}^{2} O_{R j \beta}^{2} + O_{L i \alpha}^{2} O_{L j \beta}^{2} ) }
{ \Delta_{\tilde{f}_{\alpha}}^{3} \Delta_{\tilde{f}_{\beta}}^{3}} \times
\\ \nonumber
&{}& \left\{
\Delta_{\tilde{f}_{\alpha}} \Delta_{\tilde{f}_{\beta}}
( 3 \dm{4} + 2 \dm{2} \sfer{\alpha}{2} + \dm{2} \sfer{\beta}{2}
- 2 \sfer{\alpha}{2} \sfer{\beta}{2})
- 16 \dm{8} - 16 \dm{6} \sfer{\alpha}{2} \right. \\ \nonumber
&{}& \left. 
- 16 \dm{6} \sfer{\beta}{2}
-8 \dm{4} \sfer{\alpha}{4} -8 \dm{4} \sfer{\beta}{4}
\right\},
\end{eqnarray}
where $\Delta_{\tilde{f}_{\alpha,\beta}} \equiv \dm{2} + 
m_{\tilde{f}_{\alpha,\beta}}^{2} $,
the subscript $i,j$ correspond to the generation of the SM particles
whose the mass is below the freeze-out temperature and 
the subscript $\alpha, \beta$ corresponds to the generations
of sfermions which is regarded as the mediator between the DM
and the SM particles in our model.
The dependence of $a$ and $b$ on the DM and sfermion masses is roughly
given by
\begin{eqnarray}
 a, b \sim \frac{ O^{4} }{ m_{\tilde{f}}^{2} }
           \frac{ m_{\chi_{\Phi}}^{2} }{ m_{\tilde{f}}^{2} }
\sim \frac{ y^{4} }{ m_{\tilde{f}}^{2} }
           \frac{ m_{\chi_{\Phi}}^{2} }{ m_{\tilde{f}}^{2} }
,
\end{eqnarray}
where $O$ means the coupling constants between the DM and sfermions
and $y$ is a Yukawa coupling between them.

%%%%%%%%%%
\section{Analytic formulae for Higgs mass and muon $g-2$} \label{sec:analytic}
%%%%%%%%%%

%%%%%%%%%%%%
\subsection{Higgs mass} \label{sec:higgsanaly}
%%%%%%%%%%%%
The one-loop correction to the lightest neutral Higgs mass $\Delta m^2_{h}$ is given by \cite{Martin:2009bg}
\begin{eqnarray}
\Delta m_{h}^2 & = 
\Bigg[
\dfrac{\sin^2 \beta}{2} \left( \dfrac{\partial^2}{\partial v_u^2} -
\dfrac{1}{v_u} \dfrac{\partial}{\partial v_u} \right) +
\dfrac{\cos^2 \beta}{2} \left( \dfrac{\partial^2}{\partial v_d^2} -
\dfrac{1}{v_d} \dfrac{\partial}{\partial v_d} \right)  \nonumber \\ 
& \qquad + 
\sin\beta \cos\beta \dfrac{ \partial^2 }{\partial v_u \partial v_d}
\Bigg] \Delta V_H
\label{eq:higgsmasscal}
\end{eqnarray}
where $\Delta V_{H}$ is the one-loop corrections to the Higgs potential and is defined as
\begin{eqnarray}
\qquad \Delta V_H = \sum_{X=u, d, e}\,\sum_{i=1}^{10} 
2 N_c \left[
F( m_{\tilde{X}_i}^2 ) - F ( m_{X_i}^2 )
\right], \hspace{7mm} N_{c} =
\begin{cases}
3 & (X=u,d) \\
1 & (X=e)
\end{cases}
\end{eqnarray}
where $m_{X_i}^2$ and $m_{\tilde{X}_i}^2$ are the squared-mass
eigenvalues of fermions and scalars, respectively, which are obtained
by diagonalizing (\ref{uflavor})-(\ref{eflavor}) for
fermions ($M_u^\dagger M_u$ and $M_uM_u^\dagger$, etc.).
For diagonalization of scalars mass matrices, we use the scalar mass matrices which are defined in Eq.\ (B.1)--(B.3) in \cite{Nishida:2016lyk}. The function $F$ is defined as \cite{Martin:2009bg}
\begin{eqnarray}
F(x) = \frac{x^2}{64\pi^2} \left[
\ln \left( \frac{x} {\mu^2} \right) - \frac{3}{2} \right] ,
\end{eqnarray}
where $\mu$ represents the renormalization scale which is set to be 
$M_{\text{SUSY}}$ in evaluating the Higgs mass.

%%%%%%%
\subsection{Muon g-2}
\label{subsec:muong-2}
%%%%%%%
We show the SUSY contributions $\Delta a_{\mu}^{\rm SUSY}$ and non-SUSY contributions $\Delta a_{\mu}^{\rm non-SUSY}$ in Eq.\ (\ref{eq:formulag2}). First, we consider the SUSY contributions. In order to calculate the SUSY contributions, we use the mass eigenstate basis for gauginos, charged leptons, charged sleptons and neutral sleptons. The analytic formula of SUSY contributions is the same as the previous paper \cite{Nishida:2016lyk}. Let us define the diagonalization matrix for neutralinos, charginos, sneutrinos, in order to evaluate the SUSY contributions $\Delta a_{\mu}^{\rm SUSY}$ of the muon $g-2$. In the basis of \{$\tilde{B}, \tilde{W}^{0}, \tilde{H}^{0}_{d},\tilde{H}^{0}_{u}$\}, the neutralino mass matrix $M_{\chi^{0}}$ is given by
\begin{eqnarray}
M_{\chi^0} = \left(
\begin{array}{cccc}
M_1  &  0  &  -g_1 v_d/\sqrt{2}  &  g_1 v_u/\sqrt{2}  \\
0  &  M_2  &  g_2 v_d/\sqrt{2}  &  -g_2 v_u/\sqrt{2}  \\
-g_1 v_d/\sqrt{2}  &  g_2 v_d/\sqrt{2}  &  0  &  -\mu_H  \\
g_1 v_u/\sqrt{2}  &  -g_2 v_u/\sqrt{2}  &  -\mu_H  &  0
\end{array}
\right) . \label{neutralinomassmat} 
\end{eqnarray}
In the basis of \{$\tilde{W}^{-}, \tilde{H}_{d}^{-}$\} and
\{$\tilde{W}^{+}, \tilde{H}_{u}^{+}$\}, the chargino mass matrix
$M_{\chi^{\pm}}$ is given by
\begin{eqnarray}
M_{\chi^\pm} = \left(
\begin{array}{cc}
M_2  &  \sqrt{2} g v_u  \\
\sqrt{2} g v_d  &  \mu_H
\end{array}
\right),  \label{charginomassmat}
\end{eqnarray}
where the charged wino $\tilde{W}^{\pm}$ are defined as
\begin{eqnarray}
\tilde{W}^\pm = \frac{i}{\sqrt{2}}
( \tilde{W}^1 \mp i \tilde{W}^2 ) .
\end{eqnarray}
By using the neutralino mixing matrix $N$ and the chargino mixing matrices $J, K$, the mass matrix in Eq.\ (\ref{neutralinomassmat}) and (\ref{charginomassmat}) are diagonalized by
\begin{align}
N M_{\chi^0} N^\dagger & = {\rm diag} \big(\,
m_{\chi^0_1}, m_{\chi^0_2}, m_{\chi^0_3}, m_{\chi^0_4}
\big),  \label{nuediagonalize} \\
J M_{\chi^\pm} K^\dagger & = {\rm diag} \big(  
m_{\chi^\pm_1}, m_{\chi^\pm_2}
\big) ,  \label{chardiagonalize}
\end{align}
where $m_{\chi^0_x}$ ($x=1, \ldots, 4$) are the positive mass eigenvalues\ ($m_{\chi^0_x} < m_{\chi^0_y}$, if $x < y$),\ and $m_{\chi_x}^\pm$ ($x = 1,2$) are the positive mass eigenvalues ($m_{\chi^\pm_1} < m_{\chi^\pm_2}$). The diagonalization of neutral sleptons are defined by
\begin{align}
( U_{\tilde{\nu}} M^2_{\tilde{\nu}} U_{\tilde{\nu}}^\dagger )_{\alpha\beta} 
& = m^2_{\tilde{N}_\alpha} \delta_{\alpha\beta}  \hspace{5mm} 
(\alpha,\beta = 1,\dots,5), \label{eq:nuscaldia}
\end{align}
where $M_{\nu}^{2}$ is the neutral slepton mass matrix defined in \cite{Nishida:2016lyk}.

The SUSY contributions to the muon $g-2$ are divided into three parts: neutralinos ($\chi^{0}$), charginos ($\chi^{\pm}$) and singlino ($\chi_{\Phi}$). The singlino contribution is calculated by the replacement of $\chi^{0}$ with $\chi_{\Phi}$ in the neutralino diagram (with appropriate replacement of coefficients). The SUSY contributions are given by
\begin{equation}
 \Delta a_{\mu}^{\rm SUSY} = \Delta a_{\mu}^{\chi^{0}} + \Delta a_{\mu}^{\chi^{\pm}} + \Delta a_{\mu}^{\chi_{\Phi}}, \label{eq:g2susycont}
\end{equation}
where 
\begin{align}
\Delta a_\mu^{\chi^0} & = \sum_{a,x} 
\frac{1}{16\pi^2} \bigg[
\frac{m_\mu m_{\chi^0_x }}{m^2_{\tilde{E}_a}} n_{2ax}^L n_{2ax}^R F_2^N(r_{1ax})
-\frac{m_\mu^2}{6m^2_{\tilde{E}_a}} \big( n_{2ax}^L n_{2ax}^{L} 
+ n_{2ax}^R n_{2ax}^R \big) F_1^N(r_{1ax}) \bigg] ,  \label{neu}  \\
\Delta a_\mu^{\chi^\pm} & = \sum_{\alpha,x}
\frac{1}{16\pi^2} \bigg[
\frac{-3 m_\mu m_{\chi_x}^\pm}{ m^2_{\tilde{\nu}_a} } 
c_{2 \alpha x}^L c_{2 \alpha x}^R  F_2^C(r_{2\alpha x})
+\frac{m_\mu^2}{3 m^2_{\tilde{\nu}_\alpha}} 
\big( c_{2 \alpha x}^L c_{2 \alpha x}^L + c_{2 \alpha x}^R c_{2 \alpha x}^R \big) 
F_1^C(r_{2\alpha x})  \bigg] ,  \label{char}  \\
\Delta a_\mu^{\chi_\Phi} & = \sum_a
\frac{1}{16\pi^2} \bigg[
\frac{m_\mu m_{\chi_\Phi}}{m^2_{\tilde{E}_a}} s_a^L s_a^R F_2^N(r_{3a})
-\frac{m_\mu^2}{6 m^2_{\tilde{E}_a}} 
\big( s_{2a}^L s_{2a}^L + s_{2a}^R s_{2a}^R \big) F_1^N(r_{3a}) \bigg] ,  \label{phi} 
\end{align}
with $r_{1ax} = m^2_{\chi^0_x} / m^2_{\tilde{E}_a}$,
$r_{2\alpha x} = m^2_{\chi^\pm_x} / m^2_{\tilde{N}_\alpha}$,
$r_{3a} = m^2_{\chi_\Phi} / m^2_{\tilde{E}_a}$,
and $m_\mu$ is the muon mass, the function  $F_{1,2}^N$ and $F_{1,2}^C$ are defined by 
\begin{eqnarray}
&& F_1^N(x) = \frac{ 2 }{ (1-x)^4 } 
\left(
1 - 6 x^2 + 3 x^3 + 2 x^3 - 6 x^2 \ln x
\right),  \\ 
&& F_2^N(x) = \frac{ 3 }{ (1-x)^3 } 
\left(
1 - x^2 + 2 x \ln x
\right), \\
&& F_1^C(x) = \frac{ 2 }{ (1-x)^4 } 
\left(
2 + 3 x - 6 x^2 + x^3 + 6 x \ln x
\right), \\
&& F_2^C(x) = \frac{ -3 }{ (1-x)^3 } 
\left(
3 - 4 x + x^2 + 2 \ln x
\right),
\end{eqnarray}
and by using diagonalization matrices Eq.\ (\ref{eq:eferdia}), (\ref{eq:escaldia}),\ (\ref{nuediagonalize}), (\ref{chardiagonalize}) and (\ref{eq:nuscaldia}), the coefficients $n_{Iax}^{L,R}, c_{I\alpha x}^{L,R}, s_{Ia}^{L,R}$ in Eq.\ (\ref{neu})--(\ref{phi}) are defined by
\begin{align}
n^L_{I a x} & = 
-\sum_{i,j=1}^4 (y_{e})_{ij} (V_{e_R})_{iI} (U_{\tilde{e}})_{aj} N_{x3} 
+ y_{\bar{e}} (V_{e_R})_{5I} (U_{\tilde{e}})_{a,10} N_{x 4}  \nonumber \\
& \qquad 
-\sum_{i=1}^4 \sqrt{2} g_1 (V_{e_R})_{iI} (U_{\tilde{e}})_{a, i+5} N_{x1}
-\frac{g_2}{\sqrt{2}} (V_{e_R})_{5I} (U_{\tilde{e}})_{a5} N_{x2}  \nonumber \\
& \qquad 
-\frac{g_1}{\sqrt{2}} (V_{e_R})_{5I} (U_{\tilde{e}})_{a5} N_{x1}
\label{nlax},   \\
n^R_{I a x} & = 
\sum_{i,j=1}^4 (y_{e})_{ij} (V_{e_L})_{jI} (U_{\tilde{e}})_{a,i+5} N_{x3}
-y_{\bar{e}} (V_{e_L})_{5I} (U_{\tilde{e}})_{a5} N_{x4}  \nonumber
\label{nrax}  \\
& \qquad 
+\sum_{i=1}^4 \bigg[
\frac{g_2}{\sqrt{2}} (V_{e_L})_{iI} (U_{\tilde{e}})_{ai} N_{x2}
+\frac{g_1}{\sqrt{2}} (V_{e_L})_{iI} (U_{\tilde{e}})_{ai} N_{x1}
\bigg]  \nonumber  \\[1mm]
& \qquad 
+\sqrt{2} g_1 (V_{e_L})_{5I} (U_{\tilde{e}})_{a,10} N_{x1},  \\
c^L_{Iax} & = 
-\sum_{i,j=1}^4 (y_{e})_{ij} (V_{e_R})_{iI} (U_{\tilde{\nu}})_{aj} J_{x2}
+ g_2 (V_{e_R})_{5I} ( U_{\tilde{\nu}} )_{a5} J_{x1},  \label{eq:cl}  \\
c^R_{Iax} & = 
y_{\bar{e}} (V_{e_L})_{5I} (U_{\tilde{\nu}})_{a5} K_{x2}  
-\sum_{i=1}^4 g_2 (V_{e_L})_{iI} (U_{\tilde{\nu}})_{ai} K_{x1},
\label{eq:cr}  \\
s_{Ia}^{L} & = \sum_{i=1}^4 \Big[ 
-{y_e}_i (V_{e_R})_{iI} (U_{\tilde{e}})_{a,10}
-{y_L}_i (V_{e_R})_{5I} (U_{\tilde{e}})_{ai}
\Big],  \label{eq:sla}  \\
s^R_{Ia} & = \sum_{i=1}^4 \Big[
-{y_e}_i (V_{e_L})_{5I} (U_{\tilde{e}})_{a,i+5}
-{y_L}_i (V_{e_L})_{iI} (U_{\tilde{e}})_{a5}
\Big].  \label{eq:sra}
\end{align} 

Next, we show the non-SUSY contributions. The contributions from vector-like fermions to the muon $g-2$ are investigated in detail in \cite{Dermisek:2013gta} and we derive $\Delta a_{\mu}^{\rm non-SUSY}$ in accordance with \cite{Dermisek:2013gta}. We use the mass eigenstate basis for charged leptons, neutral leptons and CP-even neutral Higgs bosons. Let us define the diagonalization matrix for neutral leptons and CP-even Higgs boson in order to evaluate the non-SUSY contributions of the muon $g-2$ ($\Delta a_{\mu}^{\rm non-SUSY}$). Neutral lepton mass matrix $M_{\nu}$ can be read in the superpotential (\ref{superpotential44bar}) and is given by
\begin{eqnarray}
M_{\nu} & = &
\bordermatrix{
   & \nu_{1R} & \nu_{2R} & \nu_{3R} & \nu_{4R} & \nu_{5R} \cr
 \nu_{1L} &   &   &   &  & Y_{L_{1}} \epsilon^{4} V   \cr             
 \nu_{2L} &   &   &   &  & Y_{L_{2}} \epsilon^{3} V \cr
 \nu_{3L} &   &   &   &  & Y_{L_{3}} \epsilon^{2} V   \cr 
 \nu_{4L} &   &   &   &  & Y_{L_{4}} \epsilon^{1} V   \cr 
 \nu_{5L} &   &   &   &  & \cr 
},\label{nuflavor}
\end{eqnarray}
where blank elements mean zero. The diagonalization of this matrix is
defined by
\begin{equation}
 ( V_{\nu_{R}} M_{\nu} V^{\dagger}_{\nu_{L}} )_{ij} = m_{\nu_{i}} \delta_{ij},
\hspace{3mm} (i,j = 1,\cdots,5) \label{eq:neulepfermidiamat}
\end{equation} 
where only $m_{\nu_{5}}$ is finite value and other masses ($i = 1,\cdots,4$) are zero.
The CP-even neutral Higgs mass matrix $M_{h^{0}}^{2}$ is given by
\begin{equation}
 M_{h^{0}}^{2} = \left( 
\begin{array}{cc}
  M_{A}^{2} \sin^{2}\beta + M_{Z}^{2} \cos^{2}\beta  & 
- \left( M_{A}^{2}  + M_{Z}^{2}  \right) \sin\beta \cos\beta \\
- \left( M_{A}^{2}  + M_{Z}^{2}  \right) \sin\beta \cos\beta &
  M_{A}^{2} \cos^{2}\beta + M_{Z}^{2} \sin^{2}\beta
\end{array}
\right),
\end{equation}
where $M_{A} = 2b / \sin(2\beta)$ is the CP-odd neutral Higgs boson mass as in the MSSM.
This diagonalization of this matrix is defined by 
\begin{equation}
( U_{h^{0}} M_{h^{0}}^{2} U^{\dagger}_{h^{0}} )_{XY} = m^{2}_{h^{0}_{X}} \delta_{XY},
\hspace{3mm} (X,Y = 1,2) \label{eq:higgsdiamat}
\end{equation}
where the mass eigenvalues are ordered as $m_{h^{0}_{1}} < m_{h^{0}_{2}}$.

 The non-SUSY contributions are divided into 3 parts: $W$-boson, $Z$-boson and Higgs bosons. Then, $\Delta a_{\mu}^{\rm non-SUSY}$ is given by
\begin{equation}
 \Delta a_{\mu}^{\rm non-SUSY} = \Delta a_{\mu}^{Z} + 
 \Delta a_{\mu}^{W}  + \Delta a_{\mu}^{h}, \label{eq:g2nonsusycont}
\end{equation}
where
\begin{align}
 \Delta a_{\mu}^{Z} & = - \frac{m_{\mu}}{8 \pi^{2} M_{Z}^{2}}
 \sum_{a=4,5} 
\left[
\left(
(g_{2a}^{ZL})^{2} + (g^{ZR}_{2a})^{2}
\right) m_{\mu} F_{Z} (x_{Za}) + 
g_{2a}^{ZL} g_{2a}^{ZR} m_{E_{a}} G_{Z}(x_{Za})
\right], \label{eq:Z} \\
%%%
 \Delta a_{\mu}^{W} & = - \frac{m_{\mu}}{16 \pi^{2} M_{W}^{2}}
\left[
\left(
(g^{WL}_{52})^{2} + (g^{WR}_{52})^{2}
\right) m_{\mu} F_{W} (x_{W}) + 
g^{WL}_{52} g^{WR}_{52} m_{\nu_{5}} G_{W}(x_{W})
\right], \label{eq:W} \\
\Delta a_{\mu}^{h} & =  - \sum_{X=1,2} \sum_{a=4,5} 
\frac{m_{\mu}}{32 \pi^{2} m_{h^{0}_{X}}^{2}}
\left[
\left(
( \lambda_{2 a X}  )^{2} + ( \lambda_{a 2 X}  )^{2}
\right) m_{\mu} F_{h} (x_{h^{0}aX}) + 
\lambda_{2 a X} \lambda_{a 2 X} m_{E_{a}} G_{h} (x_{h^{0}aX})
\right], \label{eq:h}
\end{align}
with $x_{Za} = m_{E_{a}}^{2} / M_{Z}^{2}$, $x_{W} = m_{\nu_{5}}^{2} / M_{W}^{2}$, $x_{h^{0}aX} = m_{E_{a}}^{2} / m_{h^{0}_{X}}^{2}$, and $M_{W}$ is the $W$-boson mass, $m_{E_{a}}$ is the mass eigenvalues of charged lepton defined in Eq.\ (\ref{eq:eferdia}), the function $F_{Z}, F_{W}, F_{h}, G_{Z}, G_{W}, G_{h}$ are defined by \cite{Dermisek:2013gta}
% $m_{\nu_{5}}$ and $m_{h^{0}_{X}}$ is the mass eigenvalues of neutral leptons and CP-even Higgs boson defined in Appendix \ref{sec:g2formula}
\begin{eqnarray}
&& F_Z(x) = \frac{ 12 }{ (1-x)^4 } 
\left(
8 - 38 x + 39 x^2 - 14 x^3 + 5 x^4 - 18 x^2 \ln x
\right),  \\ 
&& F_W(x) = \frac{ -6 }{ (1-x)^4 } 
\left(
10 - 43 x + 78 x^2 -49 x^3 + 4 x^4 + 18 x^3 \ln x
\right), \\
&& F_h(x) = \frac{ 12 }{ (1-x)^4 } 
\left(
8 - 38 x + 39 x^2 -14 x^3 + 5 x^4 - 18 x^2 \ln x
\right), \\
&& G_Z(x) = \frac{ 2 }{ (1-x)^3 } 
\left(
-4 + 3 x + x^3 - 6 x \ln x
\right), \\
&& G_W(x) = \frac{ -1 }{ (1-x)^3 } 
\left(
-4 + 15 x -12 x^2 + x^3 + 6 x^2 \ln x
\right), \\
&& G_h(x) = \frac{ 1 }{ (1-x)^3 } 
\left(
3 -4 x + x^2 + 2 \ln x
\right).
\end{eqnarray}
 and by using mixing matrices Eq.\  (\ref{eq:eferdia}), (\ref{eq:neulepfermidiamat}) and (\ref{eq:higgsdiamat}), the coefficients $g^{ZL,ZR}_{2a}, g^{WL,WR}_{52}, \lambda_{2a}, \lambda_{a2}$ are defined by
\begin{align}
 g^{ZL}_{xy} & = \sum_{i=1}^{4} \frac{g_{2}}{\cos\theta_{W}} 
\left(
- \frac{1}{2} + \sin^{2}\theta_{W} ( V_{e_{L}} )_{x i} (V_{e_{L}})_{y i}
\right)
+ \frac{g_{2}}{\cos\theta_{W}} \sin^{2}\theta_{W}
 ( V_{e_{L}} )_{x 5} (V_{e_{L}})_{y 5}, \label{eq:gzl} \\
%%%%%%%%%%
 g^{ZR}_{xy} & = \sum_{i=1}^{4} \frac{ - g_{2}}{\cos\theta_{W}} 
\sin^{2}\theta_{W} ( V_{e_{R}} )_{x i} (V_{e_{R}})_{y i}
+ \frac{g_{2}}{\cos\theta_{W}}
\left( 
- \frac{1}{2} +  \sin^{2}\theta_{W}
\right)
 ( V_{e_{R}} )_{x 5} (V_{e_{R}})_{y 5}, \label{eq:gzr} \\
%%%%%%%%%%
 g^{WL}_{xy} & = \sum_{i=1}^{4} \frac{ g_{2}}{\sqrt{2}} 
 ( V_{\nu_{L}} )_{x i} (V_{e_{L}})_{y i}, \label{eq:gwl} \\
%%%%%%%%%%
 g^{WR}_{xy} & = \frac{ g_{2}}{\sqrt{2}} 
 ( V_{\nu_{L}} )_{x 5} (V_{e_{L}})_{y 5}, \label{eq:gwr} \\
%%%%%%%%%%
 \lambda_{xyX} & = \sum_{i,j=1}^{4} (V_{e_{R}})_{x i} (y_{e})_{ij}
(V_{e_{L}})_{y j} (U_{h^{0}})_{X1} +
y_{\bar{e}} (V_{e_{R}})_{x 5} (V_{e_{L}})_{y 5} (U_{h^{0}})_{X2}
\end{align}
where $\theta_{W}$ is the Weinberg angle.

\section{Form factors for lepton flavor violation}
Following~\cite{Ibrahim:2012ds,Ibrahim:2015hva}, we summarize the form factors which are necessary to evaluate the branching ratio of lepton flavor violating processes as shown in (\ref{eq:branchingtau}) and (\ref{eq:branchingmu}).
The form factors are divided into four parts: neutralinos, charginos, Z-boson and W-boson parts as
\begin{align}
 F_{2}^{\l_{i} l_{j}}(0) &= F_{2 \chi^{0}}^{l_{i} l_{j}} + F_{2 \chi^{+}}^{l_{i} l_{j}} + F_{2 Z}^{l_{i} l_{j}} + F_{2 W}^{l_{i} l_{j}}, \label{eq:form2} \\
 F_{3}^{\l_{i} l_{j}}(0) & = F_{3 \chi^{0}}^{l_{i} l_{j}} + F_{3 \chi^{+}}^{l_{i} l_{j}} + F_{3 Z}^{l_{i} l_{j}} + F_{3 W}^{l_{i} l_{j}} \label{eq:form3}.
\end{align}
The neutralino contributions are given by 
\begin{align}
 F^{l_{i} l_{j}}_{2 \chi^0} &= \sum_{a=1}^{10} \sum_{x=1}^4 \bigg[\frac{-m_{l_{i}}(m_{l_{i}} +m_{l_{j}})}{192 \pi^2 m^2_{\chi_x^0}}
\{n^L_{iax} n^L_{jax} + n^R_{iax} n^{R}_{jax} \} F_1 \left(\frac{M^2_{\tilde{E}_a}}{m^2_{\chi_{x}^0}}\right) \nonumber \\
& \hspace{18mm} - \frac{(m_{l_{i}} +m_{l_{j}})}{64 \pi^2 m_{\chi_x^0}}
\{ n^L_{iax} n^R_{jax} + n^R_{iax} n^L_{jax} \} F_2 \left(\frac{M^2_{\tilde{E}_{a}}}{m^2_{\chi_x^0}}\right)\bigg]\,, \\
F^{l_{i} l_{j}}_{3 \chi^0} &= \sum_{a=1}^{10} \sum_{x=1}^{4} \frac{(m_{l_{i}} +m_{l_{j}})m_{\chi_x^0} }{32 \pi^2 M^2_{\tilde{E}_{a}}}
\{ n^L_{iax} n^R_{jax} - n^R_{iax} n^L_{jax} \} 
F_3\left(\frac{m^2_{\chi_x^0}}{M^2_{\tilde{E}_{a}}}\right)\,,
\end{align}
where the functions are defined by 
\begin{align}
 F_1(x)&= \frac{1}{(x-1)^4} \{- x^3 +6x^2 -3x -2 -6x \ln x \}, \\
 F_2(x)&= \frac{1}{(x-1)^3} \{-x^2 +1 +2x\ln x \},\, \\
 F_3(x)&= \frac{1}{2(x-1)^2} \{x +1 + \frac{2 x \ln x}{1-x} \},\,
\end{align}
and $n_{i a x}^{L,R}$ are given in Eq.\ (\ref{nlax}) and (\ref{nrax}).

The chargino contributions are given by 
\begin{align}
 F^{l_{i} l_{j}}_{2 \chi^+}&=\sum_{\alpha=1}^5 \sum_{x=1}^2 \bigg[ \frac{m_{l_{i}}(m_{l_{i}} +m_{l_{j}})}{64 \pi^2 m^2_{\chi_x^+}}
\{c^{L}_{i \alpha x} c^{L}_{j \alpha x} + c^{R}_{i \alpha x} c^{R}_{j \alpha x} \} F_4 \left(\frac{M^2_{\tilde{N}_\alpha}}{m^2_{\chi_x^+}}\right) \nonumber \\
& \hspace{18mm} + \frac{(m_{l_{i}} +m_{l_{j}})}{64 \pi^2 m_{\chi_x^+}}
\{ c^{L}_{i \alpha x} c^{R}_{j \alpha x} + c^{R}_{i \alpha x} c^{L}_{j \alpha x}  \} F_5\left(\frac{M^2_{\tilde{N}_\alpha}}{m^2_{\chi_x^+}}\right)\bigg] \,, \\
F^{l_{i} l_{j}}_{3 \chi^+}&= \sum_{\alpha=1}^5 \sum_{x=1}^2 \frac{(m_{l_{i}} +m_{l_{j}})m_{\chi_x^+} }{32 \pi^2 M^2_{\tilde{N}_\alpha}}
\{ c^{L}_{i \alpha x} c^{R}_{j \alpha x} - c^{R}_{i \alpha x} c^{L}_{j \alpha x} \} 
F_6\left(\frac{m^2_{\chi_x^+}}{M^2_{\tilde{N}_\alpha}}\right)\,,
\end{align}
where the functions are defined by 
\begin{align}
 F_4(x)&= \frac{1}{3(x-1)^4} \{-2 x^3 -3x^2 +6x -1 +6x^2 \ln x \}, \\
 F_5(x)&= \frac{1}{(x-1)^3} \{3x^2 -4x +1 -2x^2 \ln x \},\,  \\
 F_6(x)&= \frac{1}{2(x-1)^2} \{-x +3 + \frac{2\ln x}{1-x} \},\,
\end{align}
and $c_{i \alpha x}^{L,R}$ are given in Eq.\ (\ref{eq:cl}) and (\ref{eq:cr}).

The contributions from the Z-boson exchange are given by
\begin{align}
 F^{l_{i} l_{j}}_{2 Z} &=  \sum_{a=1}^{5} \frac{{m_{l_{i}}(m_{l_{i}} +m_{l_{j}})}}{64 \pi^2 m^2_Z}  \{g^{ZL}_{ia} g^{ZL}_{ja} 
+ g^{ZR}_{ia} g^{ZR}_{ja}  \}  F_Z \left(\frac{m^2_{E_{a}}}{m^2_Z}\right)\nonumber\\
 & \hspace{10mm} + \frac{{m_{E_{a}}(m_{l_{i}} +m_{l_{j}})}}{64 \pi^2 m^2_Z} 
  \{ g^{ZL}_{ia} g^{ZR}_{ja} + g^{ZR}_{ia} g^{ZL}_{ja}\} G_Z \left(\frac{m^2_{E_{a}}}{m^2_Z}\right), \\
F^{l_{i} l_{j}}_{3 Z} &=   \sum_{a=1}^{5}\frac{{(m_{l_{i}} +m_{l_{j}})}}{32 \pi^2}  \frac{m_{E_{a}}}{m^2_Z}
\{ g^{ZL}_{ia} g^{ZR}_{ja} - g^{ZR}_{ia} g^{ZL}_{ja} \} 
I_1\left(\frac{m^{2}_{E_{a}}}{m^{2}_{Z}}\right) \,,
\end{align}
where the functions are defined by 
\begin{align}
 F_{Z}(x)&=\frac{1}{3(x-1)^{4}}\left[-5 x^4+14x^{3}-39 x^2+18 x^2 \ln x+ 38 x -8 \right], \\
 G_{Z}(x)&=\frac{2}{(x-1)^{3}}\left[x^3 + 3 x-6 x \ln x-4 \right], \\
 I_1(x)&=\frac{2}{(1-x)^{2}}\left[1+\frac{1}{4}x +\frac{1}{4}x^2+\frac{3 x\ln x}{2(1-x)} \right],\,
\end{align}
and $g_{xy}^{ZL,ZR}$ are given in Eq.\ (\ref{eq:gzl}) and (\ref{eq:gzr}).

The contributions from the W-boson exchange are given by
\begin{align}
 F^{l_{i} l_{j}}_{2 W} &= \frac{{m_{l_{i}}(m_{l_{i}} +m_{l_{j}})}}{32 \pi^2 m^2_W}  \{g^{WL}_{5i} g^{WL}_{5j} + g^{WR}_{5i} g^{WR}_{5j} \} F_W \left(\frac{m^2_{\nu_{5}}}{m^2_W}\right)\nonumber\\
 & \hspace{10mm} + \frac{{m_{\nu_{5}}(m_{l_{i}} +m_{l_{j}})}}{32 \pi^2 m^2_W} 
  \{ g^{WL}_{5i} g^{WR}_{5j} + g^{WR}_{5i} g^{WL}_{5j}  \} G_W \left(\frac{m^2_{\nu_{5}}}{m^2_W}\right), \\
F^{l_{i} l_{j}}_{3 W} & = - \frac{{m_{\nu_{5}}(m_{l_{i}} +m_{l_{j}})}}{32 \pi^2 m^2_W}   \{  g^{WL}_{5i} g^{WR}_{5j} - g^{WR}_{5i} g^{WL}_{5j} \} 
I_2\left(\frac{m^{2}_{\nu_{5}}}{m^{2}_{W}}\right) \,,
\end{align}
where the functions are defined by 
\begin{align}
F_{W}(x)&=\frac{1}{6(x-1)^{4}}\left[4 x^4- 49x^{3}+18 x^3 \ln x+78x^{2}-43 x +10 \right], \\
G_{W}(x)&=\frac{1}{(x-1)^{3}}\left[4 -15 x+12 x^2 - x^3-6 x^2 \ln x \right],\, \\ 
I_2(x)&=\frac{2}{(1-x)^{2}}\left[1-\frac{11}{4}x +\frac{1}{4}x^2-\frac{3 x^2\ln x}{2(1-x)} \right]\,,
\end{align}
and $g_{xy}^{WL,WR}$ are given in Eq.\ (\ref{eq:gwl}) and (\ref{eq:gwr}).

\section{RG Equations} \label{sec:RGEs}

We present the RG equations of model parameters in our model. Due to
the asymptotically non-free nature of the gauge sector, the two-loop
RG equations are used for gauge coupling constants and gaugino masses.

\subsection{Gauge couplings and gaugino masses}
\label{sec:gaugegauginoRGE}

The two-loop RG equations of gauge coupling constants $g_i$
and gaugino masses $M_i$ ($i=1,2,3$) are given by
\begin{align}
\frac{d g_i}{d ( \log \mu )} & = b_i \frac{ g_i^3 }{16 \pi^2 }
+ \frac{ g_i^3 }{ ( 16 \pi^2 )^2 } 
\bigg[ \sum_j b_{ij} g_j^2 
- \sum_{ a = u, d, e } c_{ia} \Big[ {\rm Tr} \big( 
{\bm y}_a^\dagger {\bm y}_a \big) + y_{\bar{a}}^* y_{\bar{a}}
\Big]  \nonumber \\
& \hspace{55mm} - \sum_{k=1}^4 
\sum_{x=Q,u,d,L,e} \! d_{ix}  y_{x_k}^* y_{x_k} \bigg] , \\
%%%%%%%%%%%%%%%%%%%%%%%%%%%%%%%%%%%%%%
\frac{d M_i}{d ( \log \mu )} & = 2b_i \frac{ g_i^2 M_i }{16 \pi^2 }
+ \frac{ 2 g_i^2 }{ ( 16 \pi^2 )^2 } 
\bigg[ \sum_j b_{ij} g_j^2 ( M_i + M_j ) 
+ \sum_{ a = u, d, e } c_{ia} \Big[ {\rm Tr} \big( 
{\bm y}_a^\dagger {\bm a}_a \big) +y_{\bar{a}}^* a_{\bar{a}}
\nonumber \\
& \qquad - M_i \big[ {\rm Tr} ( {\bm y}_a^\dagger {\bm y}_a ) 
+ y_{\bar{a}}^* y_{\bar{a}} \big] \Big] 
+ \sum_{k=1}^4 
\sum_{x=Q,u,d,L,e} \!\! d_{ix} ( y_{x_k}^* A_{x_k}
- M_i y_{x_k}^* y_{x_k} ) \bigg] ,
\end{align}
where the one-loop beta function coefficients are $b_i=(53/5,5,1)$,
and the coefficient matrices $b_{ij}$, $c_{ia}$, $d_{ix}$ are
\begin{eqnarray}
&& b_{ij} = \left( 
\begin{array}{ccc}
977/75 & 39/5 & 88/3 \\
13/5 & 53 & 40 \\
11/3 & 15 & 178/3 \\
\end{array} 
\right) , \\
&& c_{ia} = \bordermatrix{
& u & d & e \cr
& 26/5 & 14/5 & 18/5 \cr
& 6 & 6 & 2 \cr
& 4 & 4 & 0 \cr
} , \\
&& d_{ix} = \bordermatrix{
& Q & u & d & L & e \cr
& 2/5 & 16/5 & 4/5 & 6/5 & 12/5 \cr
& 6 & 0 & 0 & 2 & 0 \cr
& 4 & 2 & 2 & 0 & 0 \cr
} .
\end{eqnarray}

\subsection{Yukawa couplings and bilinear terms}

The RG equations of Yukawa couplings and the bilinear terms are given by
\begin{align}
\frac{d (y_u)_{ij}}{d ( \log \mu )} & =
( \gamma_u {\bm y}_u )_{ij} +
( {\bm y}_u \gamma_Q )_{ij} +
\gamma_{H_u}  (y_u)_{ij} , \\
%%%%%%%%%%%%%%%%%%%%%%%%%%%%%%%%%%%%%% 
\frac{d (y_d)_{ij}}{d ( \log \mu )} & =
( \gamma_d {\bm y}_d )_{ij} +
( {\bm y}_d \gamma_Q )_{ij} +
\gamma_{H_d} (y_d)_{ij} , \\
%%%%%%%%%%%%%%%%%%%%%%%%%%%%%%%%%%%%%%
\frac{d (y_e)_{ij}}{d ( \log \mu )} & =
( \gamma_e {\bm y}_e )_{ij} +
( {\bm y}_e \gamma_L )_{ij} +
\gamma_{H_d} (y_e)_{ij} , \\
%%%%%%%%%%%%%%%%%%%%%%%%%%%%%%%%%%%%%%
\frac{d y_{\bar{u}}}{d ( \log \mu )} & =
(\gamma_{\bar{u}} +
\gamma_{\bar{Q}} +
\gamma_{H_d}
) y_{\bar{u}} , \\
%%%%%%%%%%%%%%%%%%%%%%%%%%%%%%%%%%%%%%
\frac{d y_{\bar{d}}}{d ( \log \mu )} & =
(\gamma_{\bar{d}} +
\gamma_{\bar{Q}} +
\gamma_{H_u}
) y_{\bar{d}} , \\
%%%%%%%%%%%%%%%%%%%%%%%%%%%%%%%%%%%%%%
\frac{d y_{\bar{e}}}{d ( \log \mu )} & =
(\gamma_{\bar{e}} +
\gamma_{\bar{L}} +
\gamma_{H_u}
) y_{\bar{e}} , \\
%%%%%%%%%%%%%%%%%%%%%%%%%%%%%%%%%%%%%%
\frac{d y_{Q_i}}{d ( \log \mu )} & =
( y_Q \gamma_Q )_i +
(\gamma_{\bar{Q}} + \gamma_\Phi ) y_{Q_i} , \\
%%%%%%%%%%%%%%%%%%%%%%%%%%%%%%%%%%%%%%
\frac{d y_{u_i}}{d ( \log \mu )} & =
( \gamma_u y_u )_i +
(\gamma_{\bar{u}} + \gamma_\Phi ) y_{u_i} , \\
%%%%%%%%%%%%%%%%%%%%%%%%%%%%%%%%%%%%%% 
\frac{d y_{d_i}}{d ( \log \mu )} & =
( \gamma_d y_d )_i +
(\gamma_{\bar{d}} + \gamma_\Phi ) y_{d_i} , \\
%%%%%%%%%%%%%%%%%%%%%%%%%%%%%%%%%%%%%% 
\frac{d y_{L_i}}{d ( \log \mu )} & =
( y_L \gamma_L )_i +
(\gamma_{\bar{L}} + \gamma_\Phi ) y_{L_i} , \\
%%%%%%%%%%%%%%%%%%%%%%%%%%%%%%%%%%%%%%
\frac{d y_{e_i}}{d ( \log \mu )} & =
( \gamma_e y_e )_i +
(\gamma_{\bar{e}} + \gamma_\Phi ) y_{e_i} , \\
%%%%%%%%%%%%%%%%%%%%%%%%%%%%%%%%%%%%%%
\frac{d y}{d ( \log \mu )} & =
3 \gamma_\Phi y ,  \\
%%%%%%%%%%%%%%%%%%%%%%%%%%%%%%%%%%%%%% 
\frac{d \mu_H}{d ( \log \mu )} & =
( \gamma_{H_u} + \gamma_{H_d} )\mu_H , \\
%%%%%%%%%%%%%%%%%%%%%%%%%%%%%%%%%%%%%%
\frac{d M}{d ( \log \mu )} & =
2 \gamma_\Phi M .
%%%%%%%%%%%%%%%%%%%%%%%%%%%%%%%%%%%%%%
\end{align}
The anomalous dimensions $\gamma$'s are
\begin{align}
(\gamma_Q)_{ij} & = \frac{1}{16\pi^2} 
\left[\left(
\yukawa{u}^\dagger \yukawa{u} +
\yukawa{d}^\dagger \yukawa{d}
\right)_{ij} + y_{Q_i}^* y_{Q_j} -
\left( 
\frac{8}{3} g_3^2 +
\frac{3}{2} g_2^2 +
\frac{1}{30} g_1^2   
\right) \delta_{ij}
\right] , \\
%%%%%%%%%%%%%%%%%%%%%%%%%%%%%%%%%%%%%%%%%%
(\gamma_u)_{ij} & = \frac{1}{16\pi^2} 
\left[
2 \left(\yukawa{u} \yukawa{u}^\dagger
\right)_{ij} + y_{u_i} y_{u_j}^* -
\left( 
\frac{8}{3} g_3^2 +
\frac{8}{15} g_1^2   
\right) \delta_{ij}
\right] ,  \\
%%%%%%%%%%%%%%%%%%%%%%%%%%%%%%%%%%%%%%%%%%
(\gamma_d)_{ij} & = \frac{1}{16\pi^2} 
\left[
2 \left(\yukawa{d} \yukawa{d}^\dagger
\right)_{ij} + y_{d_i} y_{d_j}^* - 
\left(
\frac{8}{3} g_3^2 +
\frac{2}{15} g_1^2
\right) \delta_{ij}
\right] , \\
%%%%%%%%%%%%%%%%%%%%%%%%%%%%%%%%%%%%%%%%%%
(\gamma_L)_{ij} & = \frac{1}{16\pi^2} 
\left[\left(
\yukawa{e}^\dagger \yukawa{e}
\right)_{ij} + y{L_i}^* y_{L_j} -
\left( 
\frac{3}{2} g_2^2 +
\frac{3}{10} g_1^2
\right) \delta_{ij}
\right] , \\
%%%%%%%%%%%%%%%%%%%%%%%%%%%%%%%%%%%%%%%%%%      
(\gamma_e)_{ij} & = \frac{1}{16\pi^2} 
\left[
2 \left(\yukawa{e}^\dagger \yukawa{e} 
\right)_{ij} + y_{e_i} y_{e_j}^* -
\frac{6}{5} g_1^2
\delta_{ij}
\right] , \\
%%%%%%%%%%%%%%%%%%%%%%%%%%%%%%%%%%%%%%%%%% 
\gamma_{\bar{Q}} & = \frac{1}{16\pi^2} 
\left[
\sum_i y_{Q_i}^* y{Q_i} +
\yukbar{u}^* \yukbar{u} +
\yukbar{d}^* \yukbar{d} -
\left(
\frac{8}{3} g_3^2 +
\frac{3}{2} g_2^2 +
\frac{1}{30} g_1^2
\right)
\right] , \\
%%%%%%%%%%%%%%%%%%%%%%%%%%%%%%%%%%%%%%%%%% 
\gamma_{\bar{u}} & = \frac{1}{16\pi^2} 
\left[
\sum_i y_{u_i}^* y_{u_i} +
2 \yukbar{u}^* \yukbar{u} -
\left(
\frac{8}{3} g_3^2 +
\frac{8}{15} g_1^2
\right)
\right] , \\
%%%%%%%%%%%%%%%%%%%%%%%%%%%%%%%%%%%%%%%%%% 
\gamma_{\bar{d}} & = \frac{1}{16\pi^2} 
\left[
\sum_i y_{d_i}^* y_{d_i} +
2 \yukbar{d}^* \yukbar{d} -
\left(
\frac{8}{3} g_3^2 +
\frac{2}{15} g_1^2
\right)
\right] , \\
%%%%%%%%%%%%%%%%%%%%%%%%%%%%%%%%%%%%%%%%%% 
\gamma_{\bar{L}} & = \frac{1}{16\pi^2} 
\left[
\sum_i y_{L_i}^* y_{L_i} +
\yukbar{e}^* \yukbar{e} -
\left(
\frac{3}{2} g_2^2 +
\frac{3}{10} g_1^2
\right)
\right] , \\
%%%%%%%%%%%%%%%%%%%%%%%%%%%%%%%%%%%%%%%%%% 
\gamma_{\bar{e}} & = \frac{1}{16\pi^2} 
\left[
\sum_i y_{e_i}^* y_{e_i} +
2 \yukbar{e}^* \yukbar{e} -
\frac{6}{5} g_1^2
\right] , \\
%%%%%%%%%%%%%%%%%%%%%%%%%%%%%%%%%%%%%%%%%% 
\gamma_{H_u} & = \frac{1}{16\pi^2} 
\left[
3\,\text{Tr} \left( \yukawa{u} \yukawa{u}^\dagger \right) +
3 \yukbar{d}^* \yukbar{d} +
\yukbar{e}^* \yukbar{e} -
\left(
\frac{3}{2} g_2^2 +
\frac{3}{10} g_1^2
\right)
\right] , \\
%%%%%%%%%%%%%%%%%%%%%%%%%%%%%%%%%%%%%%%%%% 
\gamma_{H_d} & = \frac{1}{16\pi^2} 
\left[
\text{Tr} \left( 3 \yukawa{d} \yukawa{d}^\dagger + 
\yukawa{e} \yukawa{e}^\dagger \right) +
3 \yukbar{u}^* \yukbar{u} -
\left(
\frac{3}{2} g_2^2 +
\frac{3}{10} g_1^2
\right)
\right] , \\
%%%%%%%%%%%%%%%%%%%%%%%%%%%%%%%%%%%%%%%%%%
\gamma_\Phi & = \frac{1}{16\pi^2} 
\left[
\sum_i
\left(
6 y_{Q_i}^* y_{Q_i} +
3 y_{u_i}^* y_{u_i} +
3 y_{d_i}^* y_{d_i} +
2 y_{L_i}^* y_{L_i} +
  y_{e_i}^* y_{e_i}
\right) + y^* y
\right] . \label{eq:Yrge} 
\end{align}
%%%%%%%%%%%%%%%%%%%%%%%%%%%%%%%%%%%%%%%%%%

\subsection{$A$ and $B$ terms}

The RG equations of SUSY-breaking $A$ and $B$ terms are given by
\begin{align}
\frac{d (a_u)_{ij}}{d ( \log \mu )} & =
( \gamma_u \aterm{u} )_{ij} +
( \aterm{u} \gamma_Q )_{ij} +
\gamma_{H_u} (a_u)_{ij} +
2( \tilde{\gamma}_u {\bm y}_u )_{ij} +
2( {\bm y}_u \tilde{\gamma}_Q )_{ij} +
2\tilde{\gamma}_{H_u} (y_u)_{ij} , \\
%%%%%%%%%%%%%%%%%%%%%%%%%%%%%%%%%%%%%%%%%%
\frac{d (a_d)_{ij}}{d ( \log \mu )} & =
( \gamma_d \aterm{d} )_{ij} +
( \aterm{d} \gamma_Q )_{ij} +
\gamma_{H_d} (a_d)_{ij} +
2( \tilde{\gamma}_d {\bm y}_d )_{ij} +
2( {\bm y}_d \tilde{\gamma}_Q )_{ij} +
2\tilde{\gamma}_{H_d} (y_d)_{ij} , \\
%%%%%%%%%%%%%%%%%%%%%%%%%%%%%%%%%%%%%%%%%%
\frac{d (a_e)_{ij}}{d ( \log \mu )} & =
( \gamma_e \aterm{e} )_{ij} +
( \aterm{e} \gamma_L )_{ij} +
\gamma_{H_d} (a_e)_{ij} +
2( \tilde{\gamma}_e {\bm y}_e )_{ij} +
2( {\bm y}_e \tilde{\gamma}_L )_{ij} +
2\tilde{\gamma}_{H_d} (y_e)_{ij} , \\
%%%%%%%%%%%%%%%%%%%%%%%%%%%%%%%%%%%%%%%%%% 
\frac{d \abar{u}}{d ( \log \mu )} & =
(
\gamma_{\bar{u}} +
\gamma_{\bar{Q}} +
\gamma_{H_d}
) y_{\bar{u}} +
2(
\tilde{\gamma}_{\bar{u}} +
\tilde{\gamma}_{\bar{Q}} +
\tilde{\gamma}_{H_d}
) \abar{u} , \\
%%%%%%%%%%%%%%%%%%%%%%%%%%%%%%%%%%%%%%%%%%
\frac{d \abar{d}}{d ( \log \mu )} & =
(
\gamma_{\bar{d}} +
\gamma_{\bar{Q}} +
\gamma_{H_u}
) y_{\bar{d}} +
2(
\tilde{\gamma}_{\bar{d}} +
\tilde{\gamma}_{\bar{Q}} +
\tilde{\gamma}_{H_u}
) \abar{d} , \\
%%%%%%%%%%%%%%%%%%%%%%%%%%%%%%%%%%%%%%%%%%
\frac{d \abar{e}}{d ( \log \mu )} & =
(
\gamma_{\bar{e}} +
\gamma_{\bar{L}} +
\gamma_{H_u}
) y_{\bar{e}} +
2(
\tilde{\gamma}_{\bar{e}} +
\tilde{\gamma}_{\bar{L}} +
\tilde{\gamma}_{H_u}
) \abar{e} , \\
%%%%%%%%%%%%%%%%%%%%%%%%%%%%%%%%%%%%%%%%%%
 \frac{d \avec{Q}{i}}{d ( \log \mu )} & =
( \avec{Q}{} \gamma_Q )_i +
( \gamma_{\bar{Q}} + \gamma_\Phi ) \avec{Q}{i} +
2(y_Q \tilde{\gamma}_Q )_i +
2( \tilde{\gamma}_{\bar{Q}} + \tilde{\gamma}_\Phi ) y_{Q_i} , \\
%%%%%%%%%%%%%%%%%%%%%%%%%%%%%%%%%%%%%%%%%%
\frac{d \avec{Q}{i}}{d ( \log \mu )} & =
( \gamma_u \avec{u}{} )_i +
( \gamma_{\bar{u}} + \gamma_\Phi ) \avec{u}{i} +
2( \tilde{\gamma}_u y_u )_i +
2( \tilde{\gamma}_{\bar{u}} + \tilde{\gamma}_\Phi ) y_{u_i} , \\
%%%%%%%%%%%%%%%%%%%%%%%%%%%%%%%%%%%%%%%%%%
\frac{d \avec{d}{i}}{d ( \log \mu )} & =
( \gamma_d \avec{Q}{} )_i +
( \gamma_{\bar{d}} + \gamma_\Phi ) \avec{d}{i} +
2( \tilde{\gamma}_d y_d )_i +
2( \tilde{\gamma}_{\bar{d}} + \tilde{\gamma}_\Phi ) y_{d_i} , \\
%%%%%%%%%%%%%%%%%%%%%%%%%%%%%%%%%%%%%%%%%%
\frac{d \avec{L}{i}}{d ( \log \mu )} & =
( \avec{Q}{} \gamma_L )_i +
( \gamma_{\bar{L}} + \gamma_\Phi ) \avec{L}{i} +
2( y_L \tilde{\gamma}_L )_i +
2( \tilde{\gamma}_{\bar{L}} + \tilde{\gamma}_\Phi ) y_{L_i} , \\
%%%%%%%%%%%%%%%%%%%%%%%%%%%%%%%%%%%%%%%%%%
\frac{d \avec{e}{i}}{d ( \log \mu )} & =
( \gamma_e \avec{Q}{} )_i +
( \gamma_{\bar{e}} + \gamma_\Phi ) \avec{e}{i} +
2( \tilde{\gamma}_e y_e )_i +
2( \tilde{\gamma}_{\bar{e}} + \tilde{\gamma}_\Phi ) y_{e_i} , \\
%%%%%%%%%%%%%%%%%%%%%%%%%%%%%%%%%%%%%%%%%%
\frac{d A_y}{d ( \log \mu )} & =
3 \gamma_\Phi A_y + 6 \tilde{\gamma}_\Phi y , \\
%%%%%%%%%%%%%%%%%%%%%%%%%%%%%%%%%%%%%%%%%%
\frac{d b_H}{d ( \log \mu )} & =
( \gamma_{H_u} + \gamma_{H_d} )b_H +
2( \tilde{\gamma}_{H_u} + \tilde{\gamma}_{H_d} )\mu_H , \\
%%%%%%%%%%%%%%%%%%%%%%%%%%%%%%%%%%%%%%%%%%
\frac{d b_M}{d ( \log \mu )} & =
2 \gamma_\Phi b_M + 4 \tilde{\gamma}_\Phi M ,
\end{align}
where the definitions of $\tilde{\gamma}$'s are
\begin{align}
(\tilde{\gamma}_Q)_{ij} & = \frac{1}{16\pi^2} 
\left[\left(
\yukawa{u}^{\dagger} \aterm{u} + 
\yukawa{d}^{\dagger} \aterm{d} 
\right)_{ij} + y_{Q_i}^* \avec{Q}{j} + 
\left(
\frac{8}{3} g_3^2 M_3 +
\frac{3}{2} g_2^2 M_2 +
\frac{1}{30} g_1^2 M_1   
\right) \delta_{ij}
\right] , \\
%%%%%%%%%%%%%%%%%%%%%%%%%%%%%%%%%%%%%%%%%%
(\tilde{\gamma}_u)_{ij} & = \frac{1}{16\pi^2} 
\left[ 2\left(
\aterm{u} \yukawa{u}^\dagger
\right)_{ij} + \avec{u}{i} y_{u_j}^* + 
\left( 
\frac{8}{3} g_3^2 M_3 +
\frac{8}{15} g_1^2 M_1
\right) \delta_{ij}
\right] , \\
%%%%%%%%%%%%%%%%%%%%%%%%%%%%%%%%%%%%%%%%%%
(\tilde{\gamma}_d)_{ij} & = \frac{1}{16\pi^2} 
\left[ 2\left(
\aterm{d} \yukawa{d}^\dagger
\right)_{ij} + \avec{d}{i} y_{d_j}^* + 
\left(
\frac{8}{3} g_3^2 M_3 +
\frac{2}{15} g_1^2 M_1   
\right) \delta_{ij}
\right] , \\
%%%%%%%%%%%%%%%%%%%%%%%%%%%%%%%%%%%%%%%%%%
(\tilde{\gamma}_L)_{ij} & = \frac{1}{16\pi^2} 
\left[\left(
\yukawa{e}^\dagger \aterm{e}
\right)_{ij} + y_{L_i}^* \avec{L}{j} + 
\left(
\frac{3}{2} g_2^2 M_2 +
\frac{3}{10} g_1^2 M_1 
\right) \delta_{ij}
\right] , \\
%%%%%%%%%%%%%%%%%%%%%%%%%%%%%%%%%%%%%%%%%%      
(\tilde{\gamma}_e)_{ij} & = \frac{1}{16\pi^2} 
\left[ 2\left(
\aterm{e}^\dagger \yukawa{e} 
\right)_{ij} + \avec{e}{i} y_{e_j}^* + 
\frac{6}{5} g_1^2 M_1 \delta_{ij}
\right] , \\
%%%%%%%%%%%%%%%%%%%%%%%%%%%%%%%%%%%%%%%%%% 
\tilde{\gamma}_{\bar{Q}} & = \frac{1}{16\pi^2} 
\left( \sum_i 
y_{Q_i}^* \avec{Q}{i} +
\yukbar{u}^* \abar{u} +
\yukbar{d}^* \abar{d} +
\frac{8}{3} g_3^2 M_3 +
\frac{3}{2} g_2^2 M_2 +
\frac{1}{30} g_1^2 M_1
\right) , \\
%%%%%%%%%%%%%%%%%%%%%%%%%%%%%%%%%%%%%%%%%% 
\tilde{\gamma}_{\bar{u}} & = \frac{1}{16\pi^2} 
\left( \sum_i 
y_{u_i}^* \avec{u}{i} +
2 \yukbar{u}^* \abar{u} +
\frac{8}{3} g_3^2 M_3 +
\frac{8}{15} g_1^2 M_1
\right) , \\
%%%%%%%%%%%%%%%%%%%%%%%%%%%%%%%%%%%%%%%%%% 
\tilde{\gamma}_{\bar{d}} & = \frac{1}{16\pi^2} 
\left( \sum_i 
y_{d_i}^* \avec{d}{i} +
2 \yukbar{d}^* \abar{d} +
\frac{8}{3} g_3^2 M_3 +
\frac{2}{15} g_1^2 M_1
\right) , \\
%%%%%%%%%%%%%%%%%%%%%%%%%%%%%%%%%%%%%%%%%% 
\tilde{\gamma}_{\bar{L}} & = \frac{1}{16\pi^2} 
\left( \sum_i 
y_{L_i}^* \avec{L}{i} +
\yukbar{e}^* \abar{e} +
\frac{3}{2} g_2^2 M_2 +
\frac{3}{10} g_1^2 M_1
\right) , \\
%%%%%%%%%%%%%%%%%%%%%%%%%%%%%%%%%%%%%%%%%% 
\tilde{\gamma}_{\bar{e}} & = \frac{1}{16\pi^2} 
\left( \sum_i 
y_{e_i}^* \avec{e}{i} +
2 \yukbar{e}^* \abar{e} +
\frac{6}{5} g_1^2 M_1
\right) , \\
%%%%%%%%%%%%%%%%%%%%%%%%%%%%%%%%%%%%%%%%%% 
\tilde{\gamma}_{H_u} & = \frac{1}{16\pi^2} 
\left[
3\,\text{Tr} \left( \aterm{u} \yukawa{u}^\dagger \right) +
3 \yukbar{d}^* \abar{d} +
\yukbar{e}^* \abar{e} +
\frac{3}{2} g_2^2 M_2 +
\frac{3}{10} g_1^2 M_1
\right] , \\
%%%%%%%%%%%%%%%%%%%%%%%%%%%%%%%%%%%%%%%%%% 
\tilde{\gamma}_{H_d} & = \frac{1}{16\pi^2} 
\left[
\text{Tr} \left( 3\aterm{d} \yukawa{d}^\dagger + 
\aterm{e} \yukawa{e}^\dagger \right) +
3 \yukbar{u}^* \abar{u} +
\frac{3}{2} g_2^2 M_2  +
\frac{3}{10} g_1^2 M_1
\right] , \\
%%%%%%%%%%%%%%%%%%%%%%%%%%%%%%%%%%%%%%%%%%
\tilde{\gamma}_\Phi & = \frac{1}{16\pi^2} 
\left[ \sum_i 
\left(
6 y_{Q_i}^* \avec{Q}{i} +
3 y_{u_i}^* \avec{u}{i} +
3 y_{d_i}^* \avec{d}{i} +
2 y_{L_i}^* \avec{L}{i} +
  y_{e_i}^* \avec{e}{i}
\right) + y^* A_y
\right] .
\end{align}

\subsection{Soft scalar masses}

We define the following functions to write down the RG equations of
soft scalar masses:
\begin{align}
f (x_1, x_2, x_3; y; z) & =
\frac{1}{16 \pi^2}
\left(
x_1 y y^\dagger + y y^\dagger x_1 + y x_2 y^\dagger 
+ x_3 y y^\dagger + z z^\dagger \right) , \\[1mm]
g ( a, b, c  ) & = 
\frac{1}{16 \pi^2} \left(
\frac{32a}{3} g_3^2|M_3|^2  + 6b g_2^2|M_2|^2 +\frac{2c^2}{15} g_1^2|M_1|^2 
\right) -\frac{c}{80\pi^2}g_1^2S, \\[3mm]
S & = \text{Tr} \left( {\bf m}_Q^2 - 2{\bf m}_u^2 + {\bf m}_d^2 -
  {\bf m}_L^2 + {\bf m}_e^2 \right) + m_{H_u}^2 - m_{H_d}^2  \nonumber \\
& \hspace{3cm} - m_{\bar{Q}}^2 + 2 m_{\bar{u}}^2 - m_{\bar{d}}^2 
  + m_{\bar{L}}^2 - m_{\bar{e}}^2 ,
\end{align}
where $x_{1,2,3}$ are generally soft scalar masses in generation space
and $y$, $z$ are Yukawa couplings and $A$ parameters with generation
indices. The RG equations of soft scalar masses are given by
\begin{align}
\ddt{{\bf m}_Q^2} & =  \sum_{x = u,d}
f({\bf m}_Q^2, {\bf m}_x^2, m_{H_x}^2; {\bm y}_u^\dagger; \aterm{u}^\dagger)
+ f ({\bf m}_Q^2, m_{\bar{Q}}^2, m_\Phi^2; y_Q ; A_Q ) 
- g(1, 1, 1) , \\
%%%%%%%%%%%%%%%%%%%%%%%%%%%%%%%%%%%%%%%%%%
\ddt{{\bf m}_u^2} & = 
2 f ({\bf m}_u^2, {\bf m}_Q^2, m_{H_u}^2; {\bm y}_u; \aterm{u})
+ f ({\bf m}_u^2, m_{\bar{u}}^2, m_\Phi^2; y_u ; A_u )
- g(1, 0, -4) , \\
%%%%%%%%%%%%%%%%%%%%%%%%%%%%%%%%%%%%%%%%%%
\ddt{{\bf m}_d^2} & = 
2 f ({\bf m}_d^2, {\bf m}_Q^2, m_{H_d}^2; {\bm y}_d; \aterm{d})
+ f ({\bf m}_d^2, m_{\bar{d}}^2, m_\Phi^2; y_d ; A_d )
- g (1, 0, 2) , \\
%%%%%%%%%%%%%%%%%%%%%%%%%%%%%%%%%%%%%%%%%%
 \ddt{{\bf m}_L^2} & = 
f({\bf m}_L^2, {\bf m}_e^2, m_{H_d}^2; {\bm y}_e^\dagger; \aterm{e}^\dagger)
+ f ({\bf m}_L^2, m_{\bar{L}}^2, m_\Phi^2; y_L ; A_L )
- g (0, 1, -3) , \\
%%%%%%%%%%%%%%%%%%%%%%%%%%%%%%%%%%%%%%%%%%
\ddt{{\bf m}_e^2} & = 
2 f ({\bf m}_e^2, {\bf m}_L^2, m_{H_d}^2; {\bm y}_e; \aterm{e})
+ f ({\bf m}_e^2, m_{\bar{e}}^2, m_\Phi^2; y_e ; A_e )
- g(0, 0, 6) , \\
%%%%%%%%%%%%%%%%%%%%%%%%%%%%%%%%%%%%%%%%%%
\ddt{m_{\bar{Q}}^2} & = 
f (m_{\bar{Q}}^2, m_{\bar{u}}^2, m_{H_d}^2; \yukbar{u}^* ; \abar{u}^* )
+ f (m_{\bar{Q}}^2, m_{\bar{d}}^2, m_{H_u}^2; \yukbar{d}^* ; \abar{d}^* )
\nonumber \\
& \hspace{4cm}  + f (m_{\bar{Q}}^2, {\bf m}_Q^2, m_\Phi^2; y_Q ; A_Q )
- g(1, 1, -1) , \\
%%%%%%%%%%%%%%%%%%%%%%%%%%%%%%%%%%%%%%%%%%
\ddt{m_{\bar{u}}^2} & = 
2 f (m_{\bar{Q}}^2, m_{\bar{u}}^2, m_{H_d}^2; \yukbar{u} ; \abar{u} )
+ f (m_{\bar{u}}^2, {\bf m}_u^2, m_\Phi^2; y_u ; A_u )
- g(1, 0, 4) , \\
%%%%%%%%%%%%%%%%%%%%%%%%%%%%%%%%%%%%%%%%%%
\ddt{m_{\bar{d}}^2} & = 
2 f (m_{\bar{Q}}^2, m_{\bar{d}}^2, m_{H_u}^2; \yukbar{d} ; \abar{d} )
+ f (m_{\bar{d}}^2, {\bf m}_d^2, m_\Phi^2; y_d ; A_d )
- g(1, 0, -2) , \\
%%%%%%%%%%%%%%%%%%%%%%%%%%%%%%%%%%%%%%%%%%
\ddt{m_{\bar{L}}^2} & = 
f (m_{\bar{L}}^2, m_{\bar{e}}^2, m_{H_u}^2; \yukbar{e}^* ; \abar{e}^* )
+ f (m_{\bar{L}}^2, {\bf m}_L^2, m_\Phi^2; y_L ; A_L )
- g(0, 1, 3) , \\
%%%%%%%%%%%%%%%%%%%%%%%%%%%%%%%%%%%%%%%%%%
\ddt{m_{\bar{e}}^2} & = 
2 f (m_{\bar{L}}^2, m_{\bar{e}}^2, m_{H_u}^2; \yukbar{e} ; \abar{e} )
+ f (m_{\bar{e}}^2, {\bf m}_e^2, m_\Phi^2; y_e ; A_e )
- g(0, 0, -6) , \\
%%%%%%%%%%%%%%%%%%%%%%%%%%%%%%%%%%%%%%%%%%
\ddt{m_{H_u}^2} & = 
\text{Tr} \left[ 3
f ({\bf m}_Q^2, {\bf m}_u^2, m_{H_u}^2; {\bm y}_u^\dagger ; \aterm{u}^\dagger )
\right]
+ 3 f (m_{\bar{Q}}^2, m_{\bar{d}}^2, m_{H_u}^2; \yukbar{d}^* ; \abar{d}^* ) 
\nonumber \\
& \hspace*{4cm}  
+ f (m_{\bar{L}}^2, m_{\bar{e}}^2, m_{H_u}^2; \yukbar{e}^* ; \abar{e}^* )
- g(0, 1, 3) , \\
%%%%%%%%%%%%%%%%%%%%%%%%%%%%%%%%%%%%%%%%%%
\ddt{m_{H_{d}}^2} & = 
\text{Tr} \left[ 3 
f ({\bf m}_Q^2, {\bf m}_d^2, m_{H_d}^2; {\bm y}_d^\dagger; \aterm{d}^\dagger )
+ f ({\bf m}_L^2, {\bf m}_e^2, m_{H_d}^2; {\bm y}_e^\dagger; \aterm{e}^\dagger )
\right]  \nonumber \\
& \hspace{4cm}
+ 3 f (m_{\bar{Q}}^2, m_{\bar{u}}^2, m_{H_d}^2; \yukbar{u}^* ; \abar{u}^* )
- g (0, 1, -3) , \\
%%%%%%%%%%%%%%%%%%%%%%%%%%%%%%%%%%%%%%%%%%
 \ddt{m_\Phi^2} & = 
% \sum_{X = Q, u, d, L, e}
12 f (m_{\bar{Q}}^2, {\bf m}_Q^2, m_\Phi^2; y_Q ; A_Q ) 
+6 f (m_{\bar{u}}^2, {\bf m}_u^2, m_\Phi^2; y_u ; A_u )  \nonumber \\
& \hspace*{2cm}  
+6 f (m_{\bar{d}}^2, {\bf m}_d^2, m_\Phi^2; y_d ; A_d )
+4 f (m_{\bar{L}}^2, {\bf m}_L^2, m_\Phi^2; y_L ; A_L )   \nonumber \\
& \hspace*{2cm}  
+2 f (m_{\bar{e}}^2, {\bf m}_e^2, m_\Phi^2; y_e ; A_e ) 
+  f (m_\Phi^2, m_\Phi^2, m_\Phi^2; y; A_y ) .
\end{align}


\begin{thebibliography}{99}
\bibitem{Aad:2012tfa}
  G.~Aad {\it et al.} [ATLAS Collaboration],
  %``Observation of a new particle in the search for the Standard Model Higgs boson with the ATLAS detector at the LHC,''
  Phys.\ Lett.\ B {\bf 716} (2012) 1
  doi:10.1016/j.physletb.2012.08.020
  [arXiv:1207.7214 [hep-ex]];
  %%CITATION = doi:10.1016/j.physletb.2012.08.020;%%
  %6620 citations counted in INSPIRE as of 08 Nov 2016
  S.~Chatrchyan {\it et al.} [CMS Collaboration],
  %``Observation of a new boson at a mass of 125 GeV with the CMS experiment at the LHC,''
  Phys.\ Lett.\ B {\bf 716}, 30 (2012)
  doi:10.1016/j.physletb.2012.08.021
  [arXiv:1207.7235 [hep-ex]].
  %%CITATION = doi:10.1016/j.physletb.2012.08.021;%%
  %6462 citations counted in INSPIRE as of 08 Nov 2016

\bibitem{Aad:2015zhl} 
  G.~Aad {\it et al.} [ATLAS and CMS Collaborations],
  %``Combined Measurement of the Higgs Boson Mass in $pp$ Collisions at $\sqrt{s}=7$ and 8 TeV with the ATLAS and CMS Experiments,''
  Phys.\ Rev.\ Lett.\  {\bf 114}, 191803 (2015)
  doi:10.1103/PhysRevLett.114.191803
  [arXiv:1503.07589 [hep-ex]].
  %%CITATION = doi:10.1103/PhysRevLett.114.191803;%%
  %564 citations counted in INSPIRE as of 11 Nov 2016

\bibitem{Bennett:2006fi} 
G.~W.~Bennett {\it et al.}  [Muon g-2 Collaboration],
%``Final Report of the Muon E821 Anomalous Magnetic Moment Measurement
%at BNL,''
Phys.\ Rev.\ D {\bf 73} (2006) 072003 
[hep-ex/0602035].
%%CITATION = HEP-EX/0602035;%%

\bibitem{hagiwara:2007}
K.~Hagiwara, R.~Liao, A.~D.~Martin, D.~Nomura and T.~Teubner,
%``(g-2)_mu and alpha(M_Z^2) re-evaluated using new precise data,''
J.\ Phys.\ G {\bf 38} (2011) 085003
[arXiv:1105.3149 [hep-ph]].
%%CITATION = ARXIV:1105.3149;%%
  
\bibitem{Nishida:2016lyk} 
  M.~Nishida and K.~Yoshioka,
  %``Higgs Boson Mass and Muon g-2 with Strongly Coupled Vector-like Generations,''
  arXiv:1605.06675 [hep-ph].
  %%CITATION = ARXIV:1605.06675;%%
  %3 citations counted in INSPIRE as of 07 Nov 2016  
  
%\cite{Agashe:2014kda}
\bibitem{Agashe:2014kda} 
  K.~A.~Olive {\it et al.} [Particle Data Group],
  %``Review of Particle Physics,''
  Chin.\ Phys.\ C {\bf 38}, 090001 (2014).
  doi:10.1088/1674-1137/38/9/090001
  %%CITATION = doi:10.1088/1674-1137/38/9/090001;%%
  %5728 citations counted in INSPIRE as of 21 Jan 2017  
  

%\cite{Oh:2015xoa}
\bibitem{Oh:2015xoa} 
  S.~H.~Oh {\it et al.},
  %``High-resolution mass models of dwarf galaxies from LITTLE THINGS,''
  Astron.\ J.\  {\bf 149}, 180 (2015)
  doi:10.1088/0004-6256/149/6/180
  [arXiv:1502.01281 [astro-ph.GA]].
  %%CITATION = doi:10.1088/0004-6256/149/6/180;%%
  %26 citations counted in INSPIRE as of 13 Nov 2016
  
%\cite{Markevitch:2001ri}
\bibitem{Markevitch:2001ri} 
  M.~Markevitch, A.~H.~Gonzalez, L.~David, A.~Vikhlinin, S.~Murray, W.~Forman, C.~Jones and W.~Tucker,
  %``A Textbook example of a bow shock in the merging galaxy cluster 1E0657-56,''
  Astrophys.\ J.\  {\bf 567}, L27 (2002)
  doi:10.1086/339619
  [astro-ph/0110468].
  %%CITATION = doi:10.1086/339619;%%
  %301 citations counted in INSPIRE as of 13 Nov 2016

%\cite{Ade:2015xua}
\bibitem{Ade:2015xua} 
  P.~A.~R.~Ade {\it et al.} [Planck Collaboration],
  %``Planck 2015 results. XIII. Cosmological parameters,''
  Astron.\ Astrophys.\  {\bf 594}, A13 (2016)
  doi:10.1051/0004-6361/201525830
  [arXiv:1502.01589 [astro-ph.CO]].
  %%CITATION = doi:10.1051/0004-6361/201525830;%%
  %2397 citations counted in INSPIRE as of 13 Nov 2016

\bibitem{Froggatt:1978nt} 
  C.~D.~Froggatt and H.~B.~Nielsen,
  %``Hierarchy of Quark Masses, Cabibbo Angles and CP Violation,''
  Nucl.\ Phys.\ B {\bf 147}, 277 (1979).
  doi:10.1016/0550-3213(79)90316-X
  %%CITATION = doi:10.1016/0550-3213(79)90316-X;%%
  %1431 citations counted in INSPIRE as of 07 Nov 2016

%\cite{Leurer:1993gy}
\bibitem{Leurer:1993gy} 
  M.~Leurer, Y.~Nir and N.~Seiberg,
  %``Mass matrix models: The Sequel,''
  Nucl.\ Phys.\ B {\bf 420}, 468 (1994)
  doi:10.1016/0550-3213(94)90074-4
  [hep-ph/9310320];
  %%CITATION = doi:10.1016/0550-3213(94)90074-4;%%
  %414 citations counted in INSPIRE as of 13 Nov 2016
  %\cite{Leurer:1992wg}
%\bibitem{Leurer:1992wg} 
  M.~Leurer, Y.~Nir and N.~Seiberg,
  %``Mass matrix models,''
  Nucl.\ Phys.\ B {\bf 398}, 319 (1993)
  doi:10.1016/0550-3213(93)90112-3
  [hep-ph/9212278].
  %%CITATION = doi:10.1016/0550-3213(93)90112-3;%%
  %315 citations counted in INSPIRE as of 13 Nov 2016

 \bibitem{Olive:2016xmw} 
  C.~Patrignani {\it et al.} [Particle Data Group Collaboration],
  %``Review of Particle Physics,''
  Chin.\ Phys.\ C {\bf 40}, no. 10, 100001 (2016).
  doi:10.1088/1674-1137/40/10/100001
  %%CITATION = doi:10.1088/1674-1137/40/10/100001;%%
  %44 citations counted in INSPIRE as of 10 Nov 2016

 %\cite{Green:1984sg}
\bibitem{Green:1984sg} 
  M.~B.~Green and J.~H.~Schwarz,
  %``Anomaly Cancellation in Supersymmetric D=10 Gauge Theory and Superstring Theory,''
  Phys.\ Lett.\  {\bf 149B}, 117 (1984).
  doi:10.1016/0370-2693(84)91565-X
  %%CITATION = doi:10.1016/0370-2693(84)91565-X;%%
  %2474 citations counted in INSPIRE as of 11 Nov 2016

%\cite{Fischler:1981zk}
\bibitem{Fischler:1981zk} 
  W.~Fischler, H.~P.~Nilles, J.~Polchinski, S.~Raby and L.~Susskind,
  %``Vanishing Renormalization of the D Term in Supersymmetric U(1) Theories,''
  Phys.\ Rev.\ Lett.\  {\bf 47}, 757 (1981).
  doi:10.1103/PhysRevLett.47.757
  %%CITATION = doi:10.1103/PhysRevLett.47.757;%%
  %196 citations counted in INSPIRE as of 11 Nov 2016

%\cite{Dine:1987xk}
\bibitem{Dine:1987xk} 
  M.~Dine, N.~Seiberg and E.~Witten,
  %``Fayet-Iliopoulos Terms in String Theory,''
  Nucl.\ Phys.\ B {\bf 289}, 589 (1987);
  doi:10.1016/0550-3213(87)90395-6
  %%CITATION = doi:10.1016/0550-3213(87)90395-6;%%
  %644 citations counted in INSPIRE as of 11 Nov 2016
%
%\cite{Atick:1987gy}
%\bibitem{Atick:1987gy} 
  J.~J.~Atick, L.~J.~Dixon and A.~Sen,
  %``String Calculation of Fayet-Iliopoulos d Terms in Arbitrary Supersymmetric Compactifications,''
  Nucl.\ Phys.\ B {\bf 292}, 109 (1987);
  doi:10.1016/0550-3213(87)90639-0
  %%CITATION = doi:10.1016/0550-3213(87)90639-0;%%
  %556 citations counted in INSPIRE as of 11 Nov 2016
%
%\cite{Dine:1987gj}
%\bibitem{Dine:1987gj} 
  M.~Dine, I.~Ichinose and N.~Seiberg,
  %``F Terms and d Terms in String Theory,''
  Nucl.\ Phys.\ B {\bf 293}, 253 (1987).
  doi:10.1016/0550-3213(87)90072-1
  %%CITATION = doi:10.1016/0550-3213(87)90072-1;%%
  %393 citations counted in INSPIRE as of 11 Nov 2016


%\cite{Blumenhagen:2006ci}
\bibitem{Blumenhagen:2006ci} 
  R.~Blumenhagen, B.~Kors, D.~Lust and S.~Stieberger,
  %``Four-dimensional String Compactifications with D-Branes, Orientifolds and Fluxes,''
  Phys.\ Rept.\  {\bf 445}, 1 (2007)
  doi:10.1016/j.physrep.2007.04.003
  [hep-th/0610327].
  %%CITATION = doi:10.1016/j.physrep.2007.04.003;%%
  %623 citations counted in INSPIRE as of 11 Nov 2016

%\cite{Dreiner:2003hw}
\bibitem{Dreiner:2003hw} 
  H.~K.~Dreiner and M.~Thormeier,
  %``Supersymmetric Froggatt-Nielsen models with baryon and lepton number violation,''
  Phys.\ Rev.\ D {\bf 69}, 053002 (2004)
  doi:10.1103/PhysRevD.69.053002
  [hep-ph/0305270];
  %%CITATION = doi:10.1103/PhysRevD.69.053002;%%
  %60 citations counted in INSPIRE as of 11 Nov 2016
%
%\cite{Dreiner:2003yr}
%bibitem{Dreiner:2003yr} 
  H.~K.~Dreiner, H.~Murayama and M.~Thormeier,
  %``Anomalous flavor U(1)(X) for everything,''
  Nucl.\ Phys.\ B {\bf 729}, 278 (2005)
  doi:10.1016/j.nuclphysb.2005.08.047
  [hep-ph/0312012].
  %%CITATION = doi:10.1016/j.nuclphysb.2005.08.047;%%
  %60 citations counted in INSPIRE as of 11 Nov 201

%\cite{Blumenhagen:2003jy}
\bibitem{Blumenhagen:2003jy}
  R.~Blumenhagen, D.~Lust and S.~Stieberger,
  %``Gauge unification in supersymmetric intersecting brane worlds,''
  JHEP {\bf 0307}, 036 (2003)
  doi:10.1088/1126-6708/2003/07/036
  [hep-th/0305146].
  %%CITATION = doi:10.1088/1126-6708/2003/07/036;%%
  %103 citations counted in INSPIRE as of 12 Nov 2016


%\cite{Kachru:2003aw}
\bibitem{Kachru:2003aw}
  S.~Kachru, R.~Kallosh, A.~D.~Linde and S.~P.~Trivedi,
  %``De Sitter vacua in string theory,''
  Phys.\ Rev.\ D {\bf 68}, 046005 (2003)
  doi:10.1103/PhysRevD.68.046005
  [hep-th/0301240].
  %%CITATION = doi:10.1103/PhysRevD.68.046005;%%
  %2334 citations counted in INSPIRE as of 12 Nov 2016

%\cite{Choi:2006xt}
\bibitem{Choi:2006xt}
  K.~Choi,
  %``Dynamical gauge coupling unification from moduli stabilization,''
  Phys.\ Lett.\ B {\bf 642}, 404 (2006)
  doi:10.1016/j.physletb.2006.10.006
  [hep-th/0606104].
  %%CITATION = doi:10.1016/j.physletb.2006.10.006;%%
  %4 citations counted in INSPIRE as of 12 Nov 2016

%\cite{Bando:1996in}
\bibitem{Bando:1996in} 
  M.~Bando, J.~Sato, T.~Onogi and T.~Takeuchi,
  %``Predictions of m(b) / m(tau) and m(t) in an asymptotically nonfree theory,''
  Phys.\ Rev.\ D {\bf 56}, 1589 (1997)
  doi:10.1103/PhysRevD.56.1589
  [hep-ph/9612493];
  %%CITATION = doi:10.1103/PhysRevD.56.1589;%%
  %9 citations counted in INSPIRE as of 13 Nov 2016
%  \bibitem{Bando:1997dg} 
M.~Bando, J.~Sato and K.~Yoshioka,
%``Infrared fixed points in an asymptotically nonfree theory,''
Prog.\ Theor.\ Phys.\ {\bf 98} (1997) 169 
[hep-ph/9703321];
%%CITATION = HEP-PH/9703321;%%
M.~Bando, T.~Kobayashi, T.~Noguchi and K.~Yoshioka,
%``Yukawa hierarchy from extra dimensions and infrared fixed points,''
Phys.\ Lett.\ B {\bf 480} (2000) 187 
[hep-ph/0002102];
%%CITATION = HEP-PH/0002102;%%
%M.~Bando, T.~Kobayashi, T.~Noguchi and K.~Yoshioka,
%``Fermion mass hierarchies and small mixing angles from extra dimensions,''
Phys.\ Rev.\ D {\bf 63} (2001) 113017 
[hep-ph/0008120].
%%CITATION = HEP-PH/0008120;%%

\bibitem{Xing:2007fb} 
  Z.~z.~Xing, H.~Zhang and S.~Zhou,
  %``Updated Values of Running Quark and Lepton Masses,''
  Phys.\ Rev.\ D {\bf 77}, 113016 (2008)
  doi:10.1103/PhysRevD.77.113016
  [arXiv:0712.1419 [hep-ph]].
  %%CITATION = doi:10.1103/PhysRevD.77.113016;%%
  %281 citations counted in INSPIRE as of 10 Nov 2016

%\cite{Ema:2016ops}
\bibitem{Ema:2016ops} 
  Y.~Ema, K.~Hamaguchi, T.~Moroi and K.~Nakayama,
  %``Flaxion: a minimal extension to solve puzzles in the standard model,''
  arXiv:1612.05492 [hep-ph].
  %%CITATION = ARXIV:1612.05492;%%
  %1 citations counted in INSPIRE as of 11 Jan 2017

%\cite{Adler:2008zza}
\bibitem{Adler:2008zza} 
  S.~Adler {\it et al.} [E949 and E787 Collaborations],
  %``Measurement of the K+ --> pi+ nu nu branching ratio,''
  Phys.\ Rev.\ D {\bf 77}, 052003 (2008)
  doi:10.1103/PhysRevD.77.052003
  [arXiv:0709.1000 [hep-ex]].
  %%CITATION = doi:10.1103/PhysRevD.77.052003;%%
  %61 citations counted in INSPIRE as of 06 Jan 2017


%\cite{Turner:1985si}
\bibitem{Turner:1985si} 
  M.~S.~Turner,
  %``Cosmic and Local Mass Density of Invisible Axions,''
  Phys.\ Rev.\ D {\bf 33}, 889 (1986).
  doi:10.1103/PhysRevD.33.889
  %%CITATION = doi:10.1103/PhysRevD.33.889;%%
  %376 citations counted in INSPIRE as of 11 Jan 2017

\bibitem{Roszkowski:1994tm} 
  L.~Roszkowski,
  %``A Simple way of calculating cosmological relic density,''
  Phys.\ Rev.\ D {\bf 50}, 4842 (1994)
  % doi:10.1103/PhysRevD.50.4842
  [hep-ph/9404227, hep-ph/9404227];
  %%CITATION = doi:10.1103/PhysRevD.50.4842;%%
  %18 citations counted in INSPIRE as of 07 Nov 2016
  J.~D.~Wells,
  %``Annihilation cross-sections for relic densities in the low velocity limit,''
  hep-ph/9404219.
  %%CITATION = HEP-PH/9404219;%%
  %11 citations counted in INSPIRE as of 07 Nov 2016

%%%%%% Effective potential Higgs mass
%\cite{Coleman:1973jx} 
\bibitem{Coleman:1973jx} 
S.~R.~Coleman and E.~J.~Weinberg,
%``Radiative Corrections as the Origin of Spontaneous Symmetry Breaking,''
Phys.\ Rev.\ D {\bf 7} (1973) 1888.
%%CITATION = PHRVA,D7,1888;%%

%%%%%%MSSM Higgs mass correction
%\cite{Okada:1990gg}
\bibitem{Okada:1990gg}
Y.~Okada, M.~Yamaguchi and T.~Yanagida,
%``Renormalization group analysis on the Higgs mass in the softly
%broken supersymmetric standard model,'' 
Phys.\ Lett.\ B {\bf 262} (1991) 54.
% doi:10.1016/0370-2693(91)90642-4
%%CITATION = doi:10.1016/0370-2693(91)90642-4;%%  

%%%%%%%%
%%%%%%%% vector-generation correction
%\cite{Moroi:1991mg}
\bibitem{Moroi:1991mg}
T.~Moroi and Y.~Okada,
%``Radiative corrections to Higgs masses in the supersymmetric model
%with an extra family and antifamily,'' 
Mod.\ Phys.\ Lett.\ A {\bf 7} (1992) 187;
%%CITATION = MPLAE,A7,187;%%
%T.~Moroi and Y.~Okada,
%``Upper bound of the lightest neutral Higgs mass in extended
%supersymmetric Standard Models,''
Phys.\ Lett.\ B {\bf 295} (1992) 73;
%%CITATION = PHLTA,B295,73;%%
K.~S.~Babu, I.~Gogoladze and C.~Kolda,
%``Perturbative unification and Higgs boson mass bounds,''
hep-ph/0410085;
%%CITATION = HEP-PH/0410085;%%
K.~S.~Babu, I.~Gogoladze, M.~U.~Rehman and Q.~Shafi,
%``Higgs Boson Mass, Sparticle Spectrum and Little Hierarchy Problem
%in Extended MSSM,'' 
Phys.\ Rev.\ D {\bf 78} (2008) 055017
[arXiv:0807.3055 [hep-ph]];
%%CITATION = ARXIV:0807.3055;%%

\bibitem{Martin:2009bg} 
S.~P.~Martin,
%``Extra vector-like matter and the lightest Higgs scalar boson mass
%in low-energy supersymmetry,'' 
Phys.\ Rev.\ D {\bf 81} (2010) 035004
[arXiv:0910.2732 [hep-ph]].
%%CITATION = ARXIV:0910.2732;%%

% %%%%%%MSSM Higgs mass correction
% %\cite{Okada:1990gg}
% \bibitem{Okada:1990gg}
% Y.~Okada, M.~Yamaguchi and T.~Yanagida,
% %``Renormalization group analysis on the Higgs mass in the softly
% %broken supersymmetric standard model,'' 
% Phys.\ Lett.\ B {\bf 262} (1991) 54.
% % doi:10.1016/0370-2693(91)90642-4
% %%CITATION = doi:10.1016/0370-2693(91)90642-4;%%  

% %%%%%%%%
% %%%%%%%% vector-generation correction
% %\cite{Moroi:1991mg}
% \bibitem{Moroi:1991mg}
% T.~Moroi and Y.~Okada,
% %``Radiative corrections to Higgs masses in the supersymmetric model
% %with an extra family and antifamily,'' 
% Mod.\ Phys.\ Lett.\ A {\bf 7} (1992) 187;
% %%CITATION = MPLAE,A7,187;%%
% %T.~Moroi and Y.~Okada,
% %``Upper bound of the lightest neutral Higgs mass in extended
% %supersymmetric Standard Models,''
% Phys.\ Lett.\ B {\bf 295} (1992) 73;
% %%CITATION = PHLTA,B295,73;%%
% K.~S.~Babu, I.~Gogoladze and C.~Kolda,
% %``Perturbative unification and Higgs boson mass bounds,''
% hep-ph/0410085;
% %%CITATION = HEP-PH/0410085;%%
% K.~S.~Babu, I.~Gogoladze, M.~U.~Rehman and Q.~Shafi,
% %``Higgs Boson Mass, Sparticle Spectrum and Little Hierarchy Problem
% %in Extended MSSM,'' 
% Phys.\ Rev.\ D {\bf 78} (2008) 055017
% [arXiv:0807.3055 [hep-ph]];
% %%CITATION = ARXIV:0807.3055;%%
% %\bibitem{Martin:2009bg} 
% S.~P.~Martin,
% %``Extra vector-like matter and the lightest Higgs scalar boson mass
% %in low-energy supersymmetry,'' 
% Phys.\ Rev.\ D {\bf 81} (2010) 035004
% [arXiv:0910.2732 [hep-ph]].
% %%CITATION = ARXIV:0910.2732;%%

%%%%%%%MSSM
%\cite{Lopez:1993vi}
\bibitem{Lopez:1993vi} 
J.~L.~Lopez, D.~V.~Nanopoulos and X.~Wang,
%``Large (g-2)-mu in SU(5) x U(1) supergravity models,''
Phys.\ Rev.\ D {\bf 49} (1994) 366 
[hep-ph/9308336];
%%CITATION = HEP-PH/9308336;%%
T.~Moroi,
%``The Muon anomalous magnetic dipole moment in the minimal
%supersymmetric standard model,'' 
Phys.\ Rev.\ D {\bf 53} (1996) 6565 
%[Phys.\ Rev.\ D {\bf 56}, 4424 (1997)]
[hep-ph/9512396];
%%CITATION = HEP-PH/9512396;%%
M.~Carena, G.~F.~Giudice and C.~E.~M.~Wagner,
%``Constraints on supersymmetric models from the muon anomalous
%magnetic moment,''
Phys.\ Lett.\ B {\bf 390} (1997) 234 
[hep-ph/9610233].
%%CITATION = HEP-PH/9610233;%%
%\bibitem{Maiani:1977cg} 
L.~Maiani, G.~Parisi and R.~Petronzio,
%``Bounds on the Number and Masses of Quarks and Leptons,''
Nucl.\ Phys.\ B {\bf 136} (1978) 115 ;
%%CITATION = NUPHA,B136,115;%%
S.~Theisen, N.~D.~Tracas and G.~Zoupanos,
%``Unification of Coupling Constants Without a Covering {GUT},''
Z.\ Phys.\ C {\bf 37} (1988) 597;
%%CITATION = ZEPYA,C37,597;%%
D.~Ghilencea, M.~Lanzagorta and G.~G.~Ross,
%``Strong unification,''
Phys.\ Lett.\ B {\bf 415} (1997) 253 
[hep-ph/9707462].
%%CITATION = HEP-PH/9707462;%%

%% vector-like non-SUSY contributions (g-2)_mu
%\cite{Dermisek:2013gta}
\bibitem{Dermisek:2013gta}
  R.~Dermisek and A.~Raval,
  %``Explanation of the Muon g-2 Anomaly with Vectorlike Leptons and its Implications for Higgs Decays,''
  Phys.\ Rev.\ D {\bf 88} (2013) 013017
  doi:10.1103/PhysRevD.88.013017
  [arXiv:1305.3522 [hep-ph]].
  %%CITATION = doi:10.1103/PhysRevD.88.013017;%%
  %33 citations counted in INSPIRE as of 11 Nov 2016

% %\cite{Dermisek:2013gta}
% \bibitem{Dermisek:2013gta}
%   R.~Dermisek and A.~Raval,
%   %``Explanation of the Muon g-2 Anomaly with Vectorlike Leptons and its Implications for Higgs Decays,''
%   Phys.\ Rev.\ D {\bf 88} (2013) 013017
%   doi:10.1103/PhysRevD.88.013017
%   [arXiv:1305.3522 [hep-ph]].
%   %%CITATION = doi:10.1103/PhysRevD.88.013017;%%
%   %33 citations counted in INSPIRE as of 11 Nov 2016
%   %\cite{Ishiwata:2013gma}
% %\bibitem{Ishiwata:2013gma} 
%   K.~Ishiwata and M.~B.~Wise,
%   %``Phenomenology of heavy vectorlike leptons,''
%   Phys.\ Rev.\ D {\bf 88}, no. 5, 055009 (2013)
%   doi:10.1103/PhysRevD.88.055009
%   [arXiv:1307.1112 [hep-ph]].
%   %%CITATION = doi:10.1103/PhysRevD.88.055009;%%
%   %16 citations counted in INSPIRE as of 13 Nov 2016

  %%%%%%%%%%%%%%%%%%% MSUGRA
%\cite{Chamseddine:1982jx}
\bibitem{Chamseddine:1982jx}
A.~H.~Chamseddine, R.~L.~Arnowitt and P.~Nath,
%``Locally Supersymmetric Grand Unification,''
Phys.\ Rev.\ Lett.\  {\bf 49} (1982) 970;
%doi:10.1103/PhysRevLett.49.970
%%CITATION = doi:10.1103/PhysRevLett.49.970;%%
R.~Barbieri, S.~Ferrara and C.~A.~Savoy,
%``Gauge Models with Spontaneously Broken Local Supersymmetry,''
Phys.\ Lett.\ B {\bf 119} (1982) 343;
%doi:10.1016/0370-2693(82)90685-2
%%CITATION = doi:10.1016/0370-2693(82)90685-2;%%
L.~J.~Hall, J.~D.~Lykken and S.~Weinberg,
%``Supergravity as the Messenger of Supersymmetry Breaking,''
Phys.\ Rev.\ D {\bf 27} (1983) 2359.
%doi:10.1103/PhysRevD.27.2359
%%CITATION = doi:10.1103/PhysRevD.27.2359;%%

%%%%% Higgs mass two loop correction
%\cite{Heinemeyer:1998yj}
\bibitem{Heinemeyer:1998yj} 
  S.~Heinemeyer, W.~Hollik and G.~Weiglein,
  %``FeynHiggs: A Program for the calculation of the masses of the neutral CP even Higgs bosons in the MSSM,''
  Comput.\ Phys.\ Commun.\  {\bf 124}, 76 (2000)
  doi:10.1016/S0010-4655(99)00364-1
  [hep-ph/9812320].
  %%CITATION = doi:10.1016/S0010-4655(99)00364-1;%%
  %1045 citations counted in INSPIRE as of 11 Feb 2017
 
  %\cite{Heinemeyer:1998np}
\bibitem{Heinemeyer:1998np} 
  S.~Heinemeyer, W.~Hollik and G.~Weiglein,
  %``The Masses of the neutral CP - even Higgs bosons in the MSSM: Accurate analysis at the two loop level,''
  Eur.\ Phys.\ J.\ C {\bf 9}, 343 (1999)
  doi:10.1007/s100529900006, 10.1007/s100520050537
  [hep-ph/9812472].
  %%CITATION = doi:10.1007/s100529900006, 10.1007/s100520050537;%%
  %1007 citations counted in INSPIRE as of 11 Feb 2017
  
%\cite{Carena:2000dp}
\bibitem{Carena:2000dp} 
  M.~Carena, H.~E.~Haber, S.~Heinemeyer, W.~Hollik, C.~E.~M.~Wagner and G.~Weiglein,
  %``Reconciling the two loop diagrammatic and effective field theory computations of the mass of the lightest CP - even Higgs boson in the MSSM,''
  Nucl.\ Phys.\ B {\bf 580}, 29 (2000)
  doi:10.1016/S0550-3213(00)00212-1
  [hep-ph/0001002].
  %%CITATION = doi:10.1016/S0550-3213(00)00212-1;%%
  %364 citations counted in INSPIRE as of 11 Feb 2017  


%\cite{Degrassi:2002fi}
\bibitem{Degrassi:2002fi} 
  G.~Degrassi, S.~Heinemeyer, W.~Hollik, P.~Slavich and G.~Weiglein,
  %``Towards high precision predictions for the MSSM Higgs sector,''
  Eur.\ Phys.\ J.\ C {\bf 28}, 133 (2003)
  doi:10.1140/epjc/s2003-01152-2
  [hep-ph/0212020].
  %%CITATION = doi:10.1140/epjc/s2003-01152-2;%%
  %860 citations counted in INSPIRE as of 11 Feb 2017

%\cite{Brignole:2002bz}
\bibitem{Brignole:2002bz} 
  A.~Brignole, G.~Degrassi, P.~Slavich and F.~Zwirner,
  %``On the two loop sbottom corrections to the neutral Higgs boson masses in the MSSM,''
  Nucl.\ Phys.\ B {\bf 643}, 79 (2002)
  doi:10.1016/S0550-3213(02)00748-4
  [hep-ph/0206101].
  %%CITATION = doi:10.1016/S0550-3213(02)00748-4;%%
  %232 citations counted in INSPIRE as of 21 May 2017

%\cite{Frank:2006yh}
\bibitem{Frank:2006yh} 
  M.~Frank, T.~Hahn, S.~Heinemeyer, W.~Hollik, H.~Rzehak and G.~Weiglein,
  %``The Higgs Boson Masses and Mixings of the Complex MSSM in the Feynman-Diagrammatic Approach,''
  JHEP {\bf 0702}, 047 (2007)
  doi:10.1088/1126-6708/2007/02/047
  [hep-ph/0611326].
  %%CITATION = doi:10.1088/1126-6708/2007/02/047;%%
  %643 citations counted in INSPIRE as of 11 Feb 2017  


%\cite{Hahn:2013ria}
\bibitem{Hahn:2013ria} 
  T.~Hahn, S.~Heinemeyer, W.~Hollik, H.~Rzehak and G.~Weiglein,
  %``High-Precision Predictions for the Light CP -Even Higgs Boson Mass of the Minimal Supersymmetric Standard Model,''
  Phys.\ Rev.\ Lett.\  {\bf 112}, no. 14, 141801 (2014)
  doi:10.1103/PhysRevLett.112.141801
  [arXiv:1312.4937 [hep-ph]].
  %%CITATION = doi:10.1103/PhysRevLett.112.141801;%%
  %189 citations counted in INSPIRE as of 11 Feb 2017
  
%\cite{Bagnaschi:2014rsa}
\bibitem{Bagnaschi:2014rsa} 
  E.~Bagnaschi, G.~F.~Giudice, P.~Slavich and A.~Strumia,
  %``Higgs Mass and Unnatural Supersymmetry,''
  JHEP {\bf 1409}, 092 (2014)
  doi:10.1007/JHEP09(2014)092
  [arXiv:1407.4081 [hep-ph]].
  %%CITATION = doi:10.1007/JHEP09(2014)092;%%
  %64 citations counted in INSPIRE as of 11 Feb 2017  
  
  %\cite{Vega:2015fna}
\bibitem{Vega:2015fna} 
  J.~Pardo Vega and G.~Villadoro,
  %``SusyHD: Higgs mass Determination in Supersymmetry,''
  JHEP {\bf 1507}, 159 (2015)
  doi:10.1007/JHEP07(2015)159
  [arXiv:1504.05200 [hep-ph]].
  %%CITATION = doi:10.1007/JHEP07(2015)159;%%
  %57 citations counted in INSPIRE as of 11 Feb 2017

%\cite{Yanagida:2016kag}
\bibitem{Yanagida:2016kag} 
  T.~T.~Yanagida, W.~Yin and N.~Yokozaki,
  %``Nambu-Goldstone Boson Hypothesis for Squarks and Sleptons in Pure Gravity Mediation,''
  JHEP {\bf 1609}, 086 (2016)
  doi:10.1007/JHEP09(2016)086
  [arXiv:1608.06618 [hep-ph]].
  %%CITATION = doi:10.1007/JHEP09(2016)086;%%
  %1 citations counted in INSPIRE as of 11 Feb 2017


%\cite{Choudhury:2017fuu}
\bibitem{Choudhury:2017fuu} 
  A.~Choudhury, L.~Darmé, L.~Roszkowski, E.~M.~Sessolo and S.~Trojanowski,
  %``Muon g-2 and related phenomenology in constrained vector-like extensions of the MSSM,''
  arXiv:1701.08778 [hep-ph].
  %%CITATION = ARXIV:1701.08778;%%

%\cite{Aubert:2009ag}
\bibitem{Aubert:2009ag} 
  B.~Aubert {\it et al.} [BaBar Collaboration],
  %``Searches for Lepton Flavor Violation in the Decays tau+- ---> e+- gamma and tau+- ---> mu+- gamma,''
  Phys.\ Rev.\ Lett.\  {\bf 104}, 021802 (2010)
  doi:10.1103/PhysRevLett.104.021802
  [arXiv:0908.2381 [hep-ex]].
  %%CITATION = doi:10.1103/PhysRevLett.104.021802;%%
  %300 citations counted in INSPIRE as of 11 Feb 2017

%\cite{Mori:2016vwi}
\bibitem{Mori:2016vwi} 
  T.~Mori [MEG Collaboration],
  %``Final Results of the MEG Experiment,''
  arXiv:1606.08168 [hep-ex].
  %%CITATION = ARXIV:1606.08168;%%
  %3 citations counted in INSPIRE as of 21 Jan 2017

%\cite{Kitano:2000zw}
\bibitem{Kitano:2000zw} 
  R.~Kitano and K.~Yamamoto,
  %``Lepton flavor violation in the supersymmetric standard model with vector like leptons,''
  Phys.\ Rev.\ D {\bf 62}, 073007 (2000)
  doi:10.1103/PhysRevD.62.073007
  [hep-ph/0003063].
  %%CITATION = doi:10.1103/PhysRevD.62.073007;%%
  %8 citations counted in INSPIRE as of 20 Jan 2017

%\cite{Ibrahim:2012ds}
\bibitem{Ibrahim:2012ds} 
  T.~Ibrahim and P.~Nath,
  %``$\tau\to \mu \gamma$ decay in extensions with a vectorlike generation,''
  Phys.\ Rev.\ D {\bf 87}, no. 1, 015030 (2013)
  doi:10.1103/PhysRevD.87.015030
  [arXiv:1211.0622 [hep-ph]].
  %%CITATION = doi:10.1103/PhysRevD.87.015030;%%
  %11 citations counted in INSPIRE as of 11 Feb 2017

%\cite{Ibrahim:2015hva}
\bibitem{Ibrahim:2015hva} 
  T.~Ibrahim, A.~Itani and P.~Nath,
  %``$\mu\to e \gamma$ decay in an MSSM extension,''
  Phys.\ Rev.\ D {\bf 92}, no. 1, 015003 (2015)
  doi:10.1103/PhysRevD.92.015003
  [arXiv:1503.01078 [hep-ph]].
  %%CITATION = doi:10.1103/PhysRevD.92.015003;%%
  %5 citations counted in INSPIRE as of 11 Feb 2017

\bibitem{Kaplan:1999ac} 
  D.~E.~Kaplan, G.~D.~Kribs and M.~Schmaltz,
  %``Supersymmetry breaking through transparent extra dimensions,''
  Phys.\ Rev.\ D {\bf 62}, 035010 (2000)
  doi:10.1103/PhysRevD.62.035010
  [hep-ph/9911293].
  %%CITATION = doi:10.1103/PhysRevD.62.035010;%%
  %437 citations counted in INSPIRE as of 11 Nov 2016

%\cite{Ibrahim:2014oia}
\bibitem{Ibrahim:2014oia} 
  T.~Ibrahim, A.~Itani and P.~Nath,
  %``Electron electric dipole moment as a sensitive probe of PeV scale physics,''
  Phys.\ Rev.\ D {\bf 90}, no. 5, 055006 (2014)
  doi:10.1103/PhysRevD.90.055006
  [arXiv:1406.0083 [hep-ph]].
  %%CITATION = doi:10.1103/PhysRevD.90.055006;%%
  %13 citations counted in INSPIRE as of 21 Jan 2017

%\cite{Baron:2013eja}
\bibitem{Baron:2013eja} 
  J.~Baron {\it et al.} [ACME Collaboration],
  %``Order of Magnitude Smaller Limit on the Electric Dipole Moment of the Electron,''
  Science {\bf 343}, 269 (2014)
  doi:10.1126/science.1248213
  [arXiv:1310.7534 [physics.atom-ph]].
  %%CITATION = doi:10.1126/science.1248213;%%
  %300 citations counted in INSPIRE as of 21 Jan 2017

%\cite{Belanger:2008sj}
\bibitem{Belanger:2008sj}
  G.~Belanger, F.~Boudjema, A.~Pukhov and A.~Semenov,
  %``Dark matter direct detection rate in a generic model with micrOMEGAs 2.2,''
  Comput.\ Phys.\ Commun.\  {\bf 180}, 747 (2009)
  doi:10.1016/j.cpc.2008.11.019
  [arXiv:0803.2360 [hep-ph]].
  %%CITATION = doi:10.1016/j.cpc.2008.11.019;%%
  %511 citations counted in INSPIRE as of 13 Nov 2016

%\cite{Aprile:2012zx}
\bibitem{Aprile:2012zx}
  E.~Aprile [XENON1T Collaboration],
  %``The XENON1T Dark Matter Search Experiment,''
  Springer Proc.\ Phys.\  {\bf 148} (2013) 93
  doi:10.1007/978-94-00707241-0-14
%  doi:10.1007/978-94-007-7241-0_14
  [arXiv:1206.6288 [astro-ph.IM]].
  %%CITATION = doi:10.1007/978-94-007-7241-0_14;%%
  %367 citations counted in INSPIRE as of 08 Nov 2016
    
  %\cite{Aalbers:2016jon}
\bibitem{Aalbers:2016jon} 
  J.~Aalbers {\it et al.} [DARWIN Collaboration],
  %``DARWIN: towards the ultimate dark matter detector,''
  arXiv:1606.07001 [astro-ph.IM].
  %%CITATION = ARXIV:1606.07001;%%
  %10 citations counted in INSPIRE as of 08 Nov 2016 
  
%\cite{Ahnen:2016qkx}
\bibitem{Ahnen:2016qkx} 
  M.~L.~Ahnen {\it et al.} [MAGIC and Fermi-LAT Collaborations],
  %``Limits to dark matter annihilation cross-section from a combined analysis of MAGIC and Fermi-LAT observations of dwarf satellite galaxies,''
  JCAP {\bf 1602}, no. 02, 039 (2016)
  doi:10.1088/1475-7516/2016/02/039
  [arXiv:1601.06590 [astro-ph.HE]].
  %%CITATION = doi:10.1088/1475-7516/2016/02/039;%%
  %29 citations counted in INSPIRE as of 08 Nov 2016


  %\cite{Ackermann:2015zua}
\bibitem{Ackermann:2015zua} 
  M.~Ackermann {\it et al.} [Fermi-LAT Collaboration],
  %``Searching for Dark Matter Annihilation from Milky Way Dwarf Spheroidal Galaxies with Six Years of Fermi Large Area Telescope Data,''
  Phys.\ Rev.\ Lett.\  {\bf 115}, no. 23, 231301 (2015)
  doi:10.1103/PhysRevLett.115.231301
  [arXiv:1503.02641 [astro-ph.HE]].
  %%CITATION = doi:10.1103/PhysRevLett.115.231301;%%
  %316 citations counted in INSPIRE as of 08 Nov 2016
  
  %\cite{Wood:2013taa}
\bibitem{Wood:2013taa} 
  M.~Wood, J.~Buckley, S.~Digel, S.~Funk, D.~Nieto and M.~A.~Sanchez-Conde,
  %``Prospects for Indirect Detection of Dark Matter with CTA,''
  arXiv:1305.0302 [astro-ph.HE].
  %%CITATION = ARXIV:1305.0302;%%
  %29 citations counted in INSPIRE as of 08 Nov 2016 
  
%%%%%%%%%%%%%%%%%%%%%%%%%%%%%%%%%%%%%%%
%%%This part is replaced to the first part of the reference

%%%Singlino dark matter detection
%%%Singlino dark matter detection
%\cite{Cerdeno:2004xw}
\bibitem{Cerdeno:2004xw} 
  D.~G.~Cerdeno, C.~Hugonie, D.~E.~Lopez-Fogliani, C.~Munoz and A.~M.~Teixeira,
  %``Theoretical predictions for the direct detection of neutralino dark matter in the NMSSM,''
  JHEP {\bf 0412}, 048 (2004)
  doi:10.1088/1126-6708/2004/12/048
  [hep-ph/0408102].
  %%CITATION = doi:10.1088/1126-6708/2004/12/048;%%
  %71 citations counted in INSPIRE as of 07 Jan 2017

%\cite{Ellwanger:2009dp}
\bibitem{Ellwanger:2009dp} 
  U.~Ellwanger, C.~Hugonie and A.~M.~Teixeira,
  %``The Next-to-Minimal Supersymmetric Standard Model,''
  Phys.\ Rept.\  {\bf 496}, 1 (2010)
  doi:10.1016/j.physrep.2010.07.001
  [arXiv:0910.1785 [hep-ph]].
  %%CITATION = doi:10.1016/j.physrep.2010.07.001;%%
  %703 citations counted in INSPIRE as of 07 Jan 2017

%\cite{Cerdeno:2007sn}
\bibitem{Cerdeno:2007sn} 
  D.~G.~Cerdeno, E.~Gabrielli, D.~E.~Lopez-Fogliani, C.~Munoz and A.~M.~Teixeira,
  %``Phenomenological viability of neutralino dark matter in the NMSSM,''
  JCAP {\bf 0706}, 008 (2007)
  doi:10.1088/1475-7516/2007/06/008
  [hep-ph/0701271 [HEP-PH]].
  %%CITATION = doi:10.1088/1475-7516/2007/06/008;%%
  %68 citations counted in INSPIRE as of 07 Jan 2017
%%%Singlino dark matter detection
%%%Singlino dark matter detection

\end{thebibliography}
\end{document}